Ministère de l'Enseignement Supérieur et de la Recherche Scientifique

Université Tunis El Manar

Institut Supérieur d'Informatique ISI# MEMOIRE DE MASTERE

Présenté en vue de l'obtention du diplôme de Mastère de Recherche

En Informatique

Option : Génie Logiciel

Par

## Mouadh AYACHI

### Etude de la Distribution de Calculs Creux sur une Grappe Multi-coeurs

Soutenu le 17/12/2015

| | |
|---|---|
| M<sup>me</sup> Najiba Mrabet Bellaaj | Présidente |
| M. Heithem Abbes | Rapporteur |
| M<sup>me</sup> Olfa Hamdi Larbi | Encadrante |

Réalisé au sein de l'unité de Recherche URAPOP

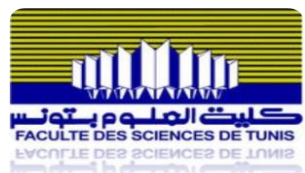

Unité de Recherche en Algorithmique Parallèle et Optimisations (URAPOP)

2015-2016

# Table des matières

# Liste des Figures

# Liste des Tableaux

# Introduction Générale

De nos jours, le calcul haute performance s'impose de plus en plus dans différents domaines de la recherche et de l'industrie, tels que l'imagerie et les diagnostics médicaux, les mathématiques financières ainsi que l'exploration pétrolière. Il fait référence au calcul intensif dans certaines applications où on a besoin d'utiliser un grand nombre de ressources de calcul (puissance de calcul, débit mémoire, espace de stockage, etc).

Ainsi, il s'avère nécessaire dans ce cas d'exécuter ces applications sur des architectures parallèles faisant coopérer plusieurs calculateurs et fonctionnant au-dessus de $10^{15}$ opérations à virgule flottante par seconde (ou un pétaflops).

Le calcul parallèle consiste à exécuter un ou plusieurs programmes, simultanément, par plusieurs processeurs. Nous avons, en général, deux manières pour réaliser un calcul parallèle. La première consiste à découper le programme en plusieurs taches de calcul puis, exécuter toutes ces taches en parallèle par différents processeurs. La seconde nécessite le partitionnement des données, de manière à ce que chaque partie des données soit attribuée à un processeur différent. Ensuite, tous les processeurs exécutent en parallèle les instructions du même programme mais en opérant sur des données différentes [Kek14]. Cette dernière méthode, appelée le parallélisme de données, est celle retenue dans ce mémoire.

L'analyse numérique est l'un des domaines où s'impose l'utilisation de plateformes parallèles, en particulier, dans le cas où les méthodes numériques traitent des matrices creuses. Une matrice creuse est une matrice de très grande taille qui contient une faible proportion d'éléments non nuls. Ces matrices peuvent être issues de différents domaines tels que la simulation de la mécanique des structures, le traitement d'image/signal, l'étude de la dynamique des fluides etc.

Ainsi, pour le traitement de ces matrices creuses de grandes tailles par des méthodes numériques, les systèmes distribués, tels que les grappes de calculs, peuvent prouver leur efficacité.

De nos jours, avec l'avènement des architectures multicœurs, la performance d'un processeur devient liée principalement au nombre de cœurs qu'il renferme. D'où, au niveau des grappes de calcul, la tendance n'est plus d'augmenter le nombre de nœuds uniquement, mais aussi d'accroitre le nombre de cœurs par processeur à l'intérieur de chaque nœud. Cette évolution induit une hiérarchisation croissante de la mémoire afin d'éviter l'importante contention sur le bus mémoire.

Le Produit matrice-Vecteur Creux (PMVC) constitue un noyau pour plusieurs méthodes numériques. La problématique traitée revient donc à distribuer ce noyau sur une grappe de nœuds multicœurs. Pour le faire, on est confronté à deux principaux problèmes : l'équilibrage des charges entre les unités de calcul (nœuds ou cœurs) et l'optimisation des communications. Une corrélation entre ces deux critères fait que généralement, l'optimisation de l'un peut dégrader l'autre. L'objectif est donc de trouver un compromis pour satisfaire ces deux critères. Ainsi, pour la distribution du PMVC, nous étudions une solution basée sur la combinaison de deux approches différentes pour la décomposition de la matrice creuse : i) le modèle hypergraphe qui optimise le volume des communications et ii) l'algorithme NEZGT (Nombre Equilibré de Zéros, Généralisé, Trié) qui assure l'équilibrage des charges entre les unités de calcul du système parallèle.

Notre travail décrivant cette mémoire s'articule autour de quatre chapitres comme suit :

Dans le premier chapitre, nous présentons d'abord les concepts de base des matrices creuses, leurs structures et les formats de compression.

Dans le second chapitre sont présentées les différentes architectures parallèles et distribuées telles que les grappes de PC.

Le troisième chapitre est consacré à la présentation du noyau de calcul le PMVC ainsi que d'un état de l'art sur les travaux connexes pour la distribution des données. Un intérêt particulier sera donné aux deux algorithmes sur lesquels se base notre travail, à savoir, Hypergraph et $NEZGT_{LIGNE}$ (Nombre Equilibré de nonZeros Généralisé Trié). Les deux algorithmes sont combinés pour une distribution efficace de la matrice creuse sur les nœuds de la grappe puis sur les cœurs au niveau de chaque nœud.

Dans le dernier chapitre nous proposons une combinaison de l'Hypergraphe avec une autre variante de l'algorithme NEZGT, à savoir, NEZGT$_{colonne}$. L'étude théorique est par la suite validée par une suite d'expérimentations sur la grille européenne Grid'5000.

Enfin, nous clôturons par une conclusion générale résumant les principaux résultats et présentant les perspectives de travail futures.

# Chapitre 1

CALCULS CREUX

# 1. Introduction

Au cours de ces dernières années, le calcul haut performance devient de plus en plus un orchestre dans différents domaines de recherche, tels que l'imagerie et les diagnostics médicaux, les mathématiques financières ainsi que l'exploration pétrolière. Il fait référence aux calculs intensifs des applications nécessitant des quantités énormes en ressources de calcul (puissance de calcul, débit mémoire, espace de stockage, etc.), pour une résolution efficace et rapide de différents problèmes scientifiques ou industriels.

Ainsi, ceci se traduit par l'exécution de ces applications sur des architectures parallèles, faisant coopérer plusieurs calculateurs et fonctionnant au-dessus de $10^{15}$ opérations à virgule flottante par seconde (ou un pétaflops). [Kho13].

Nous pouvons distinguer dans le calcul deux types soit séquentiel ou parallèle. Un calcul séquentiel consiste à exécuter un programme, instruction par instruction, par un seul processeur (unité de calcul) et de façon à ce qu'une seule instruction soit exécutée à la fois. En revanche, un calcul parallèle est défini comme l'exécution d'un ou plusieurs programmes, simultanément, par plusieurs processeurs. Nous avons, en général, deux manières de réaliser un calcul parallèle.

La première consiste à découper le programme en plusieurs taches de calcul puis, exécuter toutes ces taches en parallèle par différents processeurs.

La seconde nécessite le partitionnement des données du problème à traiter, de manière à ce que chaque partie de données soit attribuée à un processeur différent. Ensuite, tous les processeurs exécutent en parallèle les instructions du même programme mais en opérant sur des données différentes [Kek14].

Cette dernière méthode, appelée le parallélisme de données, est celle retenue dans nos travaux .Nous allons présenter dans la partie suivante la matrice creuse, les formats de compression, des exemples d'applications creuses et enfin quelques méthodes de résolution de systèmes linéaires et le noyau du calcul Produit Matrice-Vecteur Creux « PMVC ».

# 2. Matrices Creuses

## 2.1. Définition

Une matrice creuse est une matrice ayant une majorité d'éléments nuls lorsqu'il devient plus avantageux pour un algorithme d'éviter de réaliser son opération sur les éléments nuls de la matrice. Selon [Kek14] une matrice de taille N×N est qualifiée de creuse si elle contienne un nombre d'éléments non nuls de l'ordre de N.

Les matrices creuses sont souvent décrites à l'aide de la fraction des éléments nuls qu'elles contiennent, appelée vacuité (sparsity). À l'inverse, la fraction des éléments non nuls est appelée densité. Nous pouvons aussi définir la matrice creuse comme suit :

Soit A une matrice réelle d'ordre N (N est supposé assez élevé). Si le nombre d'éléments non nuls de A, noté NZ, est très petit (resp. grand), soit NZ=O(N) (resp. O(N²)), alors A est dite creuse (resp. dense) [HME07].

Les matrices creuses (Sparse en anglais) sont donc des matrices de grande taille ayant un nombre faible d'éléments non nuls par rapport à celui des éléments nuls [EHM06a].

## 2.2. Structures de matrices creuses

On peut distinguer deux structures particulières de matrices creuses la première s'appelle structure régulière alors que la deuxième est la structure irrégulière.

### a. Structures régulières

Une matrice creuse possède une structure régulière si tous les éléments non nuls sont regroupés dans la matrice de telle manière qu'nous pouvons y accéder facilement [EHM06a].

Nous pouvons citer en particulier:

- Matrices triangulaires : une matrice A est triangulaire supérieure (resp. inférieure) si $a_{ij}$=0 pour tout i ≥ j (resp. i ≤ j).

$$\begin{bmatrix} 1 & & \\ \vdots & 4 & \\ 2 & \cdots & 3 \end{bmatrix}$$

**Figure 0.1 : Matrice Triangulaire inférieur**

- Matrices bande constante : une matrice bande est une matrice creuse particulière dont les coefficients non nuls se regroupent autour de la diagonale [ALR]. Une matrice bande A de largeur 2m+1 est telle que m (dit demi largeur de bande) soit le plus petit entier tel que :

$$\forall a_{ij}=0, | i-j | > m , i, j = 1.. n, m < | (n-1)/2 |.$$

$$\begin{bmatrix} 1 & & \\ & \ddots & \\ & & 2 \end{bmatrix}$$

**Figure 0.2 : Matrice Bande constante**

- Matrices Hessenberg : une matrice est dite de Hessenberg supérieure (resp. inférieure) si $a_{ij}$=0 pour tout i > j+1 (resp. pour tout i < j+1).

$$\begin{bmatrix} 1 & 2 & \\ 3 & 4 & 7 \\ 5 & 6 & 7 \end{bmatrix}$$

**Figure 0.3 : Matrice Hessenberg inférieur**

- Matrices par bloc : une matrice par bloc est une matrice pouvant être divisée en un ensemble de sous-matrices de dimensions inférieures.

$$\begin{bmatrix} \begin{array}{cc|cc} 1 & 2 & 5 & 6 \\ 3 & 4 & 1 & 2 \\ \hline & & 7 & 5 \\ & & 3 & 4 \end{array} \end{bmatrix}$$

**Figure 0.4 : Matrice par Bloc**

## b. Structures irrégulières

Une matrice possède une structure irrégulière si les éléments sont éparpillés dans la matrice [EHM06a]. Nous pouvons citer comme exemple :

- Matrice bande variable : soit A une matrice N × N et soient $l_i$ et $u_j$.

Pour 1 ≤ i ≤ N, 1 ≤ j ≤ N définis par :

$$l_i=\text{i-min } \{j \,/a_{ij}\neq 0\}$$

$$u_j = j - \min\{i \,/\, a_{ij} \neq 0\}$$

Chaque $l_i$ et $u_j$ indique respectivement la (longueur de) demi-bande inférieure et supérieure dans la ligne i et la colonne j : $a_{ij} \neq 0$ (-uj ≤ i-j ≤ li).

La forme bande variable d'une matrice est définie par les vecteurs $\underline{u} = (u_i)$ et $\underline{l} = (l_i)$.

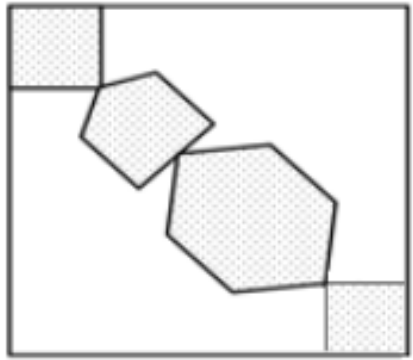

**Figure 0.5 : Matrice bande variable**

- Matrices quelconque : une matrice est dite quelconque si les éléments non nuls sont éparpillés de manière quelconque et ne forment pas une structure particulière.

$$\begin{bmatrix} 1 & 2 & \square & 1 & 2 & \square \\ 3 & \square & 7 & 3 & \square & 7 \\ 5 & 6 & \square & 5 & \square & 7 \\ \square & 2 & \square & 1 & 2 & \square \\ 3 & 4 & 7 & \square & \square & 7 \\ 5 & \square & 7 & 5 & 6 & \square \end{bmatrix}$$

**Figure 0.6 : Matrice quelconque**

## 2.3. Format de compression

Pour des raisons de taille mémoire et de performances de calcul, l'utilisateur est obligé d'exploiter la nature creuse de ces matrices et à utiliser un format de stockage des éléments non nuls uniquement.

Nous détaillons dans ce qui suit les formats COO, CSR, et CSC. En effet, ceux sont les formats les plus simples et les plus utilisés. Soit A une matrice creuse d'ordre N et contenant NNZ éléments non nuls.

COO (Coordinate) est un format de stockage particulièrement simple. Il permet de représenter une matrice creuse sous forme d'une structure de données composée de trois tableaux de tailles NNZ: un tableau de réels, Val, contenant tous les éléments non nuls de A et deux tableaux d'entiers, Lig et Col, contenant respectivement leurs indices de lignes et de colonnes.

$$A = \begin{bmatrix} a_{00} & 0 & 0 & a_{03} \\ 0 & 0 & a_{12} & 0 \\ a_{20} & a_{21} & a_{22} & 0 \\ 0 & a_{31} & 0 & a_{33} \end{bmatrix}$$

$$Val = [\ a_{00}\ \ a_{03}\ \ a_{12}\ \ a_{20}\ \ a_{21}\ \ a_{22}\ \ a_{31}\ \ a_{33}\ ]$$
$$Lig = [\ 0\ \ \ \ 0\ \ \ \ 1\ \ \ \ 2\ \ \ \ 2\ \ \ \ 2\ \ \ \ 3\ \ \ \ 3\ ]$$
$$Col = [\ 0\ \ \ \ 3\ \ \ \ 2\ \ \ \ 0\ \ \ \ 1\ \ \ \ 2\ \ \ \ 1\ \ \ \ 3\ ]$$

**Figure 0.7 : Format de Compression COO**

CSR (Compressed Sparse Row) et CSC (Compressed Sparse Column) sont les deux formats les plus connus et les plus utilisés pour le stockage des matrices creuses, vu qu'ils ne posent aucune condition sur la structure de celle-ci et ne permettent de stocker que les informations nécessaires sur la matrice. Comme COO, Ils permettent de représenter la matrice creuse sous forme de trois tableaux de données. Le format CSR (respectivement, CSC) stocke les « NNZ » éléments non nuls de la matrice et leurs indices de colonnes (respectivement, leurs indices de lignes) ligne par ligne (respectivement, colonne par colonne) dans deux tableaux distincts qu'on nomme « Val » et « Col » (respectivement, Val et Lig). Le troisième tableau « Ptr » d'une taille N+1, est utilisé pour stocker la position du premier élément non nul de chaque ligne de la matrice (respectivement, de chaque colonne de la matrice) dans val.

$$A = \begin{bmatrix} a_{00} & 0 & 0 & a_{03} \\ 0 & 0 & a_{12} & 0 \\ a_{20} & a_{21} & a_{22} & 0 \\ 0 & a_{31} & 0 & a_{33} \end{bmatrix}$$

$$CSR : \begin{cases} Val & = & [\ a_{00} \ \ a_{03} \ \ a_{12} \ \ a_{20} \ \ a_{21} \ \ a_{22} \ \ a_{31} \ \ a_{33} \ ] \\ Col & = & [\ \ 0 \ \ \ \ 3 \ \ \ \ 2 \ \ \ \ 0 \ \ \ \ 1 \ \ \ \ 2 \ \ \ \ 1 \ \ \ \ 3 \ \ ] \\ Ptr & = & [\ \ 0 \ \ \ \ \ \ \ \ \ \ 2 \ \ \ \ 3 \ \ \ \ \ \ \ \ \ \ \ \ 6 \ \ \ \ \ \ \ \ 8 \ ] \end{cases}$$

$$CSC : \begin{cases} Val & = & [\ a_{00} \ \ a_{20} \ \ a_{21} \ \ a_{31} \ \ a_{12} \ \ a_{22} \ \ a_{03} \ \ a_{33} \ ] \\ Lig & = & [\ \ 0 \ \ \ \ 2 \ \ \ \ 2 \ \ \ \ 3 \ \ \ \ 1 \ \ \ \ 2 \ \ \ \ 0 \ \ \ \ 3 \ \ ] \\ Ptr & = & [\ \ 0 \ \ \ \ \ \ \ \ \ \ 2 \ \ \ \ \ \ \ \ \ \ \ \ 4 \ \ \ \ \ \ \ \ 6 \ \ \ \ \ \ \ \ 8 \ ] \end{cases}$$

**Figure 0.8 : Format de Compression CSR et CSC**

Il existe aussi une variété de formats de stockage de matrices creuses tels que : BCCS (Block Compressed Column Storage format), BCRS (Block Compressed Row Storage format), BND (Linpack Banded format) ,BSR (Block Sparse Row format ) ,CCS (Compressed Column Storage format), COO (Coordinate format), CRS (Compressed Row Storage format), CSC (Compressed Sparse Column format) ,CSR (Compressed Sparse Row format), DIA (Diagonal format) , DNS (Dense format) , ELL (Ellpack-Itpack generalized diagonal format), JAD (Jagged Diagonal format) , LNK (Linked list storage format) , MSR (Modified Compressed Sparse Row format), NSK (Nonsymmetric Skyline format) ,SSK (Symmetric Skyline format), SSS (Symmetric Sparse Skyline format), USS (Unsymmetric Sparse Skyline format), VBR (Variable Block Row format)[Kho13].

## 3. Applications creuses

Les matrices creuses se trouvent au cœur d'un ensemble varié d'applications dans divers domaines. Citons à titre d'exemple : le calcul scientifique, l'ingénierie, la modélisation économique, la recherche de l'information, la simulation de la mécanique des structures, l'étude de la dynamique des fluides, le traitement d'images etc. Nous présentons ci-dessous, un exemple célèbre dans le domaine de la recherche d'information, à savoir la matrice des liens dans Google ,le projet apache, l'exemple du chimie théorique et celle du calcul formelle .

## 3.1. Matrice de Google

À un moment donné, on peut considérer que le Web est une collection de N ∈ N pages, avec N très grand (de l'ordre de $10^{10}$ en octobre 2005). La plupart de ces pages incluent des liens hypertextes vers d'autres pages. On dit qu'elles pointent vers ces autres pages. L'idée de base utilisée par les moteurs de recherche pour classer les pages par ordre de pertinence décroissante consiste à considérer que plus une page est la cible de liens venant d'autres pages, c'est-à-dire plus il y a de pages qui pointent vers elle, plus elle a de chances d'être fiable et intéressante pour l'utilisateur final, et réciproquement. Il s'agit donc de quantifier cette idée, c'est-à-dire d'attribuer un rang numérique ou score de pertinence à chaque page.

On se donne donc un ordre arbitraire sur l'ensemble des pages que l'on numérote ainsi d'i = 1 à i = N. La structure de connectivité du Web peut alors être représentée par une matrice C de taille N×N telle que $c_{ij}$= 1 si la page j pointe sur la page i, $c_{ij}$=0 sinon. Les liens d'une page sur elle-même ne sont pas significatifs, on pose donc $c_{ii}$=0. On observe que la ligne i contient tous les liens significatifs qui pointent sur la page i, alors que la colonne j contient tous les liens significatifs présents sur la page j.

On souhaite attribuer à chaque page i un score $r_i \in R_+^*$ de façon à pouvoir classer l'ensemble des pages par score décroissant et présenter à l'utilisateur une liste ainsi classée des pages correspondant à sa requête. L'algorithme PageRank part du principe qu'un lien de la page j pointant sur la page i contribue positivement au score de cette dernière, avec une pondération par le score $r_j$ de la page dont est issu le lien — une page ayant un score élevé a ainsi plus de poids qu'une n'ayant qu'un score médiocre — et par le nombre total de liens présents sur ladite page $N_j = \sum_{k=1}^{N} c_{kj}$.

On introduit donc la matrice Q définie par $q_{ij}$ = c/N j si N j≠0, $q_{ij}$=0 sinon. La somme des coefficients des colonnes non nulles de Q vaut toujours 1. L'application des principes ci-dessus conduit donc à une équation pour le vecteur r ∈ $R^N$ des scores des pages de la forme :

$$r_i = \sum_{j=1}^{N} q_{ij} r_j \text{ c'est-à-dire r = Qr.}$$

Le problème du classement des pages du Web se trouve ainsi ramené à la recherche d'un vecteur propre d'une énorme matrice, associé à la valeur propre 1 ! [Ham09].

## 3.2. Projet APACHE

Ce projet apache consiste à développer des applications dans différents domaines en coopération avec des équipes de recherche .En d'autre termes ils cherchent les difficultés existantes ou des problèmes et ils essayent de proposer des approches théoriques et les validés par une suite d'expérimentations.

Dans ce contexte, avec un modèle simple de turbulence comme cas test, un harnais parallèle pour les méthodes de décomposition de domaine en 2D a été écrit en ATHAPASCAN. Il permet une mise en œuvre aisée d'un partionnement de maillage conforme, de raffinement de maillage, de diverses formes de recollement aux frontières (méthode de Schur, méthodes de Schwartz, *etc.*), de schémas synchrones ou asynchrones, de traitement d'un ou plusieurs domaines par nœud de calcul. Des mesures qui ont été prise afin de tester les performances de ce harnais. Sa flexibilité a été testée sur un code de convection écrit par des chercheurs de l'IRIT. [PAP]

## 3.3. Chimie théorique

Dans le cadre de la théorie des semi-conducteurs, les valeurs propres (énergies) solutions minimisent le quotient de Rayleigh associé à l'opérateur de Schrödinger : les techniques usuelles employées dans la discipline sont souvent basées sur des algorithmes d'optimisation : on se donne une base de fonctionnelles paramétrées, on projette dessus l'opérateur, et on tente de minimiser la valeur propre la plus basse (par exemple) en faisant varier les paramètres. Outre le fait de devoir connaître une base, cette technique présente le défaut de conduire à des diagonalisations de matrices souvent denses.

La méthode proposée par ce travail est différente, basée sur la discrétisation par un schéma aux différences finies de l'opérateur. La matrice creuse obtenue est ensuite diagonalisée directement par une méthode itérative *ad hoc* (algorithme de Lanczos). Une version parallèle (en MPI) de cet algorithme est utilisée sur le Cray T3E du CEA. Une version de ce programme en ATHAPASCAN est en cours de réalisation, afin de comparer cette programmation à MPI sur un problème physiquement intéressant. L'intérêt de ce travail est double : comparaison d'ATHAPASCAN à MPI sur le T3E pour la partie parallélisme et obtention de résultats physiques pour la partie semi-conducteurs de ce travail. [PAP].

## 3.4. Calcul formel

L'implantation d'algorithmes du calcul formel dans la bibliothèque GIVARO a permis de tester sur des applications conséquentes la validité et les performances du noyau exécutif ATHAPASCAN. En continuation de ce travail et en collaboration avec l'ETH à Zürich, le noyau exécutif ATHAPASCAN a été partiellement intégré comme support exécutif dans une bibliothèque pour la programmation d'algorithmes parallèles avec le langage ALDOR (NAG), utilisé en calcul formel. Ce travail a montré expérimentalement un bon comportement de la bibliothèque sur quelques schémas algorithmiques simples. Ces travaux ont amené à proposer conjointement avec le projet SAFIR et le LMC-IMAG l'action de recherche coopérative PAPOOSE, dont l'objectif est d'étudier le développement d'un environnement pour la programmation en calcul formel parallèle, intégrant complètement ATHAPASCAN avec les langages ALDOR et C++.

Dans le cadre de l'action incitative NSF-CNRS n5926 en collaboration avec le LMC-IMAG, l'université du Delaware et l'université de Caroline du nord, l'étude sur l'implantation d'algorithmes parallèles efficaces en algèbre linéaire formelle creuse sur des corps finis est lancé. Les algorithmes étudiés concernent le calcul du rang de grandes matrices creuses par des méthodes d'élimination et des méthodes itératives probabilistes. Le développement concerne le prototypage d'un ensemble de structures C++ pour la manipulation parallèle efficace de matrices représentées par des "boites noires" (*black-box*), étendant les structures proposées par *Givaro*.

Une stratégie de parallélisation de l'algorithme de Cholesky pour la factorisation de matrices creuses a été proposée. Cet algorithme utilise intensivement les BLAS de niveau 3 pour réduire les indirections sur les accès aux données et un placement statique des calculs, réalisé à partir du graphe d'élimination, pour la régulation de charge. Cet algorithme parallèle a été implémenté dans le modèle de programmation par échange de message (MPI), et comparé avec les algorithmes existants. Les performances expérimentales de cet algorithme améliorent, pour de nombreuses matrices creuses, les performances des algorithmes existants.

Cet algorithme a ensuite été implémenté en ATHAPASCAN-1 qui intègre des mécanismes de régulation de charge dynamique. Cette réalisation est actuellement en cours d'évaluation et sera comparée avec l'implantation réalisée sur MPI. [PAP]

# 4. Méthodes de résolution de système linéaires RSL et calcul d'éléments propres

Beaucoup des problèmes de l'algèbre linéaire se ramènent à deux problèmes fondamentaux qui sont la résolution de systèmes linéaire et le calcul des valeurs et vecteurs propres.

## 4.1. Résolution de système linéaire

Un système linéaire est défini par une matrice $A \in R^{N \times N}$ et un vecteur $b \in R^n$, nous cherchons $x \in R^n$ tel que $Ax = b$. Dans la littérature il existe 2 méthodes : les méthodes directes et itératives.

## 4.2. Calcul des valeurs et des valeurs et de vecteurs propres (CVP)

Etant donné une matrice carrée A, trouver ses valeurs propres, ou seulement certaines d'entre elles et, éventuellement, les vecteurs propres correspondants, revient à chercher des scalaires $\lambda$ et des vecteurs v tels que : $v \neq 0$, $Av = \lambda v$.

**<u>Méthodes directes et méthodes itératives</u>**

Contrairement aux problèmes de calcul de valeurs et vecteurs propres où nous ne pouvons utiliser que des méthodes directes, les méthodes de résolutions des systèmes linéaires peuvent être directes ou itératives.

### a. Méthodes directes

Les méthodes directes sont les méthodes numériques les plus anciennes pour la résolution de systèmes linéaires. De plus, jusqu'aux années quatre-vingt, elles ont souvent été préférées aux méthodes itératives, en raison de leur robustesse et de leur comportement prévisible. En effet, que ce soit pour des matrices denses ou creuses, ces méthodes permettent d'obtenir, théoriquement en l'absence des erreurs d'arrondi, une solution exacte en un nombre d'opérations élémentaires fini et connu a priori. La plupart des méthodes directes sont basées, principalement, sur deux étapes de calcul : la factorisation de la matrice suivant ses propriétés (symétrie, définie positive, etc.) et la résolution par des substitutions successives [SHL06].

Dans le cas de la résolution d'un système linéaire par une méthode directe, nous effectuons des opérations élémentaires sur les lignes de la matrice augmentée (A,b) pour réduire le système à une forme plus simple facile à résoudre [EHM06a]. Nous pouvons citer comme exemples les méthodes d'élimination de Gauss, de Cholesky, de Crout, la factorisation LU, etc.

**- Méthode d'élimination de Gauss**

La méthode d'élimination de Gauss (pivot de Gauss) est une méthode directe de résolution de systèmes linéaires qui permet de transformer un système en un autre système à matrice triangulaire possédant la même solution (triangularisation). Nous résoudrons le système ainsi obtenu à l'aide d'un algorithme de remontée.

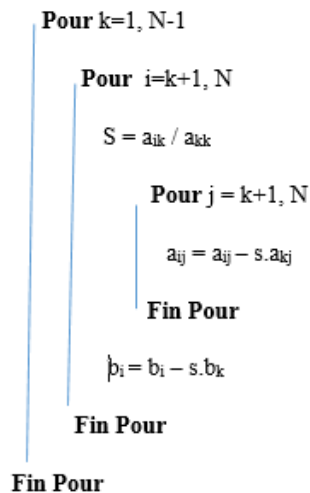

## b. Méthodes Itératives

Les méthodes itératives procèdent par itérations successives d'un même bloc d'opérations élémentaires, au cours desquelles une séquence infinie de solutions approximatives $\{x_k\}_{k \geq 0}$ est générée. A partir d'une solution initiale $x_0$ (choisie pour la première itération), une méthode itérative détermine à chaque itération $k \geq 1$, une solution approximative, $x_k$, qui converge progressivement vers la solution exacte, $x^*$ du système linéaire Ax = y, telle que :

$$x^* = \lim_{k \to \infty} x_k = A^{-1}, y \in R^n.$$

Les méthodes itératives génèrent une séquence de solutions approximatives $\{x_{(k)}\}$ qui converge sous certaines conditions, vers la solution exacte du système. Durant le calcul, la matrice

A reste intacte, elle n'est utilisée qu'à travers l'opérateur produit matrice-vecteur [EHM06a]. Parmi les méthodes itératives les plus connues, nous pouvons citer la méthode de Jacobi, celle de Gauss-Seidel, la SOR (Successive Over-relaxation), la méthode semi-itérative de Chebychev, etc.

- **Méthode de Gauss-Seidel [BIM]**

La méthode de Gauss-Seidel est une méthode qui permet de résoudre les systèmes linéaires Ax=y, où A est une matrice N×N et x,y sont des vecteurs de $R^n$.

Elle consiste en la manipulation suivante: Nous décomposons A comme A=D-E-F, où D est une matrice diagonale, -E est une matrice triangulaire inférieure, et -F est une matrice triangulaire supérieure.

$$\begin{pmatrix} & & -F \\ & D & \\ -E & & \end{pmatrix}$$

Nous pouvons alors transformer le système en :

$$Ax = y \leftrightarrow (D - E)x - Fx = y \leftrightarrow x = (D - E)^{-1} Fx + (D - E)^{-1} y$$

Nous définissons ensuite une suite de vecteurs $(X_k)$ par la formule :

$$x_{k+1} = (D - E)^{-1} F X_k + (D - E)^{-1} y$$

Nous espérons que la suite $(X_k)$ converge vers une solution de : Ax=y. Sous de bonnes hypothèses concernant la matrice A, c'est effectivement le cas.

## c. Directes ou itérative

Le choix direct ou itératif dépendra de l'efficacité théorique des algorithmes utilisés, de la structure et de la taille de la matrice ainsi que de l'architecture de l'ordinateur.

Parmi les difficultés qui doivent surmonter les «méthodes directes creuses» nous trouvons :

- La manipulation de structures de données complexes qui optimisent le stockage (profil de la matrice) mais qui complexifie l'algorithmique (pivotage…). Cela contribue à abaisser le ratio «calcul/accès aux données».
- La gestion efficace des données vis-à-vis la hiérarchie mémoire et la bascule IC/OOC19. Cette une question récurrente à beaucoup de problèmes.

- La gestion du compromis creux/dense (pour les méthodes par fronts) vis-à-vis de la consommation mémoire, de la facilité d'accès aux données et de l'efficacité des briques élémentaires d'algèbre linéaire.
- Le choix de la bonne renumérotation : c'est un problème NP-complet ! Pour les problèmes de grandes tailles, nous ne pouvons trouver pas en un temps «raisonnable» la renumérotation optimale. Nous devons se contenter d'une solution «locale».
- La gestion effective de la propagation des erreurs d'arrondi via le scaling, le pivotage et les calculs d'erreurs sur la solution (erreur directe/inverse20 et conditionnement). **[Boi]**

Du fait que les méthodes itératives n'affectent pas la matrice A, elles permettent une exploitation totale de son caractère creux. De plus, elles permettent d'éviter le problème du fill-in qui, dans le cas d'un stockage compressé de la matrice creuse (i.e. se limitant aux éléments non nuls), augmente et complique énormément le calcul. Par opposition aux méthodes itératives qui conservent le caractère creux de la matrice, les méthodes directes peuvent fournir à la fin du calcul des matrices (triangulaires) qui peuvent être denses, à l'exception du cas où A est une matrice bande [EHM06a].

Nous pouvons alors conclure que les méthodes itératives sont plus intéressantes dans le cas creux.

## 5. Produit Matrice-Vecteur Creux PMVC

Le projet de ce mémoire se concentre surtout sur l'opération de multiplication entre une matrice creuse et un vecteur (SpMV – *Sparse Matrix-Vector Multiplication*). Dans cette section nous présentons l'algorithme du PMVC ainsi qu'un exemple d'application.

Les méthodes itératives utilisent des boucles ayant comme noyau le Produit Matrice Vecteur (PMV). Tout au long des itérations, la matrice en entrée reste la même et seul le vecteur X qui change. Dans le cas où la matrice en question est dense, on parle de Produit Matrice-Vecteur dense (PMV). L'opération de multiplication entre une matrice creuse et un vecteur est à la base de plusieurs équations d'algèbre linéaire. Elle est représentée par l'équation : $\vec{y}=A\vec{x}$ où x et y représentent des vecteurs et A une matrice creuse. Pour effectuer cette opération, il faut diviser le problème sur chaque ligne de la matrice creuse, créant des équations indépendantes :

$$y[i]=\sum_{j=0}^{m} A[i,j]\ x[j].$$

Tout d'abord, il faut multiplier chaque valeur non nulle d'une même ligne avec un élément du vecteur se trouvant au même indice de colonne. Puis, il faut accumuler ces résultats temporaires dans le vecteur résultat à l'indice de la ligne en question. On présente l'algorithme de la multiplication du produit matrice vecteur.

**Algorithme : Multiplication matrice-vecteur**

**Pour** i=1, N

    **Pour** j=1, N

        y[i]=Y[i] + A[i] [j] * x[j]

    **Fin Pour**

**Fin Pour**

Dans le cas où A est creuse, la structure de l'algorithme du PMV dépend du format de stockage utilisé. L'utilisation du format de stockage CSR conduit à une version du Produit Matrice-Vecteur Creux (PMVC) représentée ci-dessous [Kho13].

**Algorithme : Multiplication matrice-vecteur avec le format CSR**

**Entrées :** Val, Col et Ptr (matrice), N (taille de la matrice), x (vecteur) ;

**Sorties :** y (vecteur) 3variables i, j, id;

**Pour** i = 0 à N − 1 faire

    y [i] ← 0;

**Fin Pour**

**Pour** i = 0 à N − 1 faire

    **Pour** id = Ptr [i] à Ptr [i + 1] − 1 faire

        j ← Col [id];

        y [i] ← y [i] + Val [id] × x [ j];

    **Fin Pour**

**Fin Pour**

# 6. Conclusion

Nous avons exposé dans ce chapitre un état de l'art sur les matrices creuses, ces structures ainsi que ces formats de compression. Ensuite, nous avons cité quelques applications creuses qui emploient des matrices de très grande taille telle que la matrice de liens de Google et le projet APACHE et puis nous avons vu les méthodes de résolution des systèmes linaire et le noyau de calcul le PMVC .

Beaucoup de ces calculs creux reviennent à des problèmes d'algèbre linaire. Ces derniers se ramènent à deux problèmes fondamentaux RSL et calcul des vecteurs (et valeurs) propres. Pour le calcul des vecteurs (ou valeurs) propres nous pouvons utiliser que les méthodes itératives alors que pour la RSL nous pouvons utiliser les deux celles de l'itérative et du directes. Nous avons montré que l'usage des méthodes itératives est le plus adéquat vue qu'il conserve le caractère creux des matrices.

Nous avons relevé aussi que ces méthodes itératives reposent sur le noyau de calcul du produit matrice vecteur et que dans notre cas d'étude, ces matrices sont des matrices creuses de grandes tailles nécessitant un espace de stockage des données ainsi qu'un volume de calcul important. Ainsi, l'utilisation d'environnements cibles parallèles et distribués offrant des performances élevées s'avère incontournable.

# Chapitre 2

ARCHITECTURES PARALLELES ET DISTRIBUES

# 1. Introduction

Après avoir une vue globale sur les différentes méthodes et algorithmes d'algèbre linéaire auxquels nous nous intéressons, nous passons en revue sur quelques systèmes parallèles et distribués. Nous nous focalisons particulièrement sur l'architecture d'une grappe multi-cœurs qui constitue notre plateforme cible de test. Nous présentons tout d'abord les architectures parallèles standards en les classant selon 3 catégories correspondantes aux types des mémoires. Après la présentation des architectures multi-cœurs, nous détaillons celle des grappes de calcul.

# 2. Les classifications

Il existe beaucoup de classification pour ce type de systèmes nous pouvons citer la classification de Kuck,

|  |  | Exécution Stream | | | |
|---|---|---|---|---|---|
|  |  | Simple Scalaire | Simple Vectoriel | Multiple Scalaire | Multiple Vectoriel |
| Instruction Stream | Simple Scalaire | SESES | SESEA | | |
|  | Simple Vectoriel | | SESEA | | |
|  | Multiple Scalaire | | | MISMES | MISMEA |
|  | Multiple Vectoriel | | | | |

**Tableau 0.1 : Classification de Kuck [CHE]**

la Classification de Treleaven,

| Contrôle Explicite ↑ | Mémoire partagée | Mémoire locale |
|---|---|---|
| Control Driven | Von Neuman | CSP |
| Pattern Driven | Logique | Acteurs |
| Demand Driven | Reduction (graphes) | Reduction (chaines) |
| Data Driven | Data Flow | Data Flow (tokens) |
| Contrôle implicite ↓ | | |

**Tableau 0.2 : Classification de Treleaven [CHE]**

la classification de Gajski porte sur les multiprocesseurs à mémoire partagée de type MIMD.

|  | Taches | Processus | Instructions |
|---|---|---|---|
| Cray X-MP | S | P | P |
| NYU | P | - | S |
| Cedar | P | P | S |

S : Séquentiel / P : Parallèle

**Tableau 0.3 : Classification de Gajski [CHE]**

la classification de Reina,

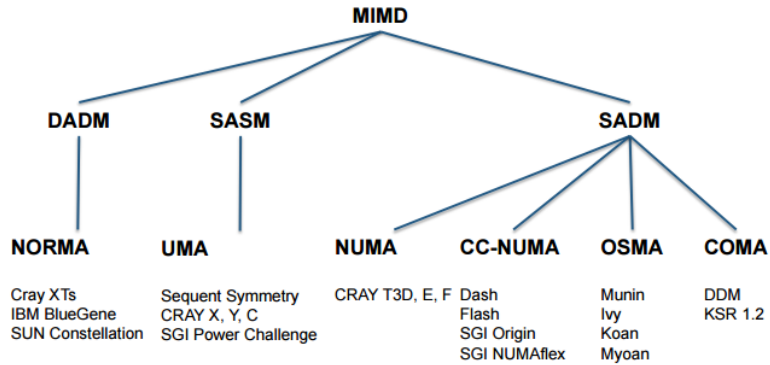

Tableau 0.4 : Classification Reina

la classification de Skillicorn qui est la généralisation de la taxonomie de Flynn. C'est une classification hiérarchique à deux niveaux :

- Niveau supérieur (unités fonctionnelles):

    • Nombre de processeurs de données et de processeurs d'instructions

    • Nombre d'unités de mémoire

    • Type de réseau d'interconnexion entre ces éléments

- Niveau inférieur (fonctionnement interne des unités):

    • Diagrammes d'états [BER]

et la classification de Flynn :

|  | flot d'instructions unique | flot d'instructions multiple |
|---|---|---|
| flot de données unique | SISD | MISD |
| flot de données multiple | SIMD | MIMD |

S: Single   M: Multiple   I: Instruction   D: Data

Tableau 0.5 : Classification de Flynn

**SISD** : une instruction - une donnée machine séquentielle (modèle de Von Neumann)

**SIMD** : plusieurs données traitées en même temps par une seule instruction. Elle est utilisée dans les gros ordinateurs vectoriels. Première machine parallèle : l'ILLIAC IV (1966)

**MISD** : une donnée unique traitée par plusieurs instructions c'est une architecture pipeline.

**MIMD** : exécution d'une instruction différente sur chaque processeur pour des données différentes.

Pour simplifier la programmation, nous exécutons la même application sur tous les processeurs. Ce mode d'exécution est appelé : **SPMD** (Single Program Multiple Data).

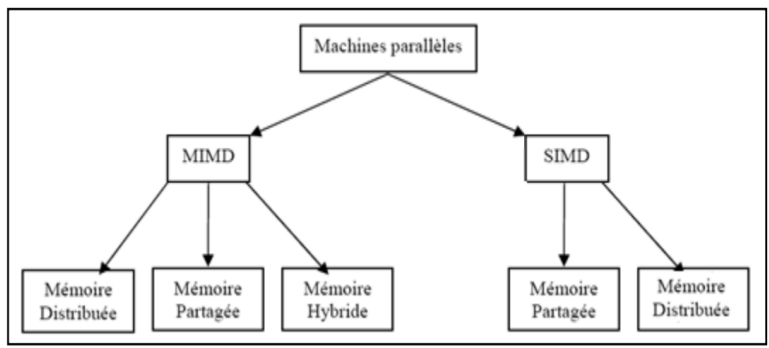

Figure 0.1 : Taxonomie de haut niveau des architectures parallèles

Cette décomposition distingue deux types de machines parallèle le MIMD et le SIMD. La première peut avoir 3 types de mémoire soit distribué, partagé ou hybride. Alors que la seconde ne peut pas avoir que la mémoire partagé et distribué. Durant notre étude on s'intéresse aux machines MIMD qu'on va détailler les types de mémoire utilisé pour ce types de machine.

# 3. Typologie de mémoire

## 3.1. Machine à mémoire partagée

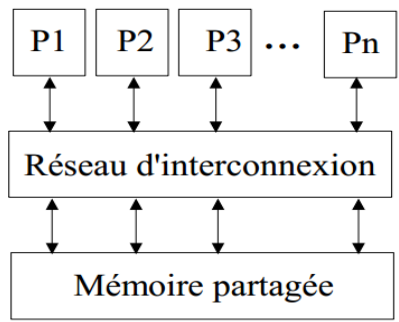

Figure 0.2 : Machine à mémoire partagée

Les machines à mémoire partagée permettent de réaliser les 2 types de parallélisme: celui de données où beaucoup de valeurs différentes sont traitées en même temps par des processeurs différents et le parallélisme de contrôle où des tâches différentes sont exécutées simultanément. Le programmeur n'a pas besoin de spécifier la distribution des données sur chaque processeur. Il définit seulement la partie du programme qui doit être parallélisée (directives) et doit gérer les synchronisations.

Parmi les caractéristiques les plus importantes pour ces machines qu'ils contiennent plusieurs processeurs avec des horloges indépendantes, il existe une seule mémoire commune à tous les processeurs et nous pouvons programmer par le standard portable OpenMP [MAR].

## 3.2. Machine à mémoire distribuée

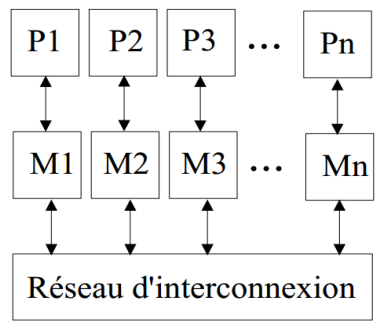

Figure 0.3 : Machine à mémoire distribuée

Ce type de machine permet d'établir une interconnexion (réseau local rapide) de systèmes indépendants nœuds (ordinateurs) avec une mémoire propre locale à chaque processeur qui exécute des instructions identiques ou non, sur des données identiques ou non . Il assure aussi le parallélisme par échange de messages [MAR].

## 3.3. Machine à mémoire hybrides

A nos jours plusieurs machines sont équipées une mémoire qui s'appelle hybride. SSHDs combinent HDD (Hard Disk Drive) et la technologie (SSD Solid State Drive) afin de réaliser un dispositif de stockage haute capacité rapide pour les ordinateurs. Un SSHD diffère d'un SSD mais il utilise la même mémoire flash à haute vitesse, mais il a aussi la capacité plus élevée associée à une norme d'entraînement de haute Disk donc faire un SSHD beaucoup plus rentable que d'un SSD [SSD].

## **Typologie architecture parallèle**

Plusieurs architectures parallèles ont été conçues pour la résolution des problèmes scientifiques, commerciaux ou d'ingénierie complexes, reconnus gourmands en ressources de calcul. Il y a, globalement, deux types d'architectures parallèles.

Ces architecture sont créés principalement afin d'augmenter la puissance de calcul d'une machine nous pouvons intervenir sur :

- ❖ Les temps de commutation des circuits de base
- ❖ La vitesse d'horloge
- ❖ Les transferts d'information (bus d'interconnexion)
- ❖ Les accès mémoire
- ❖ Les temps d'exécution des instructions élémentaire
- ❖ La mise en place de mémoire cache pour alimenter les processeurs en infos
- ❖ La réalisation parallèle de tâches (OS, SGBD) : du time sharing au parallélisme de processus (Unix)

Processeurs RISC (à jeu réduit d'instructions) : 35 à 100 MIPS (Millions d'instruction par seconde) [**SAP**].

Depuis 2004 les CPU sont multi-cœurs (suite à la réinterprétation de la loi de Moore) :

- ❖ De plus en plus de cœurs par « socket » (processeur)
- ❖ La puissance d'un cœur reste stable (n'évolue presque plus), sauf en utilisant des unités vectorielles au sein de chaque cœur
- ❖ De plus en plus de cœurs concourent pour accéder à la RAM
- ❖ Une hiérarchie de mémoires cache apparait de plus en plus et les vitesses du cache et de la RAM ne cessent de s'écarter

**Remarque** : Beaucoup d'autres composants existent aussi dans un processeur multi-cœurs.

Tous les fournisseurs informatiques ont annoncé des puces avec plusieurs cœurs, on citera quelques exemples.

**Intel :** La marque d'Intel Core i7 est utilisée pour ses microprocesseurs grands publics haut de gamme depuis novembre 2008. Les marques Core i5 et Core i3 sont apparues ensuite. Techniquement, les Core i7 peuvent appartenir aux familles Nehalem, Westmere, Sandy Bridge, Ivy Bridge et Haswell. Les Core i7 ont de deux à huit cœurs on peut citer à titre d'exemple le dernier processeur Intel® Core™ i7-5950HQ Processor qui a 4 cœurs. De nos jours, Intel a lancé sa nouvelle gamme de processeurs, destinée pour serveurs, le Intel® Xeon® Processor E7-4850 v3 (35M Cache, 2.20 GHz) offrant jusqu'à 14 cœurs et le Intel® Xeon® Processor E7-8890 v3 (45M Cache, 2.50 GHz) qui offre 18 cœurs [GIC].

**Intel Core i7 - Nehalem :**

Des cœurs hyper-threadés Des mémoires caches Un sheduler de threads Des pipelines ….

Des unités de calcul vectorielles : « SSE ». Un processeur contient beaucoup de composants différents, pouvant fonctionner en parallèle, interconnectés et ayant tous besoin d'être alimentés en données : très difficile d'exploiter à l'optimum. Les « optimisations sérielles » : une démarche essentielle.

**AMD** : En Juillet 2015 AMD lance un nouveau processeur A10-7870K et A10 PRO-7850B qui ont chacun d'eux 12 cœurs 4 CPU Cœurs et 8 GPU cœurs .AMD dispose aussi d'une gamme de processeurs, destinée pour serveurs tels que les Processeurs AMD Opteron™ 6386 SE qui disposent de nombre de cœurs allant jusqu'à 16 cœurs.

**IBM** : IBM a été le premier fournisseur à embrasser la technologie multi-cœurs. Elle a rendu cette technologie disponible pour ses clients par l'introduction des systèmes POWER4 (à 2 cœurs) en 2001 [DRI][POW]. Elle est arrivée aujourd'hui à créer son processeur POWER8 avec 12 cœurs.

**TILERA**: Une start-up américaine spécialisée dans les processeurs, qui développe des puces avec des centaines de cœurs à l'intérieur. Ses puces servent à alimenter des serveurs orientés basse consommation utilisés dans le Cloud ou dans d'autres environnements à haute performance. Elle vient de lancer sa TILE-Mx100 Hecta-core network processor avec le CPU 100 ARM Cortex A53 cœurs.

**FREESCALE SEMICONDUCTOR** : est une entreprise américaine dans le domaine des semi-conducteurs. Précédemment nommée Motorola Semiconductor, elle est issue de la branche semi-conducteurs de Motorola. Depuis 2004, Freescale est devenue indépendante de sa maison-mère. Elle a lancé un processeur S32V230 qui a un Quad ARM Cortex®-A53 cœurs et MX7D DUAL PROCESSORS qui a Dual Cortex-A7 and Cortex-M4.

**NVIDIA** : Tegra X1 est un processeur qui a un CPU 8 CPU- core, 64-bit ARM® CPU et un GPU NVIDIA Maxwell 256-core GPU, Tegra K1 avec un CPU NVIDIA 4-Plus-1™ Quad-Core ARM Cortex-A15 "r3" et un GPU 192 NVIDIA CUDA® cœurs , et le Tegra 4 Family avec le CPU ARM Cortex-A15 qui a 5 cœurs et un GPU 72 cœurs alors que pour le modèle Tegra 4i il a un GPU de 60 cœurs et le CPU ARM Cortex-A9 r4.

**Texas Instruments :** Texas Instruments (TI) est une entreprise d'électronique, fondée en 1941, basée à Dallas, renommée dans le domaine des composants électroniques passifs et des semi-conducteurs. L'un des processeurs qu'elle dispose est l'AM3715 Sitara ™ Processors qui a un CPU 1 ARM Cortex-A8. Et le AM5K2Ex avec un CPU Quad-ARM® Cortex®-A15 cœurs.

**VIA Technologies :** VIA Technologies (ou VIA) est un fabricant taïwanais de circuits intégrés (principalement des chipsets de cartes mère, de processeurs, de processeurs graphiques / GPU et de mémoires) qui fait partie du Groupe Formosa Plastics. Elle a produit de nombreux processeurs citant a titre d'exemple le WM8980 avec un CPU Dual ARM Cortex-A9.

# 4. Grappe de calcul (Clusters)

## 4.1. Définition

Une grappe de calcul, cluster en anglais, est constituée de deux ou plusieurs ordinateurs, interconnectés par un réseau local, souvent, à haut débit (par exemple, un réseau InfiniBand). Chaque calculateur faisant partie d'une grappe est appelé nœud de calcul et il possède une ou plusieurs unités de calcul et une mémoire locale. Tous les nœuds de calcul d'une grappe travaillent ensemble comme un seul calculateur parallèle. En général, une grappe de calcul dispose d'un nœud, dit frontal, qui a pour rôle la gestion des ressources et la distribution des calculs sur les nœuds. La figure 2.4 montre un exemple de grappe de calcul composée de six nœuds, ayant chacun quatre unités de calcul.

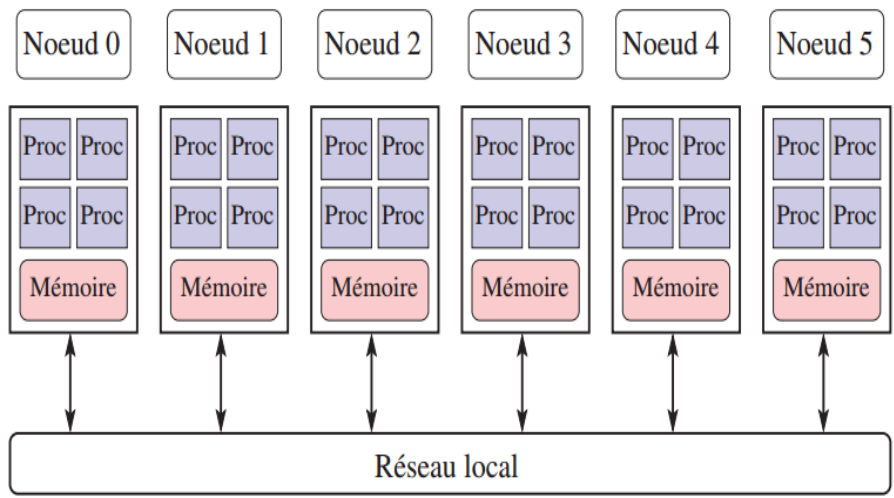

Figure 0.4 : Grappes de Calcul

## 4.2. Réseaux de connexion

Les nœuds d'un cluster peuvent être reliés entre eux par plusieurs réseaux. Dans les réseaux de communication les plus connus, avec une bande passante importante, nous trouvons **Gigabit Ethernet**, **InfiniBand** et **Myrinet**.

- **Gigabit Ethernet** : Plusieurs des premier clusters HPC ont utilisé le Fast Ethernet (100 mégabits / seconde). Actuellement, de nombreux systèmes utilisent le Gigabit Ethernet (1000 mégabits / seconde) ou 10 Gigabit Ethernet (10 000 Mégabits / seconde).Basé sur les tendances passées, 10 Gigabit Ethernet (GigE 10) est destiné à prendre le relais pour GigE comme une interconnexion de cluster. En plus d'offrir une technologie familière, il a aussi la capacité d'offrir des fonctionnalités larges, des performances accrues, et dans le même temps, un coût réduit. La transition vers 10 GigE pour le HPC est considérée par beaucoup comme similaire à la progression naturelle de Fast Ethernet à GigE. La bande passante supplémentaire et le taux de messagerie fait du 10 GigE le choix logique pour les utilisateurs de nombreuses grappes [HPC].

- **InfiniBand** : InfiniBand est une interconnexion réseau à haut débit et faible latence qui gagne en importance dans les applications de HPC. De nombreux clients, parmi les 500 les plus importants, possédant des clusters HPC, utilisent cette solution E/S très prisée et répandue. On peut ainsi relever de grandes universités et centres de recherche, les sciences humaines, la biomédecine, l'industrie pétrolière et du gaz (applications sismiques, réservoirs, et modélisme), conception et ingénierie assistée par ordinateur (CAO, IAO), Enterprise Oracle, et applications dans les finances [TST] [YAB12][YA12].

- **Myrinet** : Myrinet est une technologie récente d'interconnexion de réseau, commuté à faible latence et à haute performance, basée sur l'envoi de messages. Les produits d'interconnexion Myrinet permettent de connecter à très haute vitesse un très grand nombre d'ordinateurs afin de former des clusters très rapides [DTS] [YA13] [YAB13] [AY15].

## 5. Conclusion

Dans ce chapitre nous avons présenté les différentes architectures parallèles et distribuées. Les machines parallèles offrent de hautes performances de calcul mais à coût élevé. Dans le cas de certaines applications, nous pouvons avoir recours à un cluster (grappe). Une évolution importante conduit aujourd'hui à la complexité d'exploitation des grappes de calcul.

Notre objectif dans le chapitre suivant est d'étudier la parallélisation d'un noyau de calcul creux, à savoir le produit matrice vecteur creux (PMVC), sur ces architectures parallèles. Nous présenterons ainsi les différentes approches proposées dans la littérature pour la distribution de ce noyau et détaillerons en particulier la méthode hypergraphe et l'approche NEZGT ainsi que la méthode combinée.

# Chapitre 3

ETAT DE L'ART SUR LA DISTRIBUTION DES DONNEES ET DU PMVC SUR UNE GRAPPE MULTI COEURS

# 1. Introduction

Nous présentons dans ce chapitre le **P**roduit **M**atrice **V**ecteur **C**reux (PMVC) sa définition sa version ligne, colonne et celle du 2D. Ensuite nous présentons les bibliothèques parallèles et distribuées pour la (RSL) **R**ésolution de **S**ystèmes **L**inéaires creux. Nous exposons par la suite les travaux existants dans la littérature pour la distribution du PMVC. Nous détaillons en particulier les méthodes NEZGT$_{LIGNE}$ et Hypergraphe ainsi que la méthode combinée basée sur ces deux dernières [MeH12] [MeH13].

# 2. Produit Matrice-Vecteur Creux (PMVC)

## 2.1. Définition

Soit A une matrice carrée (N, N) et X un vecteur de $R^n$. Le problème que nous nous proposons de paralléliser dans ce paragraphe est le calcul du produit matrice-vecteur creux Y =AX sur différents types de machines parallèles.

$A_{ij}$ désigne l'élément de la matrice A en ligne i et colonne j. En séquentiel, rappelons que la forme la plus naturelle pour écrire l'algorithme est la suivante :

$$\forall\, i \in \{1,2,\ldots,N\}\ \ v_i = \sum_{j=1}^{N} A_{ij} X_j$$

Où, sous forme algorithmique :

```
Pour i←1 jusqu'à N
    V[i] ←0
    Pour j←1 jusqu'à N
        V[i] ←V[i] + A[i,j] * X[j]
```

Notons une contrainte supplémentaire lors de la parallélisation. Le produit matrice-vecteur creux est une des procédures numériques les plus utilisés. Par exemple, les méthodes itératives de résolution de systèmes linéaires. Dans ce cas, il faut enchainer les produits matrice-vecteur creux (toujours avec la même matrice) et alors, il faut que le placement sur le réseau du vecteur résultat V soit le même que celui de X.

## 2.2. PMVC Version ligne

Si A est stockée en lignes, l'organisation des calculs impose que la $i^{ème}$ composante du vecteur résultat Y soit dans le même processeur que la $i^{ème}$ ligne de A. Et dans ce cas, le placement initial du vecteur X suit celui de Y. Il y a une seule étape de communication qui consiste en un « échange total », où chaque processeur envoie un même message à tous les processeurs connaissent le vecteur X entier puisqu'il doit rencontrer toutes les composantes de la $i^{ème}$ ligne de A).s

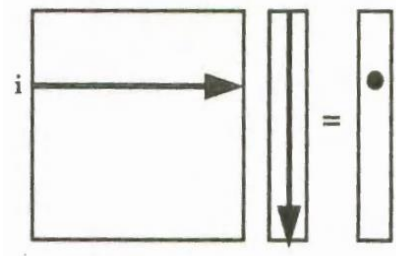

Figure 0.1 : Principe du PMVC en ligne

L'algorithme s'écrit :

```
ATA sur les N/p composantes locales du vecteur x

Pour tout processeur q de 0 jusqu'à p-1 faire en parallèle

    Pour tout k←1 jusqu'à N/p faire

        V ( q + ( k-1 ) p + 1) ← produit scalaire de la ligne q + ( k-1 ) p + 1 de A et du vecteur x
```

Le cout en calcul est exactement de N/p calculs de produits scalaires nécessitant 2N-1 opérations arithmétiques (en comptant de la même manière additions et multiplications). Soit, au total, (2N-1) N/p $\tau_a$ ($\tau_a$ représente l'unité de calcul de base). Le cout en communication est celui d'un échange total, et dépend de la topologie du réseau.

## 2.3. PMVC Version Colonne

Pour la version en colonnes réparties de manière circulaire, le calcul impose que la $j^{ème}$ colonne de A rencontre la $j^{ème}$ composante du vecteur x. Chaque processeur produit alors un vecteur résultat complet dont les composantes sont des sommes partielles (Figure 3.2). Cela revient à effectuer sur chaque processeur q les calculs locaux suivant :

$$\forall\, i \in \{1,2,\ldots,N\}\ V_i = \sum_{j=1}^{N/p} A_{i,(j-1)p+q+1}\, X_{(j-1)p+q+1}$$

Cet algorithme n'utilise qu'une seule étape de communication qui consiste en un « échange total personnalisé avec accumulation », c'est-à-dire que chaque processeur doit envoyer un message personnalisé à chacun des autres processeurs (de N/p composantes), tous les messages destinés à un même processeur étant à additionner les uns aux autres. [AAP]

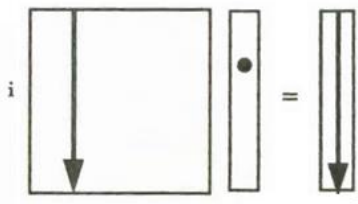

**Figure 0.2 : Principe du PMVC en colonne**

L'algorithme se réduit à :

```
Pour tour processeur q de 0 jusqu'à p-1 faire en parallèle
    Pour tout k←1 jusqu'à N/p faire
        V-provisoire ← multiplication de la colonne q + ( k+1 ) p de A par la
        composante q + ( k+ 1) p+1 du vecteur X
    ATA-personnalisé avec accumulation des vecteurs V provisoires
```

Le cout en calcul est exactement de N/p calculs de N multiplications composante par composante ; il faut lui ajouter le cout de la sommation des sommes partielles sur tout le vecteur V, soit p sommes pour chacune des N/p composantes locales du vecteur V [DTW90].

## 2.4. Version Bloc 2D

L'algorithme pour la version 2D est le suivant :

```
ATA (partiel) du vecteur X
Pour tout processeur q de 0 jusqu'à p-1 faire en parallèle
    Vecteur c partiel ← produit matrice-vecteur du bloc
    Par le vecteur partiel X que l'on vient de recevoir
ATA-personnalisé avec accumulation (partielle) des vecteurs V partiels.
```

# 3. Bibliothèques parallèles et distribués pour la (RSL)

Plusieurs applications scientifiques reviennent à la **R**ésolution de **S**ystèmes **L**inéaires (RSL) creux d'où, l'apparition d'une panoplie de bibliothèques permettant de traiter de tels systèmes sur des architectures parallèles ou distribuées.

**PETSc :** (le S est « silent ») est une suite de structures de données et de routines qui est une solution évolutive (parallèle) pour les applications scientifiques modélisés par des équations basées sur les dérivées partielles. Il prend en charge MPI, « pthreads » de mémoire partagée, et les GPU à travers CUDA ou OpenCL, ainsi que les mémoires hybrides pthreads de mémoire MPI partagés ou MPI-GPU parallélisme. La dernière version du PETSc est 3.5, réalisé en 30 Juin, 2014.

**SuperLU :** SuperLU est une bibliothèque de résolution de systèmes linéaires non symétriques sur des machines haute performance. La bibliothèque est développée en C, et appelable de C ou Fortran. La dernière version du SuperLU est 4.6.

**MUMPS :** (Multifrontal Massively Parallel sparse direct Solver) cette bibliothèque est une solution pour des grands systèmes linéaires avec des matrices symétriques et symétriques positives Version pour l'arithmétique complexe.

Elle contienne une factorisation parallèle et résoudre phases (version monoprocesseur également disponible) ainsi qu'un affinement itératif et l'analyse des erreurs en arrière .Elle dispose aussi de différents formats d'entrée de la matrice de format assemblé, distribué forme assemblée et format élémentaire. Elle permet de faire une factorisation partielle et Schur matrice du complément (centralisée ou 2D bloc-cyclique) ;

Interfaces aux MUMPS : Fortran, C, Matlab et Scilab;

Plusieurs ordres interfacés : AMD, AMF, DROP, METIS, ParMETIS, SCOTCH, PT-SCOTCH. La dernière version du MUMPS est 4.10.0.

**PaStiX** : PaStiX () est une bibliothèque scientifique qui fournit un solveur parallèle haute performance pour les très grands systèmes linéaires creux sur la base de méthodes directes.

Algorithmes numériques sont mises en œuvre en simple ou double précision (réel ou complexe) en utilisant LLt, THDV et LU, basculement statique (pour les matrices non symétriques ayant un motif symétrique). La dernière version du PaStiX 5.2.2.

**HIPS** : HIPS (Hierarchical Iterative Parallel Solver) est un solveur de systèmes linéaires creux en parallèle. HIPS prend en compte les systèmes linéaires :

- Symétriques / non symétriques,
- Réels / complexes (simple ou double précision).

Plusieurs méthodes de résolutions sont proposées en fonction du rapport mémoire/performance souhaité :

- Hybride directe / itérative,
- Itérative.

**Auteurs** : HIPS est développé par Jeremie Gaidamour et Pascal Hénon (Gaidamou et Henon à labri dot fr) en collaboration avec Yousef Saad de l'Université du Minnesota.

**Licence** : HIPS est disponible en tant que logiciel libre / libre sous licence CeCILL-C.

HIPS est réalisée au sein de l'équipe Bacchus de l'INRIA Bordeaux - Sud-Ouest et le Laboratoire Bordelais de Recherche en Informatique. La dernière version du Hips est : Hips 1.2 beta.rc5.

**Lis** : Lis (Bibliothèque de solveurs itératifs pour les systèmes linéaires, prononcés [lis]) est une bibliothèque parallèle pour résoudre des équations linéaires et problèmes aux valeurs propres qui se posent dans la solution numérique des équations aux dérivées partielles en utilisant des méthodes itératives. La dernière version du LIS est Lis-1.4.57 réalisé en 15 aout, 2014.

**PIM:** (Parallel Iterative Methods) est une collection de routines Fortran 77 qui visent à résoudre des systèmes d'équations linéaires (SLES) sur les ordinateurs parallèles en utilisant une variété de méthodes itératives. La dernière version du PIM est 2.3.

**PSBLAS** : (PSBLAS) "Sparse parallèles Basic Linear Algebra sous-programmes" projet d'essayer de fournir un cadre pour permettre des implémentations simples, efficaces et portables de solveurs itératifs pour les systèmes linéaires, tout en protégeant l'utilisateur de la plupart des détails de leur parallélisation. La dernière version du PSBLAS est 3.2.

| DIRECT SOLVERS | License | Support | Real | Complex | F77 | C | Seq | Dist | SPD | Gen |
|---|---|---|---|---|---|---|---|---|---|---|
| **SPARSE** | | | | | | | | | | |
| DSCPACK | ? | oui | X | | | X | X | M | X | |
| HSL | ? | oui | X | X | X | | X | | X | X |
| MFACT | ? | oui | X | | | X | X | M | X | |
| MUMPS | PD | oui | X | X | X | X | X | M | X | X |
| PSPASES | ? | oui | X | | X | X | | M | X | |
| SPARSE | ? | ? | X | X | | X | X | | X | X |
| SPOOLES | PD | ? | X | X | | X | X | M | | X |
| SuperLU | Own | oui | X | X | X | X | X | M | | X |
| TAUCS | Own | oui | X | X | | X | X | | X | X |
| Trilinos/Amesos | LGPL | oui | X | | | | X | M | X | X |
| UMFPACK | LGPL | oui | X | X | | X | X | | | X |
| Y12M | ? | oui | X | | X | | X | | X | X |

Tableau 0.1 : Solveurs numériques pour la RSL pour le cas creux [CSN]

# 4. Fragmentation des données

## 4.1. Définition

Généralement, la fragmentation d'une matrice creuse revient à la décomposer en blocs 1D (blocs de lignes ou blocs de colonnes) ou 2D (blocs de lignes-colonnes) . L'équilibrage des charges de calcul est le principal critère que plusieurs travaux cherchent à optimiser lors de la décomposition des données. [KGK08]

## 4.2. Méthodes de fragmentation de données

### a. Méthodes statique

L'approche proposée par [**KGK08**] consiste à appliquer un équilibrage statique basé sur le nombre des éléments non nus. Où chaque « thread » est affecté presque au même nombre d'éléments et, ainsi, le même nombre d'opérations à virgule flottante

[**KGK08**] présentent aussi 2 formats de compression pour les matrices creux CSR-DU et CSR-VI qui essayent d'améliorer les performances du PMVC.

Le CSR-U est appliqué pour « coarse grain delta encoding» pour les indices de colonnes alors que le CSR-VI utilise l'indexation indirecte des valeurs des données numérique. Ces deux formats sont applicables seulement pour des matrices qui ont un nombre de colonne qui contient des éléments non-nuls assez important. Ces formats ont prouvé une meilleure performance en les comparants avec le CSR. Durant ces études ils ont prouvés qu'ils sont stables et ne donnent pas un « no memory bound matrix » pour l'usage de 4 et 8 threads.

Le travail de [**ShU11**] focalise sur la performance des machines parallèles pour le calcul du PMVC et l'implémentation d'un solveur linéaire itérative en utilisant le format de compression SBCRS pour un cluster et la comparer avec les formats CRS et BCRS.

Ils ont montré que le format proposée « SBCRS » réduit le temps d'exécution mais cette gain de temps peut engendrer un travail adduisant de remplissage es zéros. Mais ce format proposé améliore la performance et réduit le stockage du matrice, il performe assez meilleur pour les larges blocks et réduit le nombre de chargement et le stockage des instructions pour les éléments des tableaux.

### b. Méthodes dynamiques

Il existe beaucoup de chercheurs qui ont proposés des méthodes dynamiques. On peut nommer à titre d'exemple « SeyongLee » et «Rudoef Eignmann » en [LeE08] qui ont proposé une distribution dynamique intra nœuds en blocs de lignes qui utilise le format de stockage « CRS » pour une grappe de calcul de 32 nœuds hétérogène (mono proc – mono cœur). Leur solution proposée consiste à avoir un système d'exécution adapté aux noyaux du PMVC qui résout le problème d'équilibrage de charge. Cela se fait en affectant dynamiquement les lignes de la matrice au processeur selon la mesure des charges de travail en temps réel.

Les travaux de «Gerald Schubert» et «Holger Fehske» ont montrés les avantages et la performance de la programmation hybride MPI/OpenMP grâce à l'amélioration de l'équilibrage de charge sans chevauchement de communication et ils ont montrés aussi que le déséquilibrage de charge est causé par le modèle de communication.

Les méthodes dynamiques montrent que l'équilibrage de charge peut jouer un rôle important dans la réduction des couts de communication .Ceci est dû au fait qu'il peut réduire les retards de synchronisation [WoV07].

Dans notre approche les méthodes dynamiques ne sont pas appropriées vue qu'on essaye de réduire le temps de calcul qui est l'un des critères principaux qu'on essaye de la satisfaire. Or ces méthodes dynamiques présentent un « overhead » assez important ce qui impose un temps de calcul additionnel.

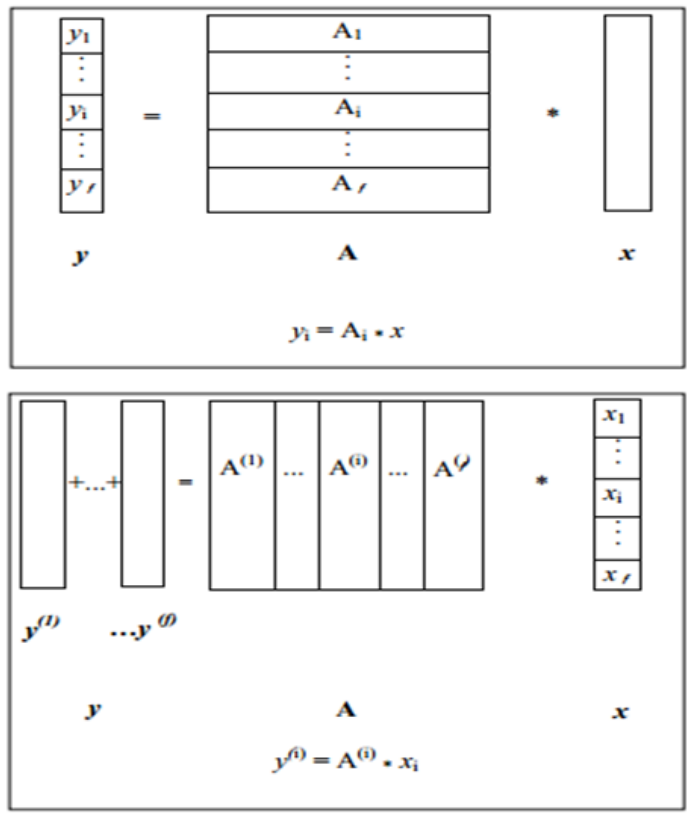

**Figure 0.3 : Fragmentation d'une matrice**

## 4.2.1. L'heuristique NEZGT LIGNE

« NEZGT » est une heuristique à trois phases. Les phases 0 et 1 correspondent à l'heuristique SPT (Shortest Processing Time) (resp. LPT (Largest Processing Time)). La phase 2 est une heuristique d'amélioration itérative. La phase 0 consiste à trier les lignes de A par nombre d'éléments non nuls croissant (resp, décroissant).

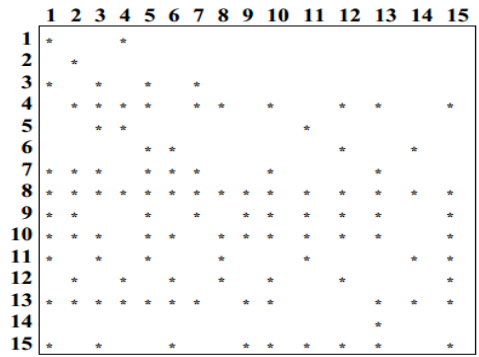

**Figure 0.4 : La matrice étudiée dans l'exemple**

**Exemple** : fragmentation NEZGT

| Indice de ligne | 1 | 2 | 3 | 4 | 5 | 6 | 7 | 8 | 9 | 10 | 11 | 12 | 13 | 14 | 15 |
|---|---|---|---|---|---|---|---|---|---|---|---|---|---|---|---|
| Nombre d'éléments non nuls par ligne | 2 | 1 | 4 | 10 | 3 | 4 | 8 | 15 | 10 | 12 | 6 | 7 | 12 | 1 | 9 |

En triant les lignes dans l'ordre décroissant du nombre d'éléments non nuls, on obtient :

| Indice de ligne | 8 | 4 | 10 | 4 | 9 | 15 | 7 | 12 | 11 | 3 | 6 | 5 | 1 | 2 | 14 |
|---|---|---|---|---|---|---|---|---|---|---|---|---|---|---|---|
| Nombre d'éléments non nuls par ligne | 15 | 12 | 12 | 10 | 10 | 9 | 8 | 7 | 6 | 4 | 4 | 3 | 2 | 1 | 1 |

**Figure 0.5 : La matrice après la phase 0**

✦ Dans la première, nous utilisons l'heuristique LS qui consiste d'abord, à affecter la ligne i (i=1…f) au fragment i. Par la suite, la ligne suivante est affectée au fragment le moins chargé et ainsi de suite. En d'autres termes, la ligne courante non affectée est toujours attribuée au fragment le moins chargé jusqu'à épuisement des lignes.

La phase 1 de l'algorithme NEZGT fournit le résultat optimal suivant:

| Fragment | Indice de ligne (Nombre d'éléments non nuls par ligne) | Nombre d'éléments non nuls par fragment |
|---|---|---|
| 1 (2 lignes) | 8(15) ; 5(3) | 18 |
| 2 (3 lignes) | 13(12) ; 6(4) ; 1(2) | 18 |
| 3 (3 lignes) | 10(12) ; 3(4) ; 14(1) | 17 |
| 4 (2 lignes) | 9(10) ; 12(7) ; | 17 |
| 5 (3 lignes) | 4(10) ; 11(6) ; 2(1) | 17 |
| 6 (2 lignes) | 15(9) ; 7(8) | 17 |

**Figure 0.6 : La matrice après la phase 1**

✚     La seconde phase, qui est une phase d'amélioration, est une heuristique itérative. Elle permet, à travers des raffinements successifs, d'améliorer la fragmentation précédente, conduisant ainsi à un meilleur équilibrage. Le critère choisi étant FD (différence entre les deux charges extrêmes), la procédure conçue consiste à effectuer des échanges ou transferts successifs de lignes entre le fragment le plus chargé et le fragment le moins chargé.

Plus précisément, soit fcmx (resp. fcmn) le fragment le plus (resp. moins) chargé, la procédure que nous adoptons consiste, soit à transférer une ligne appropriée de « fcmx » à « fcmn » ou bien à échanger une ligne du premier fragment avec une ligne du second et ce, dans le but de réduire la différence entre les deux charges, notée Diff. L'opération de transfert consiste à choisir une ligne dont le nombre d'éléments non nuls, noté nzx est inférieur à Diff.

L'opération d'échange consiste, quant à elle, à choisir deux lignes (la première appartenant à fcmx et la seconde à fcmn) telles que la différence entre le nombre des éléments non nuls du premier, noté nzx, et le nombre des éléments non nuls du second, noté nzn, soit inférieure à Diff i.e : nzx – nzn < Diff. Une meilleure alternative, toutefois plus coûteuse, consiste à optimiser le choix afin de minimiser le nouvel écart entre les charges, ce qui revient à minimiser la quantité │Diff / 2 - nzx│ en cas de transfert (resp. │Diff / 2- ( nzx – nzn ) │en cas d'échange). Cette procédure est itérée tant qu'il est possible de réduire le critère FD et/ou qu'on ne dépasse pas un nombre d'itérations fixé à l'avance [Ham10].

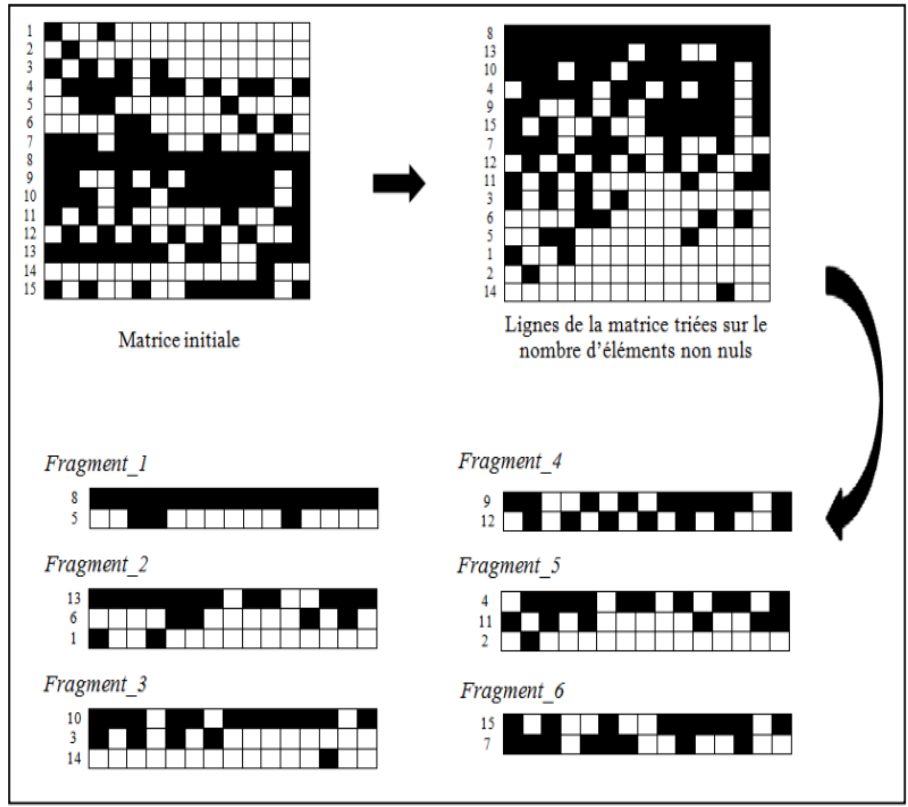

**Figure 0.7 : Exemple d'application du « NEZGT Ligne »**

### 4.2.2. Méthode de l'Hypergraph

Dans [DBH06], Devine et al présentent un hypergraphe H comme un couple (V, E) constitué d'un ensemble de sommets V et d'un ensemble d'hyperarcs E. Chaque hyperarc est un sous-ensemble de V.

$$H = (V, E)$$

Les hypergraphes ont des applications dans différents domaines dans lesquelles on utilise la théorie des graphes citons à titre d'exemple : traitement d'images, optimisation d'architecture réseaux, modélisation, calcul scientifique, etc. [TEC].

Le modèle hypergraphe est bien adapté au calcul parallèle : les sommets (V) correspondent aux données et les hyperarcs représentent les besoins de communication. Dans [ÇaA99], Catalyürek et Aykanat ont démontré que le modèle hypergraphe donne une représentation plus précise du coût de la communication (en volume) que le modèle graphique.

En particulier, pour le PMVC, le modèle hypergraphe représente exactement le volume de communications.

On peut associer à chaque sommet $v_i$ un poids $w_i$ qui est égal au nombre d'hyperarcs auxquels il est lié. La valeur de ce poids aide dans le calcul de la contrainte de partitionnement (équilibrage des charges).

Dans le cas du noyau PMVC, deux modèles de représentation d'hypergraphe peuvent être rencontrés :

 **Modèle 1D** : c'est un modèle pour une décomposition en blocs de lignes (resp. de colonnes) où chaque ligne (resp. colonne) correspond à un sommet v et chaque colonne (resp. ligne) correspond à un hyperarc [ÇaA99].

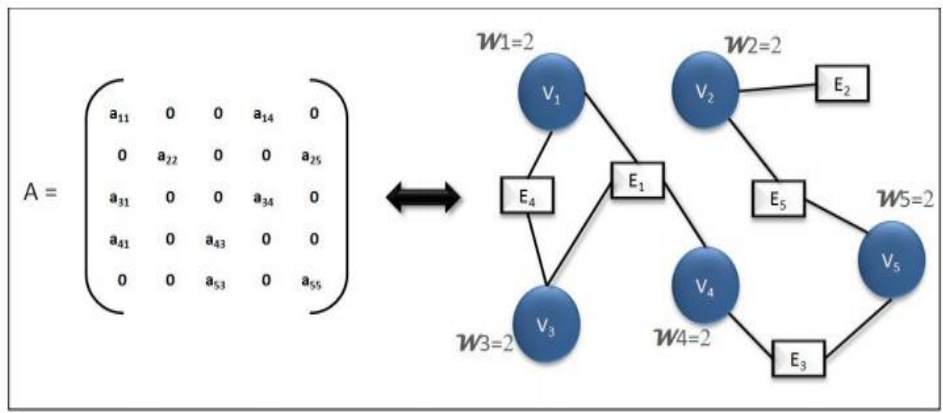

**Figure 0.8 : Hypergraph 1D**

 **Modèle 2D** : Dans la littérature on trouve le schéma de décomposition de matrice 'checkboard 2D', mais d'après Catalyürek et Aykanat, ce dernier est généralement adapté à des matrices denses ou matrices creuses avec structures régulières qui sont difficiles à exploiter. Ils ont donc proposé le modèle hypergraphe 'grain fin' où chaque élément non nul de la matrice creuse est représenté par un sommet v et chaque ligne et chaque colonne par un hyperarc e. Dans ce cas le poids de tout sommet v est égal à 2 [ÇaA01].

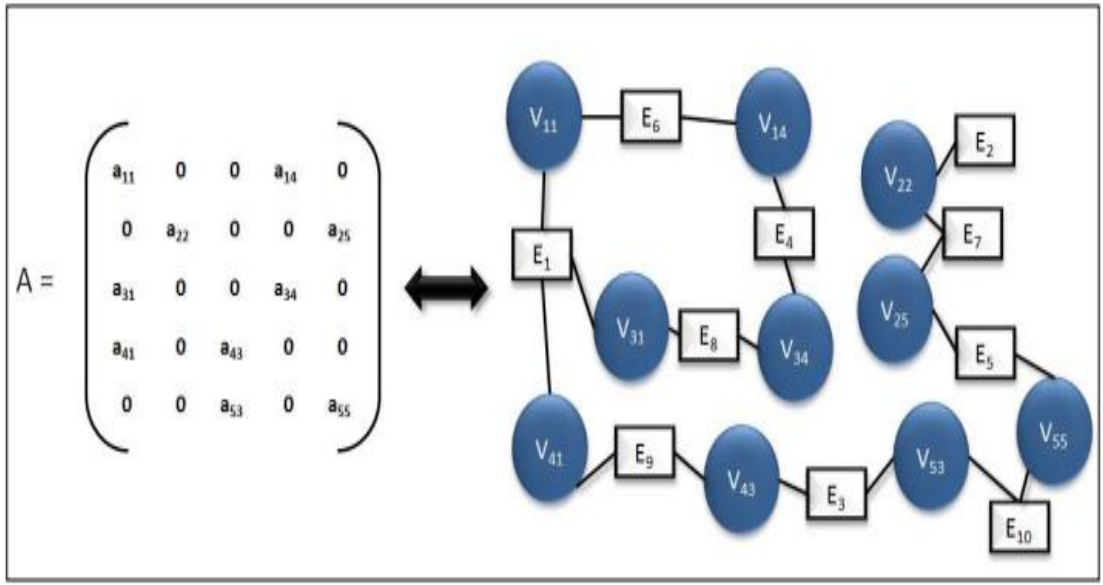

**Figure 0.9 : Hypergraph 2D**

Une étude de la scalabilité des méthodes de partitionnement des hypergraphes a montré que les méthodes de partitionnement 2D sont meilleures (scale better) que celles 1D [UçÇ10].

Le problème de partitionnement de base d'un hypergraphe est de partitionner les sommets (V) en k ensembles approximativement égaux tels que le nombre de coupes des hyperarcs (E) soit minimisé. Un hyperarc est qualifié de coupé s'il connecte au moins deux sommets appartenant à des partitions différentes.

Ces dernières années, les algorithmes de partitionnement multi-niveaux sont devenus l'approche standard pour le partitionnement des hypergraphes. Parmi les avantages qui favorisent l'utilisation de ces algorithmes, nous pouvons citer le fait qu'ils fournissent des solutions de haute qualité, qu'ils peuvent s'adapter aux hypergraphes de très grande taille, et qu'ils nécessitent une durée de temps relativement faible.

## 4.2.3. Méthode combinée

Cette méthode est désignée pour un cluster multi-cœurs vue qu'elle s'exécute sur deux niveaux inter-nœuds sur les différents nœuds d'un cluster et l'autre niveau intra-nœuds sur les cœurs du nœud.

Dans [MeH12] les auteurs ont montrés que l'adaptation d'une décomposition Hypergraphe inter-nœuds (combinaisons HYP-HYP et HYP-NEZ) permet d'optimiser les communications nécessaires pour la distribution des données. Toutefois, la combinaison NEZ-NEZ qui utilise l'algorithme NEZGT pour la distribution des données inter et intra-nœuds assure le meilleur équilibrage des charges. Par ailleurs, la combinaison HYP- NEZ fournit un meilleur Makespan. En effet, l'algorithme Hypergraphe permet de regrouper au sein d'un même nœud les éléments ayant des composantes de X et Y en commun, puis l'algorithme NEZGT assure un équilibrage des charges entre les cœurs. Une comparaison de la qualité de l'équilibrage des charges fournie par PETSc à celle ainsi que par les combinaisons étudiées a montré que l'écart est grand en faveur de ces dernières.

Cette méthode combinée a effectué une fragmentation en bloc de lignes et une fragmentation en blocs de colonnes pour l'algorithme Hypergraph et une fragmentation en blocs de lignes pour l'approche NEZGT .

|  |  | Décomposition inter-nœud | |
|---|---|---|---|
|  |  | $H_C$ | $N_L$ |
| Décomposition intra-nœud | $H_C$ | √ | √ |
|  | $N_L$ | √ | √ |

Hc : Hypergraph Colonne / Nl : NEZGT Ligne

**Tableau 0.2 : Méthode combiné**

Ce tableau contient les différentes combinaisons effectué par la méthode combinée avec une distribution de données 1D pour l'Hypergraph.

- **Communication :**

À chaque nœud k du cluster est associé un fragment $A_k$ de la matrice creuse, un vecteur $X_k$ (reçu) contenant les éléments de X dont il a besoin pour son calcul et un vecteur $Y_k$ (généré) contenant un résultat partiel de Y.

- **Communications liées au vecteur X (Fan-out):** Envoyer la composante $x_j$ de X à tous les nœuds qui vont l'utiliser pour le calcul de leur PFVC. Dans chaque nœud, $X_k$ est formé alors par l'ensemble des $x_j$ reçus.

Notons $C\_X_k$ la charge (nombre des éléments) du vecteur $X_k$. En moyenne, cette charge $C_{moy}\_X_k$ est égale à $(N+1)/f$. En effet, au pire des cas on peut avoir une charge maximale $C_{max}\_X_k$, égale à N et une charge minimale $C_{min}\_X_k$, égale à 1. Le premier cas a lieu lorsque le fragment est constitué par des éléments non nuls éparpillés sur toutes les colonnes de la matrice. Le deuxième cas, quant à lui, apparait lorsque tous les éléments non nuls du fragment appartiennent à la même colonne.

$$1 \leq C\_X_k \leq N$$

Soit $FR\_X_k$ le facteur de réduction du vecteur X pour un fragment donné $A_k$. Ce facteur représente le gain obtenu par l'envoi uniquement des éléments utiles de X (qui participeront au calcul) par rapport à l'envoi de la totalité des éléments :

$$FR\_X_k = N/C\_X_k$$

Il est clair qu'en moyenne, on peut avoir une réduction $FR_{moy}\_X=(f \times N)/(N+1)$. Au pire des cas, lorsque la totalité de X est envoyée vers un nœud k c.-à-d. $C\_X_k=N$, ce facteur est égal à 1, et au meilleur des cas on peut avoir une réduction allant jusqu'à N lorsque le fragment $A_k$ ne contient qu'une seule colonne.

$$1 \leq FR\_X_k \leq N$$

- **Communications liées au vecteur Y (Fan-in) :** Envoyer les résultats partiels de Y au nœud master. Chaque fragment $A_k$ génère un vecteur résultat $Y_k$ avec $C\_Y_k$ éléments non nuls. Il est clair qu'en moyenne Cmoy_Y est égale à N/f. On peut prouver aussi qu'au pire des cas on a $C_{max}\_Y$ est égale à N et $C_{min}\_Y$ égale à 1. Le premier cas a lieu lorsqu'il existe un fragment dont les éléments non nuls sont éparpillés sur toutes les lignes de la matrice. Le deuxième cas arrive lorsque tous les éléments d'un fragment donné appartiennent à la même ligne.

$$1 \leq C\_Y_k \leq N$$

Soit $DR_k$, les données reçues par un nœud k, l'ensemble de ces données est constitué par les éléments non nuls $A_k$ et le vecteur $X_k$.

$$DR_k = NZ_k + C\_X_k$$

Au pire des cas, la quantité des données reçues est égale à NZ-1+N et au meilleur des cas elle est égale à 2. En effet, le pire des cas a lieu lorsqu'on a f=2 : un premier fragment avec NZ-1 éléments non nuls éparpillés sur toutes les colonnes de la matrice ($NZ_k$=NZ-1 et $C\_X_k$=N) et l'autre avec un seul ($NZ_k$=1 et $C\_X_k$=1).

$$2 \leq DR_k \leq NZ-1+N$$

Chaque nœud k reçoit alors (nombre de réels) :

$$RECEPTION = DR_k = O(N+NZ)$$

Soit $DE_k$ la quantité des données envoyées par un nœud k, cette quantité est égale à la charge du vecteur $Y_k$ ($C\_Y_k$).

$$1 \leq DE_k = C\_Y_k \leq N$$

Chaque nœud k envoie alors au nœud master (nombre de réels) :

$$ENVOI = DE_k = O(N)$$

Il est à noter que la complexité des calculs ainsi que la complexité des communications sont, théoriquement, les mêmes pour les deux méthodes de fragmentation NEZGT et hypergraphe.

## 5. Conclusion

Durant notre travail nous allons proposer une décomposition en blocs de colonnes de l'approche NEZGT que nous allons appeler par la suite NEZGT$_{colonne}$. Et nous essayons d'exploiter d'autres combinaisons avec la méthode Hypergraphe. Dans le chapitre suivant, nous allons définir notre approche et les différentes combinaisons à tester et nous les validons par une suite d'expérimentations afin d'évaluer ces performances en termes de temps de calcul pour Y et équilibrage de charges pour des matrices réels choisis à partir de la collection TIM DAVIS.

# Chapitre 4

ETUDE EXPERIMENTALE DE LA DISTRIBUTION DU PMVC

# 1. Introduction

Dans ce chapitre nous présentons tout d'abord notre approche « NEZGT Colonne » qui consiste à faire une fragmentation de la matrice creuse en blocs de colonnes. Ensuite, nous s'intéressons à l'évaluation expérimentale de l'approche proposée pour la distribution du PMVC.

Nous présentons ensuite la stratégie utilisée pour la gestion des accès mémoire dans notre plateforme cible du test ainsi que les différentes bibliothèques utilisées. Enfin, nous validons notre approche par une suite d'expérimentations effectuées sur un cluster de la plate-forme GRID'5000.

# 2. Approche proposée

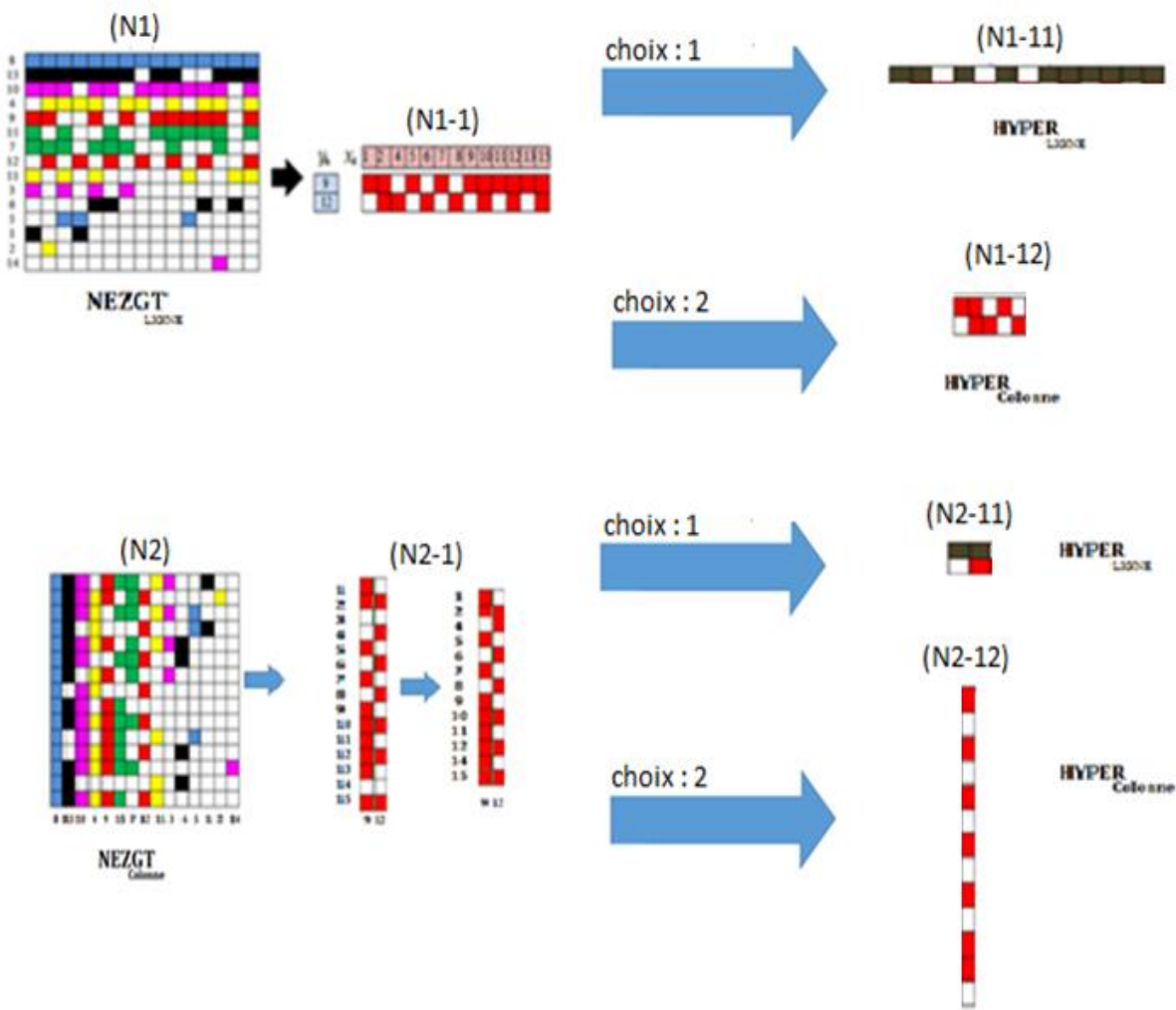

**Figure 0.1 : Figure représentative des combinaisons à tester**

|  | Type décomposition | |
|---|---|---|
|  | **Inter-nœud** | **Intra-nœud** |
| **Algorithme ou Approche Adapté** | **NEZGT Colonne** | HYPER LIGNE |
|  | **NEZGT Colonne** | HYPER COLONNE |
|  | **NEZGT LIGNE** | HYPER LIGNE |
|  | **NEZGT LIGNE** | HYPER COLONNE |

**Tableau 0.1 : Combinaison à tester**

Dans la figure 4.1 nous avons présenté les 4 combinaisons que nous allons les tester tout au long de notre étude expérimentale pour la distribution de la matrice en n fragments qui sont :

NC-HC, NC-HL, NL-HL et NL-NC.

Les matrices (N1-11), (N1-12), (N2-11) et (N2-12) représentent le résultat de l'application d'une distribution inter-nœuds en utilisant respectivement l'approche NEZGT et une distribution intra-nœuds avec la méthode Hypergraph.

Le bloc (N1-1) obtenu à partir de la première matrice en appliquant l'approche NEZGT ligne subit par la suite une décomposition en f blocs. La décomposition suivante sera avec la méthode Hypergraph :

- En appliquant l'approche NEZGT (N1-1) ainsi le fragment obtenu correspond à l'application de la première combinaison (NEZGT) décomposition inter-nœuds sur les lignes et (HYPER) décomposition intra-nœuds sur les lignes (H11).

Le bloc (N1-1) obtenu à partir de la première matrice en appliquant l'approche NEZGT ligne subit par la suite une décomposition en f blocs. La décomposition suivante sera avec la méthode Hypergraph :

- En appliquant l'approche NEZGT (N1-1) ainsi le fragment obtenu correspond à l'application de la première combinaison (NEZGT) décomposition inter-nœuds sur les lignes et (HYPER) décomposition intra-nœuds sur les colonnes (H12) qui a été déjà réalisé dans [MEH12].

Le bloc (N2-11) obtenu à partir de la première matrice en appliquant l'approche NEZGT colonne subit par la suite une décomposition en f blocs. Deux choix sont possibles pour cette décomposition :

- En appliquant L'approche NEZGT ainsi les deux fragments obtenus correspondent à l'application de la première combinaison (NEZGT) décomposition inter-nœuds sur les colonnes et (HYPER) décomposition intra-nœuds sur les colonnes (N2-11)
- Alors que la deuxième combinaison correspond à deux fragmentations inter-nœuds utilisant NEZGT sur les lignes et intra-nœuds utilisant la méthode (HYPER) sur les lignes (N2-12).

## **Approche proposée « NEZGT$_{COLONNE}$»**

« NEZGT Colonne» est une heuristique à trois phases identique à celle de la ligne mais le traitement se fait dans cette approche sur les colonnes, elle contient 3 phases aussi. Les phases 0 et 1 correspondent à l'heuristique SPT (Shortest Processing Time) (esp. LPT (Largest Processing Time)). La phase 2 est une heuristique d'amélioration itérative. La phase 0 consiste à trier les Colonnes de la matrice A par nombre d'éléments non nuls croissant (resp, décroissant).

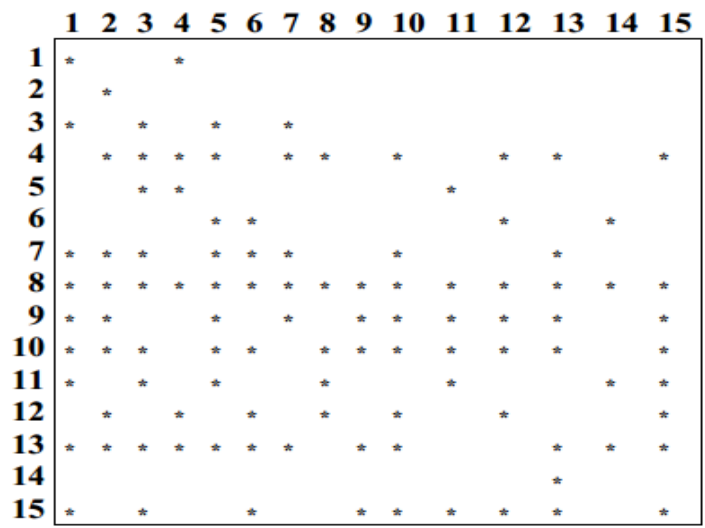

**Figure 0.2 : La matrice étudiée dans l'exemple**

**Exemple**: fragmentation NEZGT ( Colonne )

| Indice de Colonne | 1 | 2 | 3 | 4 | 5 | 6 | 7 | 8 | 9 | 10 | 11 | 12 | 13 | 14 | 15 |
|---|---|---|---|---|---|---|---|---|---|---|---|---|---|---|---|
| Nb d'éléments non nul / Colonne | 9 | 8 | 9 | 6 | 9 | 7 | 6 | 4 | 5 | 8 | 6 | 7 | 8 | 4 | 8 |

En triant les colonnes dans l'ordre décroissant du nombre d'éléments non nuls , on obtients

| Indice de Colonne | 1 | 3 | 5 | 2 | 10 | 13 | 15 | 6 | 12 | 4 | 7 | 11 | 9 | 8 | 14 |
|---|---|---|---|---|---|---|---|---|---|---|---|---|---|---|---|
| Nb d'éléments non nul / Colonne | 9 | 9 | 9 | 8 | 8 | 8 | 8 | 7 | 7 | 6 | 6 | 6 | 5 | 4 | 4 |

**Figure 0.3 : La matrice après la phase 0**

- Dans la première, nous utilisons l'heuristique LS qui consiste d'abord, à affecter la colonne i (i=1…f) au fragment i. Par la suite, la colonne suivante est affectée au fragment le moins chargé et ainsi de suite. En d'autres termes, la colonne courante non affectée est toujours attribuée au fragment le moins chargé jusqu'à épuisement des colonnes.

La phase 1 de l'algorithme NEZGT(Colonne) fournit le résultat optimal suivant:

| Fragment | Indice de ligne (Nombre d'éléments non nuls par Colonne) | Nombre d'éléments non nuls par fragment |
|---|---|---|
| 1 (3 Colonnes) | 6(7) ; 4(6) ; 9(5) | 18 |
| 2 (3 Colonnes) | 12(7) ; 7(6) ; 8(4) | 17 |
| 3 (3 Colonnes) | 15(8)) ; 11(6) ; 14(4) | 18 |
| 4 (2 Colonnes) | 2(8) ; 5(9) | 17 |
| 5 (2 Colonnes) | 10(8) ; 11(6) ; | 17 |
| 6 (2 Colonnes) | 13(8) ; 3(9) | 17 |

**Figure 00.4 : La matrice après la phase 1**

- La seconde phase, qui est une phase d'amélioration, est une heuristique itérative. Elle permet, à travers des raffinements successifs, d'améliorer la fragmentation précédente, conduisant ainsi à un meilleur équilibrage. Le critère choisi étant FD (différence entre les deux charges extrêmes), la procédure conçue consiste à effectuer des échanges ou transferts successifs de colonnes entre le fragment le plus chargé et le fragment le moins chargé.

Plus précisément, soit fcmx (resp. fcmn) le fragment le plus (resp. moins) chargé, la procédure que nous adoptons consiste, soit à transférer une colonne appropriée de « fcmx » à « fcmn » ou bien à échanger une colonne du premier fragment avec une colonne du second et ce, dans le but de réduire la différence entre les deux charges, notée Diff. L'opération de transfert consiste à choisir une colonne dont le nombre d'éléments non nuls, noté nzx est inférieur à Diff. L'opération d'échange consiste, quant à elle, à choisir deux colonnes (la première appartenant à fcmx et la seconde à fcmn) telles que la différence entre le nombre des éléments non nuls du premier, noté nzx, et le nombre des éléments non nuls du second, noté nzn, soit inférieure à Diff i.e : nzx – nzn < Diff. Une meilleure alternative, toutefois plus coûteuse, consiste à optimiser le choix afin de minimiser le nouvel écart entre les charges, ce qui revient à minimiser la quantité │Diff/2-nzx│en cas de transfert (resp. │Diff/2-(nzx-nzn)│ en cas d'échange). Cette procédure est itérée tant qu'il est possible de réduire le critère FD et/ou qu'on ne dépasse pas un nombre d'itérations fixé à l'avance.

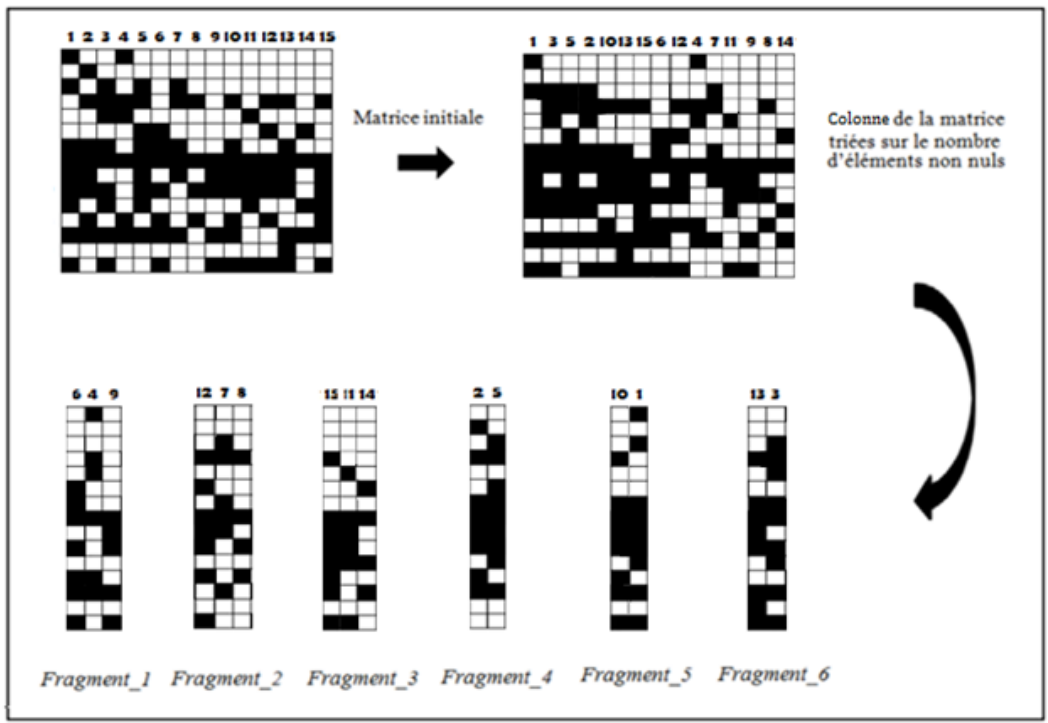

**Figure 0.5 : Exemple d'application du « NEZGT Colonne »**

# 3. Plateforme de test

Nous présentons dans cette section l'architecture NUMA utilisé par chaque nœud dans notre Plateforme de test Grid5000 que nous détaillons quelques spécifications de cette Grid par la suite.

## **Architecture NUMA**

Les plates-formes multi-cœurs avec un accès mémoire non uniforme (NUMA) sont devenues des ressources usuelles de calcul haut performance. Dans ces plates-formes, la mémoire partagée est constituée de plusieurs bancs de mémoires physiques organisés hiérarchiquement.

Cette hiérarchie est également constituée de plusieurs niveaux de mémoires caches et peut être assez complexe [RIB11].

Les machines parallèles NUMA (Non Uniform Memory Access) sont des machines multiprocesseurs à mémoire partagée, dans lesquelles la mémoire est morcelée en plusieurs bancs.

Chaque processeur accède à une zone mémoire locale rapidement, et à des zones mémoire distantes avec un temps d'accès plus long selon la proximité géographique des processeurs. Concrètement, une machine NUMA est composée d'une interconnexion de blocs basés sur l'architecture SMP, reliés entre eux par un réseau rapide ou un commutateur, comme illustré sur la figure 4.6.

Chaque bloc est constitué d'un nombre limité de processeurs et de leurs mémoires cache, connectés à un banc mémoire local. Le programmeur perçoit ces différents bancs de mémoire sous la forme d'une unique mémoire globale.

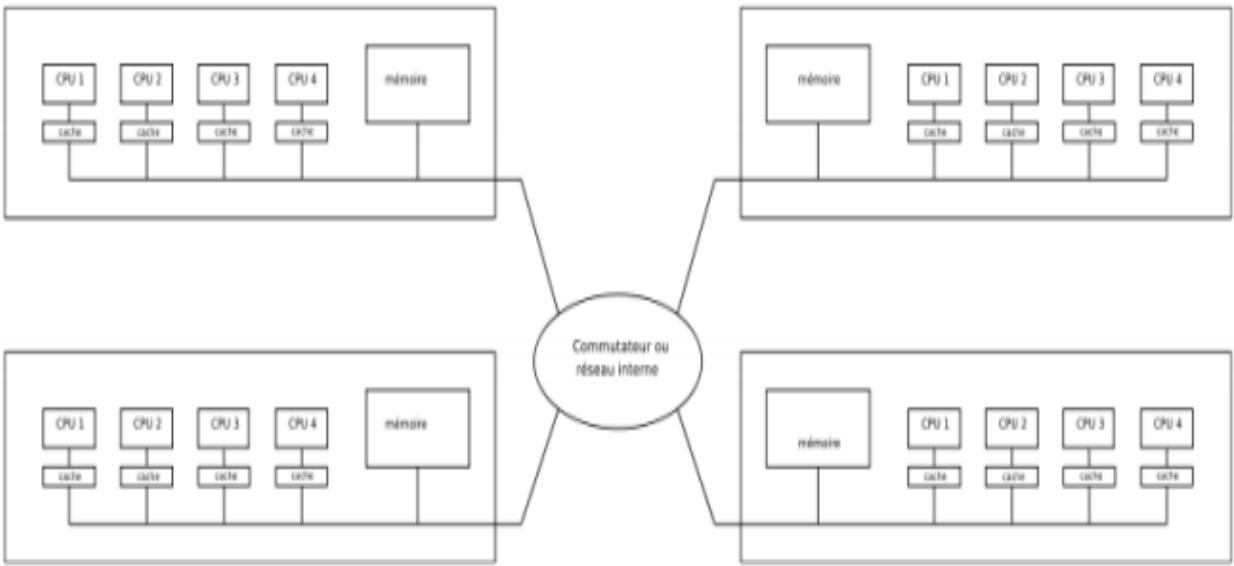

**Figure 0.6 : Machine NUMA qui a 4 nœuds chacun a 4 processeurs & 1 banc mémoire**

Comme les performances du réseau interconnectant les blocs sont limités, les variations pour l'accès aux différents bancs mémoire conduisent à la notion de facteur NUMA. Un facteur NUMA correspond au temps d'un accès à la mémoire locale, divisé par le temps d'accès à la mémoire distante. Ce facteur traduit le non uniformité d'accès à la mémoire, et contribue à l'évaluation de la machine en termes de performances. Plus celui-ci se rapproche de un, plus le coût supplémentaire, engendré par un accès distant par rapport à un accès local, est faible. On s'approche ici du modèle d'une machine accédant uniformément à une mémoire globale. Pour des machines NUMA dont la topologie est complexe, ou hiérarchique à plusieurs niveaux, plusieurs facteurs NUMA caractérisent la machine, compris aujourd'hui entre 110 et 300% selon les architectures [**JEU07**].

## 3.1. Grid'5000

Le Gri5000 est une plateforme expérimentale qui permet de développer des expériences à large échelle et reproductibles dans le contexte de la recherche en calcul parallèle et distribué. Grid'5000 est déployée à l'échelle nationale française dans plusieurs sites, les sites sont reliés entre eux par le réseau RENATER qui offre une bande passante de 10Go/s entre la plupart d'entre eux [AYCB16a][AYCB16b]. Chaque site contient un ou plusieurs clusters qui contiennent à leur tour un nombre variable de nœuds avec des caractéristiques différentes (CPU, mémoire, disques, interfaces réseaux etc.). Grid'5000 est utilisé par les plusieurs équipes tels que ALGORILLE, CARAMEL, ORPAILLEUR and VERIDIS.

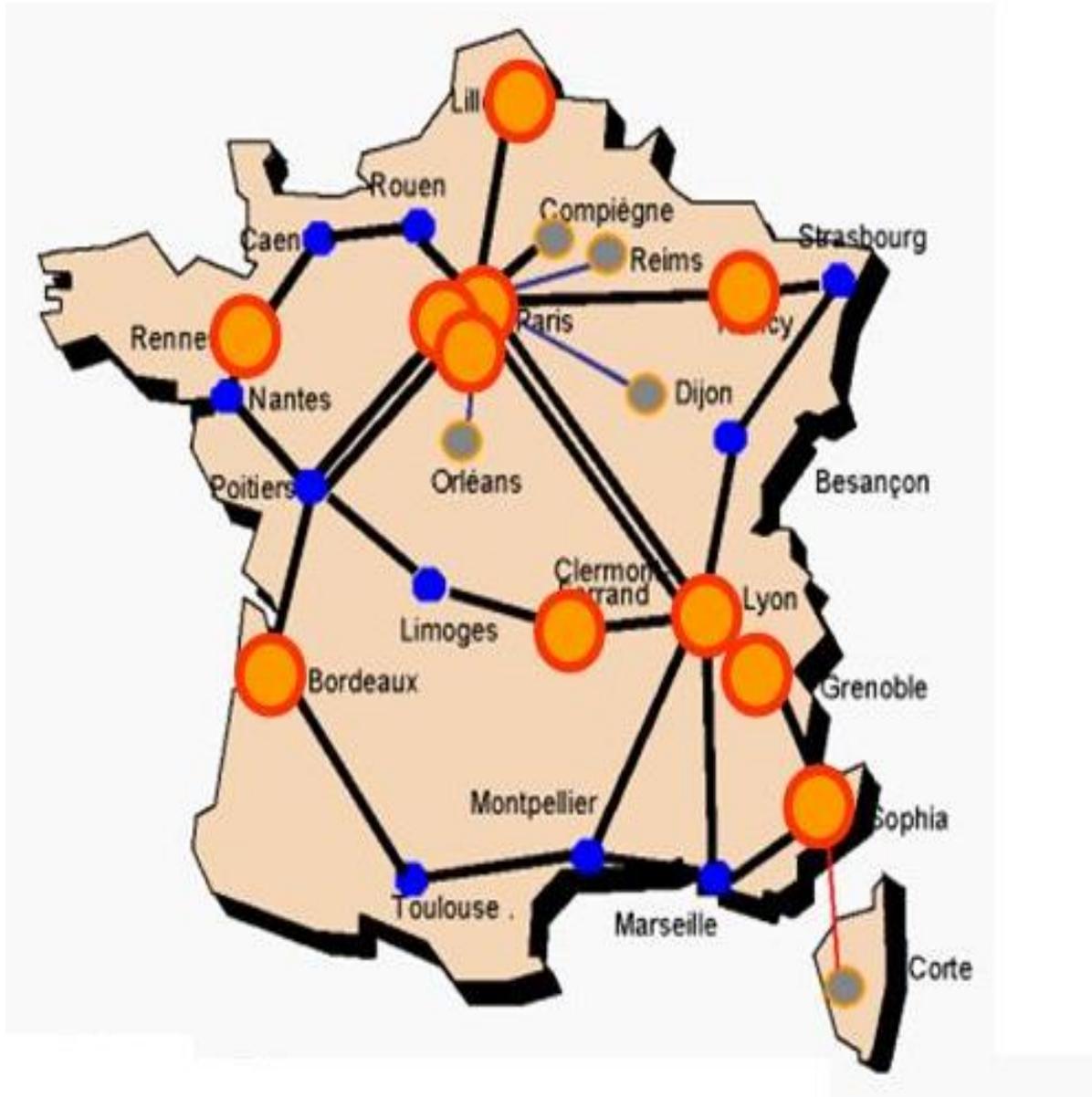

**Figure 0.7 : Les différents sites en Grid5000**

## 3.2. Bibliothèques de programmation parallèle

L'évolution de l'architecture des grappe de calcul à nœuds multi-cœurs et les contraintes du parallélisme nous oriente vers une parallélisation hybride dans le but de tirer parti des avantages des différentes approches [HCE09][RHJ09].

L'approche hybride consistant à mixer au sein d'un même code une parallélisation sur plusieurs niveaux, un niveau OpenMP au sein d'un nœud à mémoire partagée et un niveau MPI entre les différents nœuds de calcul, est une alternative intéressante qui offre beaucoup d'avantages.

 **MPI :** MPI (Message Passing Interface) est l'environnement de programmation le plus connu pour les architectures à mémoire distribuée [HCE09]. Dans un modèle de programmation parallèle par échange de messages (MPI), le programme est exécuté par plusieurs processus. Chaque processus exécute un exemplaire du programme et a accès à sa propre mémoire. De ce fait, les variables du programme deviennent des variables locales au niveau de chaque processus. De plus, un processus ne peut pas accéder à la mémoire des processus voisins. Il peut toutefois envoyer des informations à d'autres processus à condition que ces derniers (processus récepteurs) soient au courant qu'ils devaient recevoir ces informations du processus émetteur [DeS08].

La communication entre deux processus peut être de type point à point (envoi et réception d'une donnée d'un émetteur vers un destinataire), de type collectif (diffusion d'un message à un groupe de processus, opération de réduction, distribution ou redistribution des données envoyées) .

Dans un programme MPI, un commutateur permet de connaître l'ensemble des processus actifs. MPI définit des fonctions qui, à tout moment, permettent de connaître le nombre de processus gérés dans un commutateur et le rang d'un processus dans le commutateur. La bibliothèque met de nombreuses fonctions à la disposition du programmeur, lui permettant ainsi d'effectuer différents types d'envoi et de réception de messages (point à point bloquant (synchrone), point à point non bloquant (asynchrone), collectifs et des opérations globales (barrière, réduction, diffusion)).

 **OpenMP :** OpenMP est un outil de programmation à base de directives pour l'implémentation des programmes sur des systèmes partagés en utilisant le concept de multithreading. L'approche la plus simple pour paralléliser un code séquentiel avec OpenMP consiste à paralléliser les boucles. Pour le faire, il s'agit de lancer plusieurs threads, chacun exécute une ou plusieurs itérations d'une boucle [HCE09]. OpenMP possède une interface de programmation pour les langages de programmation C, C++ et Fortran, et il est supporté par de nombreux environnements (e.g. Linux ou Windows).

☐ **Programmation hybride** : La programmation hybride parallèle consiste à mélanger plusieurs paradigmes de programmation parallèle dans le but de tirer parti des avantages des différentes approches. Avec les évolutions récentes des architectures HPC, la programmation hybride MPI-OpenMP se développe et semble devenir l'approche privilégiée pour les supercalculateurs [LaW10]. L'approche hybride MPI-OpenMP consiste à mixer au sein d'un même code une parallélisation à plusieurs niveaux : un niveau OpenMP au sein d'un nœud à mémoire partagée (parallélisation entre cœurs) et un niveau MPI entre les différents nœuds de calcul [HCE09][RHJ09]. Des résultats récents ont montré qu'une telle réorganisation peut améliorer les performances sur des clusters dont les nœuds sont à mémoire partagée [RHJ09].

## a. Bibliothèques BLAS et spBLAS

Les opérations de base d'algèbre linéaire sont regroupées dans une bibliothèque numérique appelée BLAS : Basic Linear Algebra Subroutines. Cette bibliothèque est souvent fournie par le constructeur et optimisée pour une architecture donnée. Les opérations sont divisées en trois niveaux appelés BLAS1 (pour les opérations entre vecteurs), BLAS2 (pour les opérations entre matrices et vecteurs) et BLAS3 (pour les opérations entre matrices). L'optimisation concerne l'ordre des opérations et l'accès à la mémoire, pour exploiter au mieux la hiérarchie (mémoire principale, mémoire cache, etc..). Les performances (vitesse de calcul) sont d'autant meilleures que le niveau est élevé [ENP04].

La bibliothèque BLAS ne supporte qu'un certain nombre limité de structures de matrices creuses. Néanmoins, il existe une version creuse de cette bibliothèque : Sparse BLAS (spBALS) [DHR01] comportant des routines pour le traitement de matrices creuses. Elle contient également les trois niveaux des opérations exploitées dans la version dense. Toutefois, seul un petit sous-ensemble de BLAS est spécifié [SPB] :

- Niveau 1: produit scalaire creux, mise à jour de vecteur, et Gather / Scatter.
- Niveau 2: produit matrice-vecteur creux et résolution des matrices triangulaires.
- Niveau 3: produit matrice creuse-matrice dense et résolution de systèmes triangulaires avec plusieurs RHS (Right Hand Side).

Pour le calcul de notre PMVC, nous nous sommes basés alors sur le niveau 2 de spBLAS. Ainsi, au niveau de chaque cœur, il y'aura appel à la bibliothèque via la fonction csr_double_mv pour le calcul du PFVC (Produit Fragment Vecteur Creux).

## b. Bibliothèques de partionnement de l' Hypergraphe

Il existe beaucoup de bibliothèques qui permettent de faire le partitionnement d'hypergraphe voici quelques-unes :

- **PaToH** : un outil séquentiel de partitionnement séquentiel d'hypergraphes utilisant des méthodes efficaces dans le cadre du paradigme de partitionnement multi-niveaux (voir chapitre 3, section 4.2). Il a été d'abord utilisé pour paralléliser efficacement le produit matrice-vecteur par décomposition 1D et 2D de matrice creuse [PDC].

- **hMETIS** : un ensemble de programmes séquentiels pour le partitionnement d'hypergraphes, tels que ceux correspondant aux circuits VLSI. hMETIS est basée sur l'algorithme multi-niveaux de partitionnement d'hypergraphe (voir chapitre 3, section 4.2)[HCP].

- **Mondriaan**: Un programme séquentiel utilisant le modèle hypergraphe pour le partitionnement de matrices creuses en se basant sur l'algorithme multi-niveaux. Le programme est basé sur un algorithme de bi-partitionnement récursif qui coupe la matrice horizontalement et verticalement [MSP].

Cependant, tous ces packages s'exécutent en séquentiel. Pour les grandes applications parallèles, il sera plus convenable d'utiliser des bibliothèques utilisant un partitionnement exécuté en parallèle [DBH06], tels que :

- **Parkway**: bibliothèque parallèle de partitionnement d'hypergraphes utilisant l'algorithme multi-niveaux [PHP].

☐ **Zoltan**: bibliothèque fournissant un ensemble de services de gestion de données pour des applications parallèles. Parmi ces services, elle fournit un outil de partitionnement parallèle d'hypergraphe basé sur l'algorithme multi-niveaux: Zoltan-PHG [ZPP][DBH06].

Pour nos tests, nous avons choisi d'utiliser le partitionneur parallèle d'hypergraphe de Zoltan(Zoltan-PHG) décrit dans [DBH06] et qui a donné, par rapport à Parkway, une bonne accélération parallèle sur un cluster Linux avec jusqu'à 64 processeurs sans compromettre la qualité de la partition.

# 4. Résultats numériques et interprétations
## 4.1. Décomposition inter et intra-nœud

Dans notre distribution du PMVC, chaque cœur i (i=1.. fc) d'un nœud donné k est responsable du calcul d'un Produit Fragment Vecteur Creux (PFVC) $Y_{ki} = A_{ki} \times X_{ki}$. A cet effet, chaque fragment $A_{ki}$, ainsi que chaque vecteur $Y_{ki}$ et chaque vecteur $X_{ki}$ sont placés sur le banc mémoire du nœud NUMA auquel appartient le ième cœur [MeH12] [MeH13]. Nous supposons qu'on a un thread par cœur (voir figure 5.1).

Dans cette section, nous étudions les performances des quatre combinaisons que nous avons proposées pour la distribution à deux niveaux du PMVC : distributions inter et intra-nœuds. Les quatre combinaisons testées sont $NEZGT_L$-$HYPER_C$ ($NEZGT_{LIGNE}$ en inter et $HYPERRAPH_{COLONNE}$ intra-nœuds), $NEZGT_L$-$HYPER_L$ ($NEZGT_{LIGNE}$ en inter et $HYPERGRAPH_{LIGNE}$ en intra-nœuds), $NEZGT_C$-$HYPER_L$ ($NEZGT_{COLONNE}$ en inter-nœuds et $HYPERFRAPH_{LIGNE}$ en intra-nœud) et $NEZGT_C$-$HYPER_C$ ($NEZGT_{COLONNE}$ en inter-nœuds et $HYPERGRAPH_{COLONNE}$ en intra-nœud).

Les matrices de test (Tableau 4.2) sont des matrices issues de problèmes réels. Elles sont choisies parmi la collection des matrices creuses de Tim Davis, université de Florida [CTD].

Nous allons effectuer un ensemble de tests sur une grappe homogène un cluster *'paravance'* du site RENNES qui appartient à la Grid 5000 à 8 cœurs chaque nœud (2 CPU avec 8 cœurs par CPU).

| Matrice | | N | NNZ | DENSITE | Domaines D'application |
|---|---|---|---|---|---|
| 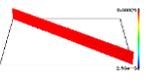 | bcsstm09 | 1083 | 1083 | 0.009% | Ingénierie structurelle |
| 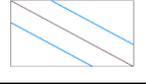 | thermal | 3456 | 66528 | 0.55% | Problème thermique |
| 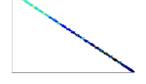 | t2dal | 4257 | 20861 | 0.11% | Problème de la réduction de modèle |
| 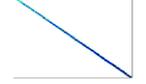 | ex19 | 12005 | 259879 | 0.18% | Problème dynamic de fluides |
| 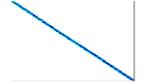 | epb1 | 14743 | 95053 | 0.04% | Problème thermique |
| 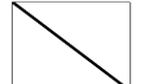 | af23560 | 23560 | 484256 | 0.08% | Analyse de stabilité transitoire d'un Navier-Stokes |
| 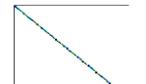 | spmsrtls | 29995 | 129971 | 0.01% | Problème mathématique |
| 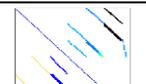 | Zhao1 | 33861 | 166453 | 0.01% | Problème de l'électromagnétisme |
| Densité = (NZ / $N^2$) * 100 | | | | | |

**Tableau 0.2 : Matrice TIM DAVIS**

Après faire les différentes expérimentations nous avons enregistré les résultats dans les tableaux suivants.

| Matrice | Nb Nœuds | $LB_{noeuds}$ | $LB_{coeurs}$ | Temps Calcul Y | Durée Scatter | Durée Gather | Durée Construction de Y | Durée Gather + Construction de Y | Temps Total Du PMVC |
|---|---|---|---|---|---|---|---|---|---|
| Af23560 | 2 | 1,09 | 2,01 | 0,000294 | 0,013487 | 0,000754 | 0,000267 | 0,001022 | 0,001316 |
| | 4 | 1,17 | 2,95 | 0,000223 | 0,018990 | 0,002087 | 0,000464 | 0,002552 | 0,002775 |
| | 8 | 1,09 | 2,35 | 0,000182 | 0,017881 | 0,004243 | 0,001024 | 0,005268 | 0,005451 |
| | 16 | 1,19 | 2,30 | 0,000125 | 0,018085 | 0,006417 | 0,001296 | 0,007714 | 0,007840 |
| | 32 | 1,66 | 4,64 | 0,000059 | 0,138148 | 0,011429 | 0,001713 | 0,013143 | 0,013212 |
| | 64 | 1,39 | 1,61 | 0,000043 | 0,101702 | 0,012314 | 0,001744 | 0,014058 | 0,014102 |
| Bcstm09 | 2 | 1,03 | 1,03 | 0,000027 | 0,000121 | 0,000065 | 0,000010 | 0,000075 | 0,000103 |
| | 4 | 1,00 | 1,06 | 0,000025 | 0,000172 | 0,000069 | 0,000010 | 0,000076 | 0,000102 |
| | 8 | 1,12 | 1,12 | 0,000020 | 0,000240 | 0,000073 | 0,000011 | 0,000084 | 0,000105 |

|  |  |  |  |  |  |  |  |  |
|---|---|---|---|---|---|---|---|---|
|  | 16 | 1,25 | 1,25 | 0,000015 | 0,001596 | 0,000080 | 0,000011 | 0,000091 | 0,000107 |
|  | 32 | 1,50 | 1,50 | 0,000014 | 0,003503 | 0,000123 | 0,000011 | 0,000134 | 0,000149 |
|  | 64 | 2,00 | 2,00 | 0,000014 | 0,008101 | 0,000156 | 0,000011 | 0,000167 | 0,000181 |
| **Epb1** | 2 | 1,03 | 1,76 | 0,000143 | 0,002465 | 0,000655 | 0,000147 | 0,000802 | 0,000945 |
|  | 4 | 1,42 | 1,79 | 0,000094 | 0,003237 | 0,001003 | 0,000224 | 0,001227 | 0,001321 |
|  | 8 | 1,31 | 1,43 | 0,000080 | 0,003647 | 0,001935 | 0,000344 | 0,002279 | 0,002360 |
|  | 16 | 1,05 | 1,07 | 0,000050 | 0,005413 | 0,002476 | 0,000404 | 0,002880 | 0,002930 |
|  | 32 | 1,09 | 1,15 | 0,000036 | 0,012443 | 0,002532 | 0,000424 | 0,002956 | 0,002992 |
|  | 64 | 1,29 | 1,34 | 0,000029 | 0,021799 | 0,002547 | 0,000430 | 0,002977 | 0,003006 |
| **Ex19** | 2 | 1,26 | 1,74 | 0,000168 | 0,006498 | 0,000441 | 0,000127 | 0,000568 | 0,000737 |
|  | 4 | 1,79 | 1,86 | 0,000151 | 0,047990 | 0,001058 | 0,000209 | 0,001268 | 0,001419 |
|  | 8 | 1,44 | 3,03 | 0,000117 | 0,049797 | 0,002185 | 0,000453 | 0,002638 | 0,002755 |
|  | 16 | 1,27 | 1,57 | 0,000076 | 0,048489 | 0,003308 | 0,000597 | 0,003905 | 0,003981 |
|  | 32 | 1,23 | 1,41 | 0,000044 | 0,128994 | 0,003322 | 0,000752 | 0,004075 | 0,004120 |
|  | 64 | 1,13 | 1,63 | 0,000043 | 0,155964 | 0,003392 | 0,001023 | 0,004416 | 0,004459 |
| **T2dal** | 2 | 1,03 | 1,78 | 0,000041 | 0,000717 | 0,000137 | 0,000091 | 0,000229 | 0,000270 |
|  | 4 | 1,06 | 1,56 | 0,000032 | 0,000724 | 0,000232 | 0,000133 | 0,000365 | 0,000398 |
|  | 8 | 1,15 | 1,33 | 0,000028 | 0,000849 | 0,000315 | 0,000168 | 0,000483 | 0,000412 |
|  | 16 | 1,18 | 1,22 | 0,000025 | 0,003468 | 0,000348 | 0,000195 | 0,000543 | 0,000569 |
|  | 32 | 1,37 | 1,48 | 0,000025 | 0,004378 | 0,000350 | 0,000205 | 0,000555 | 0,000581 |
|  | 64 | 1,80 | 2,00 | 0,000024 | 0,009409 | 0,000392 | 0,000213 | 0,000606 | 0,000630 |
| **Thermal** | 2 | 1,08 | 2,40 | 0,000080 | 0,001910 | 0,000158 | 0,000085 | 0,000243 | 0,000323 |
|  | 4 | 1,23 | 1,51 | 0,000064 | 0,002165 | 0,000314 | 0,000074 | 0,000389 | 0,000454 |
|  | 8 | 1,09 | 2,02 | 0,000033 | 0,041185 | 0,000555 | 0,000121 | 0,000677 | 0,000710 |
|  | 16 | 1,24 | 1,46 | 0,000032 | 0,041243 | 0,001115 | 0,000183 | 0,001298 | 0,001331 |
|  | 32 | 1,32 | 1,79 | 0,000031 | 0,042177 | 0,001227 | 0,000181 | 0,001409 | 0,001440 |
|  | 64 | 1,69 | 2,58 | 0,000028 | 0,042484 | 0,001832 | 0,000265 | 0,002097 | 0,002126 |
| **Spmsrtls** | 2 | 1,01 | 1,20 | 0,000175 | 0,005584 | 0,000851 | 0,000284 | 0,001135 | 0,001310 |
|  | 4 | 1,00 | 1,16 | 0,000130 | 0,005571 | 0,001847 | 0,000462 | 0,002410 | 0,002540 |
|  | 8 | 1,00 | 1,01 | 0,000101 | 0,006488 | 0,002684 | 0,000475 | 0,003160 | 0,003261 |
|  | 16 | 1,02 | 1,03 | 0,000047 | 0,009434 | 0,002699 | 0,000523 | 0,003222 | 0,003270 |
|  | 32 | 1,04 | 1,05 | 0,000041 | 0,014050 | 0,002695 | 0,000534 | 0,003230 | 0,003271 |
|  | 64 | 1,07 | 1,11 | 0,000040 | 0,016417 | 0,003667 | 0,000561 | 0,004228 | 0,004269 |
| **Zhao1** | 2 | 1,07 | 1,58 | 0,000222 | 0,006220 | 0,000918 | 0,000314 | 0,001232 | 0,001455 |
|  | 4 | 1,91 | 1,93 | 0,000181 | 0,006377 | 0,002180 | 0,000550 | 0,002731 | 0,002912 |

| | 8  | 1,02 | 1,20 | 0,000090 | 0,006455 | 0,003154 | 0,000654 | 0,003809 | 0,003900 |
| | 16 | 1,11 | 1,21 | 0,000053 | 0,012302 | 0,003998 | 0,000713 | 0,004712 | 0,004766 |
| | 32 | 1,14 | 1,21 | 0,000037 | 0,018535 | 0,003820 | 0,000716 | 0,004537 | 0,004575 |
| | 64 | 1,06 | 1,13 | 0,000031 | 0,019899 | 0,004664 | 0,000773 | 0,005438 | 0,005470 |

**Tableau 0.3 : Résultats expérimentation de la combinaison NC-HC**

| Matrice | Nb Nœuds | $LB_{noeuds}$ | $LB_{coeurs}$ | Temps Calcul Y | Durée Scatter | Durée Gather | Durée Construction de Y | Durée Gather + Construction de Y | Temps Total Du PMVC |
|---|---|---|---|---|---|---|---|---|---|
| Af23560 | 2  | 1,33 | 1,67 | 0,006691 | 0,693301 | 0,000757 | 0,000481 | 0,001238 | 0,007930 |
|         | 4  | 1,10 | 2,61 | 0,005874 | 0,731736 | 0,001874 | 0,000964 | 0,002839 | 0,008713 |
|         | 8  | 2,55 | 2,57 | 0,005480 | 0,792750 | 0,003965 | 0,000967 | 0,004933 | 0,010413 |
|         | 16 | 2,10 | 4,83 | 0,000136 | 0,823266 | 0,007364 | 0,001447 | 0,008812 | 0,008948 |
|         | 32 | 5,25 | 8,89 | 0,000092 | 0,875498 | 0,010970 | 0,001652 | 0,012622 | 0,012715 |
|         | 64 | 3,66 | 8,25 | 0,000047 | 0,932969 | 0,012461 | 0,001778 | 0,014240 | 0,014287 |
| Bcstm09 | 2  | 1,03 | 1,03 | 0,000029 | 0,000113 | 0,000062 | 0,000010 | 0,000072 | 0,000101 |
|         | 4  | 1,00 | 1,06 | 0,000027 | 0,000166 | 0,000102 | 0,000010 | 0,000112 | 0,000140 |
|         | 8  | 1,12 | 1,12 | 0,000023 | 0,000239 | 0,000139 | 0,000011 | 0,000150 | 0,000173 |
|         | 16 | 1,25 | 1,25 | 0,000022 | 0,001679 | 0,000141 | 0,000011 | 0,000153 | 0,000175 |
|         | 32 | 1,50 | 1,50 | 0,000020 | 0,003525 | 0,000195 | 0,000011 | 0,000207 | 0,000227 |
|         | 64 | 2,00 | 2,00 | 0,000019 | 0,007976 | 0,000202 | 0,000012 | 0,000215 | 0,000234 |
| Epb1    | 2  | 1,14 | 2,23 | 0,000125 | 0,002475 | 0,000608 | 0,000139 | 0,000748 | 0,000874 |
|         | 4  | 1,18 | 2,22 | 0,000080 | 0,003405 | 0,001078 | 0,000429 | 0,001507 | 0,001588 |
|         | 8  | 1,23 | 1,45 | 0,000052 | 0,003667 | 0,001928 | 0,000496 | 0,002424 | 0,002477 |
|         | 16 | 1,11 | 1,25 | 0,000048 | 0,004315 | 0,002012 | 0,000511 | 0,002523 | 0,002572 |
|         | 32 | 1,11 | 1,23 | 0,000042 | 0,010784 | 0,002522 | 0,000544 | 0,003066 | 0,003109 |
|         | 64 | 1,16 | 1,26 | 0,000028 | 0,021759 | 0,002555 | 0,000559 | 0,003114 | 0,003142 |
| Ex19    | 2  | 1,00 | 1,35 | 0,000217 | 0,006781 | 0,000399 | 0,000239 | 0,000639 | 0,000408 |
|         | 4  | 1,10 | 2,20 | 0,000161 | 0,007744 | 0,000991 | 0,000467 | 0,001458 | 0,000407 |
|         | 8  | 1,09 | 1,73 | 0,000084 | 0,010924 | 0,002021 | 0,000398 | 0,002420 | 0,000486 |
|         | 16 | 1,36 | 1,50 | 0,000065 | 0,011107 | 0,003060 | 0,000525 | 0,003586 | 0,000544 |
|         | 32 | 2,00 | 2,12 | 0,000055 | 0,050829 | 0,003052 | 0,000933 | 0,003986 | 0,004041 |
|         | 64 | 2,61 | 3,44 | 0,000053 | 0,098345 | 0,003564 | 0,001517 | 0,004862 | 0,004915 |
| T2dal   | 2  | 1,04 | 3,40 | 0,000037 | 0,000755 | 0,000169 | 0,000056 | 0,000226 | 0,000263 |
|         | 4  | 1,11 | 1,31 | 0,000027 | 0,000741 | 0,000270 | 0,000061 | 0,000331 | 0,000358 |

| Matrice | Nb | LB$_{nœuds}$ | LB$_{coeurs}$ | Temps Calcul Y | Durée Scatter | Durée Gather | Durée Construction de Y | Durée Gather + Construction de Y | Temps Total Du PMVC |
|---|---|---|---|---|---|---|---|---|---|
| | 8 | 1,24 | 1,39 | 0,000026 | 0,000800 | 0,000359 | 0,000064 | 0,000424 | 0,000450 |
| | 16 | 1,13 | 1,19 | 0,000026 | 0,002144 | 0,000632 | 0,000079 | 0,000711 | 0,000738 |
| | 32 | 1,19 | 1,36 | 0,000025 | 0,004314 | 0,000666 | 0,000103 | 0,000770 | 0,000796 |
| | 64 | 1,16 | 1,38 | 0,000021 | 0,009824 | 0,000669 | 0,000210 | 0,000879 | 0,000909 |
| **Thermal** | 2 | 1,00 | 1,64 | 0,000052 | 0,002052 | 0,000120 | 0,000068 | 0,000189 | 0,000241 |
| | 4 | 1,30 | 1,68 | 0,000041 | 0,002086 | 0,000282 | 0,000138 | 0,000421 | 0,000463 |
| | 8 | 1,27 | 1,56 | 0,000034 | 0,001956 | 0,000634 | 0,000251 | 0,000886 | 0,000920 |
| | 16 | 1,50 | 1,54 | 0,000029 | 0,002543 | 0,001109 | 0,000374 | 0,001484 | 0,001514 |
| | 32 | 1,31 | 2,15 | 0,000030 | 0,004916 | 0,001322 | 0,000449 | 0,001772 | 0,001803 |
| | 64 | 1,58 | 1,97 | 0,000027 | 0,009821 | 0,001349 | 0,000455 | 0,001804 | 0,001831 |
| **Spmsrtls** | 2 | 1,12 | 1,15 | 0,000160 | 0,003807 | 0,000861 | 0,000282 | 0,001144 | 0,001304 |
| | 4 | 1,00 | 1,02 | 0,000131 | 0,005547 | 0,002181 | 0,000378 | 0,002559 | 0,002691 |
| | 8 | 1,00 | 1,10 | 0,000077 | 0,006898 | 0,002672 | 0,000492 | 0,003165 | 0,003243 |
| | 16 | 1,01 | 1,13 | 0,000046 | 0,057451 | 0,002706 | 0,000616 | 0,003322 | 0,003369 |
| | 32 | 1,04 | 1,17 | 0,000024 | 0,087907 | 0,002822 | 0,000841 | 0,003663 | 0,003687 |
| | 64 | 1,09 | 1,18 | 0,000020 | 0,096776 | 0,002962 | 0,000891 | 0,003853 | 0,003874 |
| **Zhao1** | 2 | 1,04 | 1,49 | 0,000210 | 0,005839 | 0,001185 | 0,000298 | 0,001484 | 0,001694 |
| | 4 | 1,49 | 1,51 | 0,000177 | 0,006309 | 0,002251 | 0,000480 | 0,002731 | 0,002909 |
| | 8 | 1,00 | 1,17 | 0,000091 | 0,007405 | 0,003048 | 0,000639 | 0,003688 | 0,003779 |
| | 16 | 1,02 | 1,16 | 0,000061 | 0,086129 | 0,003595 | 0,000719 | 0,004314 | 0,004376 |
| | 32 | 1,06 | 1,22 | 0,000052 | 0,096891 | 0,003633 | 0,000742 | 0,004375 | 0,004427 |
| | 64 | 1,10 | 1,22 | 0,000050 | 0,097024 | 0,004073 | 0,000918 | 0,004992 | 0,005043 |

**Tableau 0.4 : Résultats expérimentation de la combinaison NC-HL**

| Matrice | Nb Nœuds | LB$_{nœuds}$ | LB$_{coeurs}$ | Temps Calcul Y | Durée Scatter | Durée Gather | Durée Construction de Y | Durée Gather + Construction de Y | Temps Total Du PMVC |
|---|---|---|---|---|---|---|---|---|---|
| **Af23560** | 2 | 1,14 | 1,52 | 0,000217 | 0,013034 | 0,000450 | 0,000304 | 0,000754 | 0,000972 |
| | 4 | 1,02 | 2,88 | 0,000130 | 0,015455 | 0,000487 | 0,000309 | 0,000797 | 0,000927 |
| | 8 | 1,28 | 4,59 | 0,000089 | 0,016236 | 0,000697 | 0,000317 | 0,001014 | 0,001104 |
| | 16 | 1,84 | 8,03 | 0,000081 | 0,020432 | 0,000698 | 0,000324 | 0,001023 | 0,001104 |
| | 32 | 6,65 | 8,12 | 0,000056 | 0,097450 | 0,000773 | 0,000323 | 0,001096 | 0,001153 |
| | 64 | 4,77 | 7,12 | 0,000049 | 0,150572 | 0,000824 | 0,000325 | 0,001150 | 0,001199 |
| **Bcstm09** | 2 | 1,03 | 1,03 | 0,000036 | 0,000145 | 0,000112 | 0,000011 | 0,000123 | 0,000159 |

|          | 4  | 1,00 | 1,06 | 0,000035 | 0,000197 | 0,000170 | 0,000011 | 0,000181 | 0,000216 |
|----------|----|------|------|----------|----------|----------|----------|----------|----------|
|          | 8  | 1,12 | 1,12 | 0,000028 | 0,000258 | 0,000238 | 0,000011 | 0,000250 | 0,000279 |
|          | 16 | 1,25 | 1,25 | 0,000025 | 0,002006 | 0,000241 | 0,000012 | 0,000253 | 0,000279 |
|          | 32 | 1,50 | 1,50 | 0,000024 | 0,003628 | 0,000247 | 0,000012 | 0,000259 | 0,000283 |
|          | 64 | 2,00 | 2,00 | 0,000021 | 0,008180 | 0,000254 | 0,000012 | 0,000266 | 0,000288 |
| **Epb1** | 2  | 1,15 | 2,21 | 0,000090 | 0,002509 | 0,000280 | 0,000159 | 0,000440 | 0,000530 |
|          | 4  | 1,60 | 1,72 | 0,000068 | 0,003471 | 0,000303 | 0,000159 | 0,000462 | 0,000531 |
|          | 8  | 1,54 | 2,24 | 0,000034 | 0,004106 | 0,000338 | 0,000166 | 0,000504 | 0,000539 |
|          | 16 | 1,12 | 1,21 | 0,000030 | 0,005980 | 0,000375 | 0,000167 | 0,000542 | 0,000572 |
|          | 32 | 1,11 | 1,20 | 0,000029 | 0,048401 | 0,000377 | 0,000169 | 0,000547 | 0,000576 |
|          | 64 | 1,21 | 1,27 | 0,000029 | 0,024566 | 0,000385 | 0,000178 | 0,000563 | 0,000592 |
| **Ex19** | 2  | 1,30 | 1,45 | 0,000117 | 0,005806 | 0,000207 | 0,000136 | 0,000343 | 0,000461 |
|          | 4  | 1,46 | 1,55 | 0,000082 | 0,009145 | 0,000225 | 0,000136 | 0,000362 | 0,000444 |
|          | 8  | 1,32 | 2,28 | 0,000068 | 0,009569 | 0,000374 | 0,000137 | 0,000511 | 0,000579 |
|          | 16 | 1,38 | 1,67 | 0,000056 | 0,011098 | 0,000404 | 0,000137 | 0,000542 | 0,000598 |
|          | 32 | 1,25 | 2,12 | 0,000032 | 0,049936 | 0,000526 | 0,000137 | 0,000663 | 0,000694 |
|          | 64 | 2,08 | 4,04 | 0,000031 | 0,104197 | 0,000634 | 0,000144 | 0,000779 | 0,000795 |
| **T2dal**| 2  | 1,13 | 1,95 | 0,000027 | 0,001078 | 0,000103 | 0,000043 | 0,000147 | 0,000175 |
|          | 4  | 1,56 | 1,76 | 0,000027 | 0,001851 | 0,000112 | 0,000044 | 0,000157 | 0,000184 |
|          | 8  | 1,28 | 1,37 | 0,000027 | 0,001842 | 0,000150 | 0,000045 | 0,000195 | 0,000223 |
|          | 16 | 1,13 | 1,20 | 0,000024 | 0,002252 | 0,000196 | 0,000045 | 0,000242 | 0,000267 |
|          | 32 | 1,16 | 1,29 | 0,000025 | 0,004499 | 0,000275 | 0,000046 | 0,000321 | 0,000346 |
|          | 64 | 1,22 | 1,29 | 0,000023 | 0,009517 | 0,000308 | 0,000046 | 0,000355 | 0,000378 |
| **Thermal** | 2 | 1,03 | 1,44 | 0,000055 | 0,001969 | 0,000102 | 0,000040 | 0,000142 | 0,000198 |
|          | 4  | 1,09 | 1,75 | 0,000055 | 0,002321 | 0,000103 | 0,000043 | 0,000146 | 0,000202 |
|          | 8  | 1,96 | 1,96 | 0,000032 | 0,002994 | 0,000163 | 0,000044 | 0,000207 | 0,000240 |
|          | 16 | 1,26 | 1,73 | 0,000027 | 0,003463 | 0,000169 | 0,000044 | 0,000213 | 0,000241 |
|          | 32 | 1,00 | 1,97 | 0,000025 | 0,005537 | 0,000188 | 0,000044 | 0,000233 | 0,000258 |
|          | 64 | 1,77 | 3,09 | 0,000024 | 0,010030 | 0,000134 | 0,000044 | 0,000239 | 0,000263 |
| **Spmsrtls** | 2 | 1,00 | 1,10 | 0,000103 | 0,003736 | 0,000532 | 0,000311 | 0,000843 | 0,000947 |
|          | 4  | 1,00 | 1,01 | 0,000095 | 0,003885 | 0,000563 | 0,000310 | 0,000874 | 0,000969 |
|          | 8  | 1,00 | 1,13 | 0,000067 | 0,006831 | 0,000564 | 0,000321 | 0,000886 | 0,000953 |
|          | 16 | 1,02 | 1,07 | 0,000038 | 0,007511 | 0,000745 | 0,000335 | 0,001080 | 0,001119 |
|          | 32 | 1,04 | 1,07 | 0,000027 | 0,008063 | 0,000810 | 0,000337 | 0,001147 | 0,001175 |
|          | 64 | 1,07 | 1,12 | 0,000023 | 0,008626 | 0,000845 | 0,000349 | 0,001194 | 0,001218 |

| Matrice | Nb Nœuds | LB$_{noeuds}$ | LB$_{coeurs}$ | Temps Calcul Y | Durée Scatter | Durée Gather | Durée Construction de Y | Durée Gather + Construction de Y | Temps Total Du PMVC |
|---|---|---|---|---|---|---|---|---|---|
| **Zhao1** | 2 | 1,08 | 2,06 | 0,000159 | 0,004356 | 0,000695 | 0,000407 | 0,001102 | 0,001262 |
|  | 4 | 1,45 | 1,74 | 0,000086 | 0,005572 | 0,000708 | 0,000406 | 0,001115 | 0,001201 |
|  | 8 | 1,15 | 1,39 | 0,000054 | 0,008706 | 0,000733 | 0,000408 | 0,001141 | 0,001195 |
|  | 16 | 1,24 | 2,04 | 0,000039 | 0,011965 | 0,000783 | 0,000414 | 0,001198 | 0,001237 |
|  | 32 | 1,29 | 1,88 | 0,000012 | 0,017431 | 0,000815 | 0,000415 | 0,001230 | 0,001242 |
|  | 64 | 1,06 | 1,14 | 0,000004 | 0,019770 | 0,000957 | 0,000421 | 0,001379 | 0,001387 |

**Tableau 0.5 : Résultats expérimentation de la combinaison NL-HC**

| Matrice | Nb Nœuds | LB$_{noeuds}$ | LB$_{coeurs}$ | Temps Calcul Y | Durée Scatter | Durée Gather | Durée Construction de Y | Durée Gather + Construction de Y | Temps Total Du PMVC |
|---|---|---|---|---|---|---|---|---|---|
| **Af23560** | 2 | 1,03 | 2,45 | 0,000201 | 0,016000 | 0,000393 | 0,000237 | 0,000631 | 0,000832 |
|  | 4 | 1,13 | 2,17 | 0,000126 | 0,017811 | 0,000397 | 0,000242 | 0,000640 | 0,000766 |
|  | 8 | 1,15 | 3,47 | 0,000080 | 0,095548 | 0,000566 | 0,000253 | 0,000819 | 0,000900 |
|  | 16 | 1,47 | 1,82 | 0,000052 | 0,100182 | 0,000567 | 0,000253 | 0,000820 | 0,000872 |
|  | 32 | 1,28 | 1,56 | 0,000052 | 0,337236 | 0,000719 | 0,000254 | 0,000973 | 0,001025 |
|  | 64 | 1,34 | 1,52 | 0,000032 | 0,333419 | 0,000853 | 0,000256 | 0,001109 | 0,001142 |
| **Bcstm09** | 2 | 1,03 | 1,03 | 0,000039 | 0,000109 | 0,000067 | 0,000011 | 0,000078 | 0,000117 |
|  | 4 | 1,00 | 1,06 | 0,000026 | 0,000163 | 0,000062 | 0,000011 | 0,000075 | 0,000100 |
|  | 8 | 1,12 | 1,12 | 0,000026 | 0,000267 | 0,000128 | 0,000011 | 0,000139 | 0,000165 |
|  | 16 | 1,25 | 1,25 | 0,000025 | 0,001633 | 0,000132 | 0,000011 | 0,000143 | 0,000169 |
|  | 32 | 1,50 | 1,50 | 0,000025 | 0,003443 | 0,000138 | 0,000011 | 0,000150 | 0,000175 |
|  | 64 | 2,00 | 2,00 | 0,000024 | 0,007960 | 0,000154 | 0,000011 | 0,000165 | 0,000190 |
| **Epb1** | 2 | 1,04 | 3,72 | 0,000075 | 0,002370 | 0,000193 | 0,000151 | 0,000344 | 0,000420 |
|  | 4 | 1,19 | 2,53 | 0,000068 | 0,003399 | 0,000262 | 0,000153 | 0,000416 | 0,000485 |
|  | 8 | 1,35 | 1,61 | 0,000064 | 0,003718 | 0,000339 | 0,000154 | 0,000494 | 0,000559 |
|  | 16 | 1,05 | 1,09 | 0,000028 | 0,005356 | 0,000349 | 0,000173 | 0,000523 | 0,000551 |
|  | 32 | 1,07 | 1,13 | 0,000025 | 0,048151 | 0,000357 | 0,000176 | 0,000534 | 0,000560 |
|  | 64 | 1,21 | 1,25 | 0,000025 | 0,052085 | 0,000361 | 0,000176 | 0,000638 | 0,000663 |
| **Ex19** | 2 | 2,16 | 2,17 | 0,000128 | 0,006781 | 0,000156 | 0,000124 | 0,000280 | 0,000408 |
|  | 4 | 2,74 | 2,77 | 0,000075 | 0,007603 | 0,000206 | 0,000125 | 0,000332 | 0,000407 |
|  | 8 | 1,64 | 2,65 | 0,000052 | 0,008343 | 0,000308 | 0,000126 | 0,000434 | 0,000486 |
|  | 16 | 1,32 | 1,70 | 0,000039 | 0,010539 | 0,000345 | 0,000159 | 0,000504 | 0,000544 |
|  | 32 | 1,26 | 1,56 | 0,000035 | 0,051621 | 0,000429 | 0,000160 | 0,000590 | 0,000625 |
|  | 64 | 1,37 | 1,60 | 0,000035 | 0,065904 | 0,000446 | 0,000167 | 0,000663 | 0,000669 |
| **T2dal** | 2 | 1,03 | 2,09 | 0,000052 | 0,000752 | 0,000097 | 0,000043 | 0,000141 | 0,000194 |

|          | 4  | 1,19 | 1,34 | 0,000030 | 0,000758 | 0,000108 | 0,000043 | 0,000152 | 0,000182 |
|----------|----|------|------|----------|----------|----------|----------|----------|----------|
|          | 8  | 1,12 | 1,36 | 0,000029 | 0,000799 | 0,000165 | 0,000044 | 0,000210 | 0,000239 |
|          | 16 | 1,17 | 1,20 | 0,000025 | 0,000792 | 0,000202 | 0,000044 | 0,000247 | 0,000273 |
|          | 32 | 1,37 | 1,44 | 0,000024 | 0,000805 | 0,000207 | 0,000045 | 0,000252 | 0,000277 |
|          | 64 | 1,67 | 2,00 | 0,000024 | 0,000843 | 0,000213 | 0,000045 | 0,000259 | 0,000283 |
| **Thermal** | 2  | 1,07 | 2,31 | 0,000062 | 0,001846 | 0,000077 | 0,000035 | 0,000113 | 0,000175 |
|          | 4  | 1,17 | 2,06 | 0,000059 | 0,002074 | 0,000093 | 0,000036 | 0,000129 | 0,000189 |
|          | 8  | 1,12 | 1,47 | 0,000033 | 0,039646 | 0,000147 | 0,000036 | 0,000184 | 0,000217 |
|          | 16 | 1,30 | 1,51 | 0,000033 | 0,039585 | 0,000151 | 0,000036 | 0,000186 | 0,000220 |
|          | 32 | 1,42 | 2,26 | 0,000033 | 0,043954 | 0,000186 | 0,000037 | 0,000223 | 0,000257 |
|          | 64 | 1,50 | 2,50 | 0,000031 | 0,049570 | 0,000194 | 0,000038 | 0,000232 | 0,000263 |
| **Spmsrtls** | 2  | 1,00 | 1,01 | 0,000115 | 0,003329 | 0,000519 | 0,000107 | 0,000626 | 0,000741 |
|          | 4  | 1,00 | 1,01 | 0,000067 | 0,004976 | 0,000553 | 0,000111 | 0,000665 | 0,000732 |
|          | 8  | 1,00 | 1,01 | 0,000063 | 0,006425 | 0,000565 | 0,000151 | 0,000716 | 0,000780 |
|          | 16 | 1,01 | 1,03 | 0,000032 | 0,011331 | 0,000764 | 0,000163 | 0,000927 | 0,000960 |
|          | 32 | 1,04 | 1,05 | 0,000022 | 0,018646 | 0,000802 | 0,000164 | 0,000966 | 0,000989 |
|          | 64 | 1,08 | 1,11 | 0,000015 | 0,023935 | 0,000889 | 0,000245 | 0,001135 | 0,001150 |
| **Zhao1** | 2  | 1,05 | 2,15 | 0,000139 | 0,004851 | 0,000710 | 0,000182 | 0,000892 | 0,001032 |
|          | 4  | 1,06 | 2,46 | 0,000085 | 0,005951 | 0,000839 | 0,000198 | 0,001038 | 0,001123 |
|          | 8  | 1,14 | 1,37 | 0,000050 | 0,008042 | 0,000849 | 0,000203 | 0,001053 | 0,001104 |
|          | 16 | 1,09 | 1,21 | 0,000038 | 0,027818 | 0,000930 | 0,000258 | 0,001189 | 0,001227 |
|          | 32 | 1,17 | 1,21 | 0,000029 | 0,031503 | 0,001127 | 0,000250 | 0,001338 | 0,001448 |
|          | 64 | 1,08 | 1,10 | 0,000016 | 0,040097 | 0,001179 | 0,000366 | 0,001546 | 0,001583 |

**Tableau 0.6 : Résultats expérimentation de la combinaison NL-HL**

## 4.2. Evaluation des performances en fonction du nombre de nœuds

Dans cette section, nous choisissons de varier le nombre de nœuds f pour chaque matrice testé avec f = {2, 4, 8, 16, 32, 64}. L'objectif est de comparer les quatre combinaisons au niveau de l'équilibrage des charges, la communication des données et le temps de calcul et temps calcul totale du PMVC.

**Equilibrage des charges :**

Pour l'équilibrage entre les cœurs (tous les cœurs participants au calcul), la combinaison $NEZ_L$-$HYP_L$ assure le meilleur équilibrage des charges dans 72% des cas pour les matrices que nous avons les testé, ceci est dû l'usage en premier du $NEZGT_L$ en inter-nœud et le $Hypergraph_L$ pour toute la combinaison utilisée ceci devient remarquable sur tout lors de l'augmentation du nombre de cœurs et quand nous augmentons la taille de la matrice, ceci devient plus visible. Les méthodes $NEZ_L$-$HYP_C$ avec 14% puis $NEZ_C$-$HYP_L$ avec 14% $NEZ_C$—$HYP_C$ arrivent en deuxième et troisième place, puis la méthode $NEZGT_c$ inter et $HYPER_L$ intra-nœud.

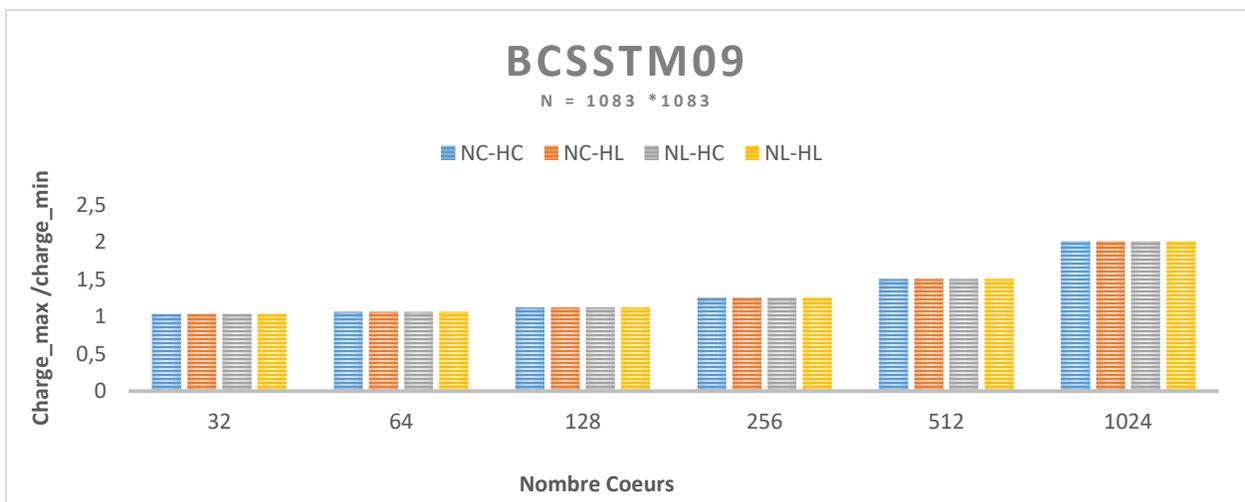

**Figure 0.8 : Equilibrage des charges au niveau des cœurs de la matrice «BCSSTM09 »**

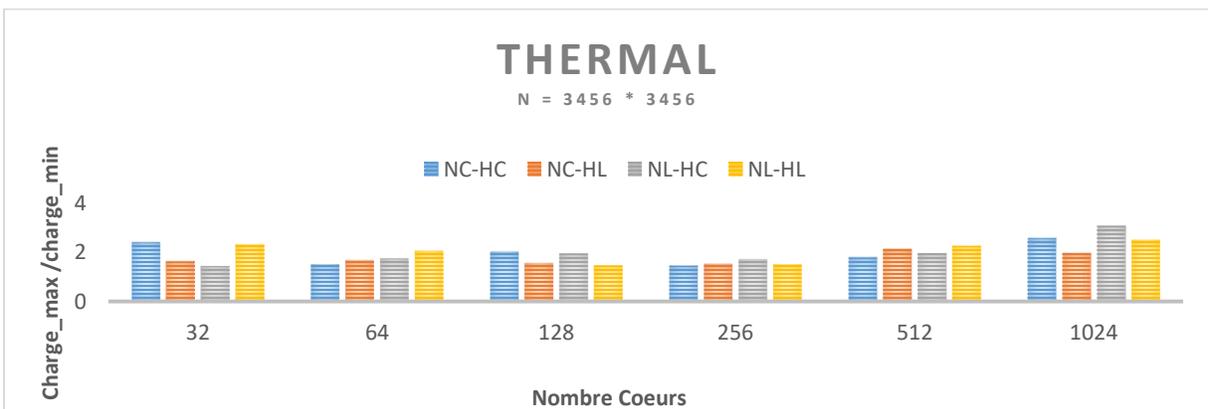

**Figure 0.9 : Equilibrage des charges au niveau des cœurs de la matrice «THERMAL »**

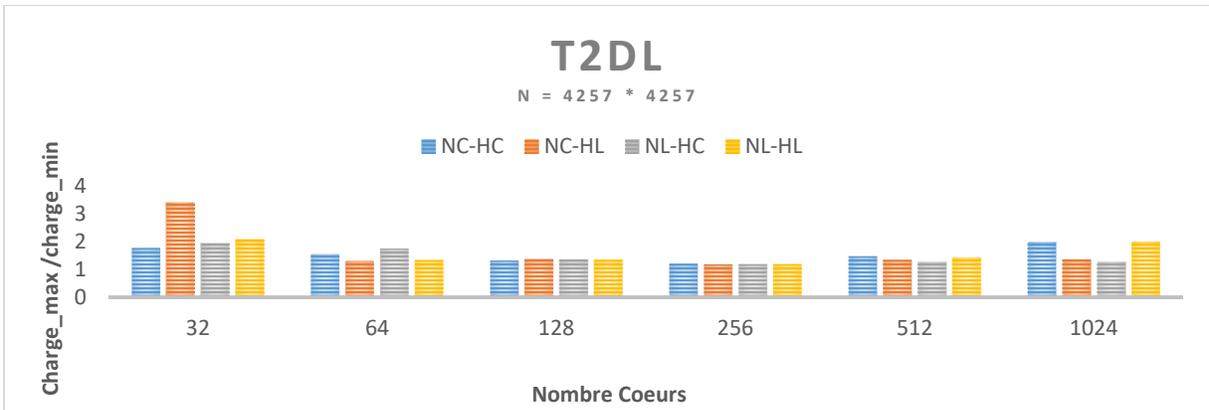

Figure 0.10 : Equilibrage des charges au niveau des cœurs de la matrice «T2DL »

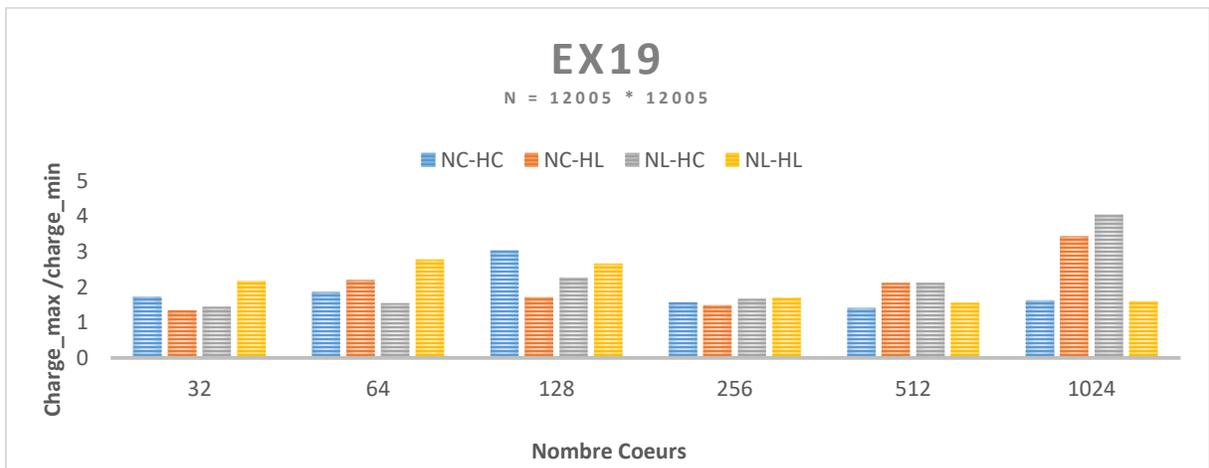

Figure 0.11 : Equilibrage des charges au niveau des cœurs de la matrice «EX19 »

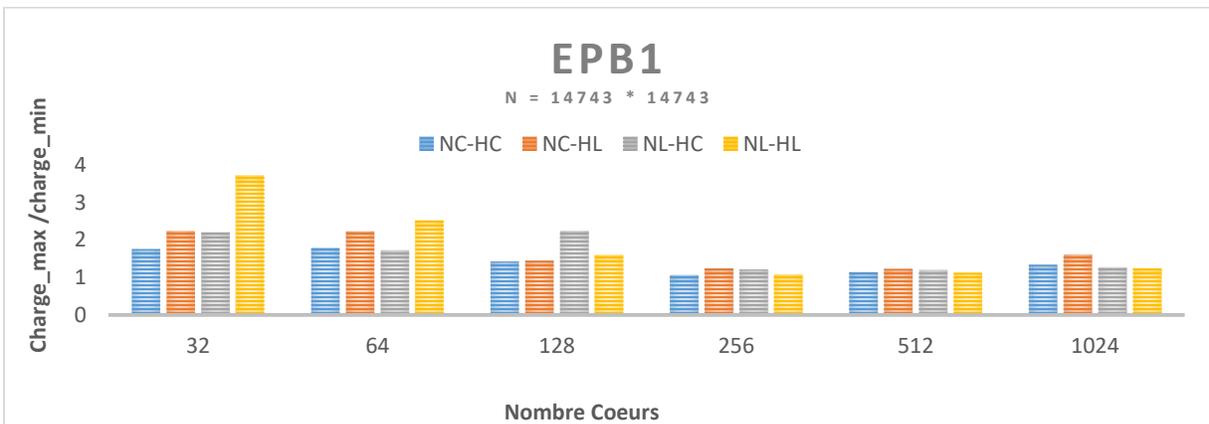

Figure 00.12 : Equilibrage des charges au niveau des cœurs de la matrice «EPB1 »

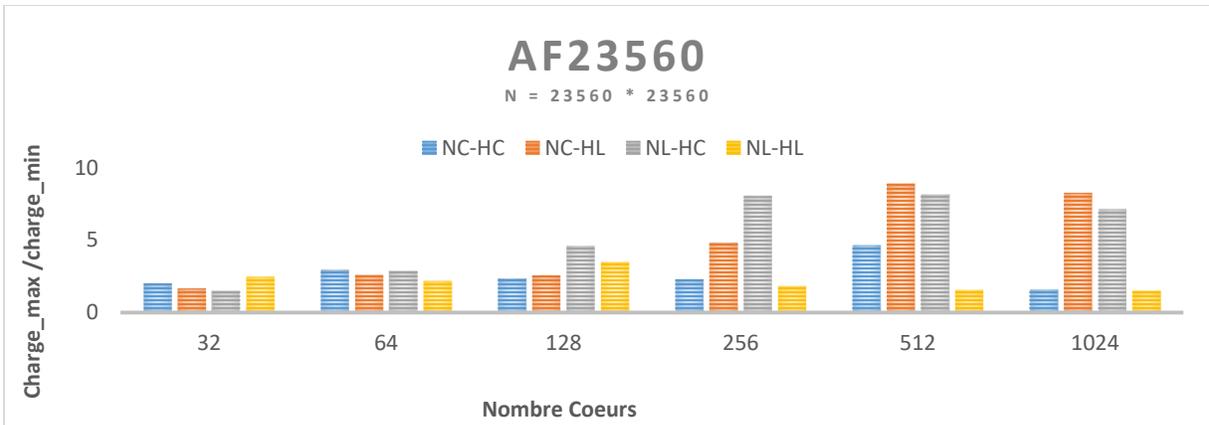

Figure 00.13 : Equilibrage des charges au niveau des cœurs de la matrice «AF23560»

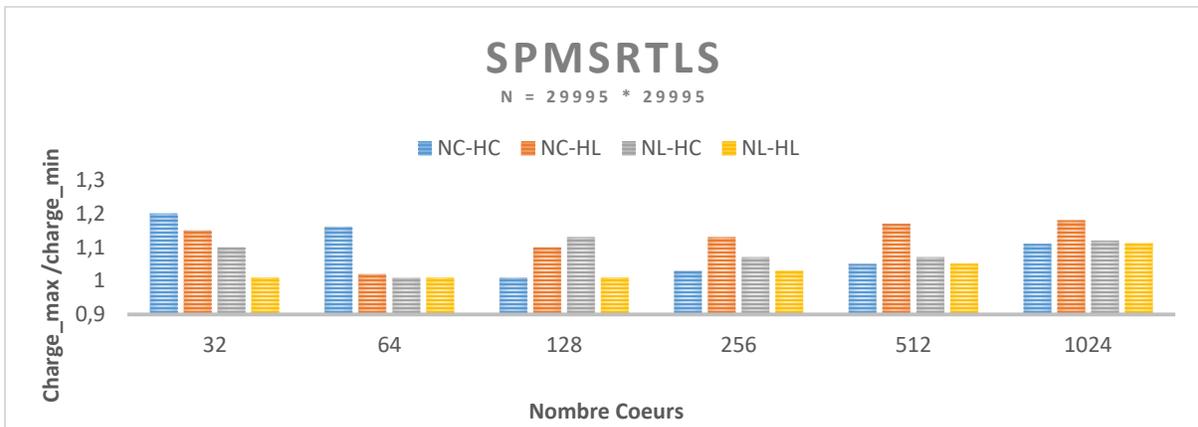

Figure 0.14 : Equilibrage des charges au niveau des cœurs de la matrice «SPMSRTLS»

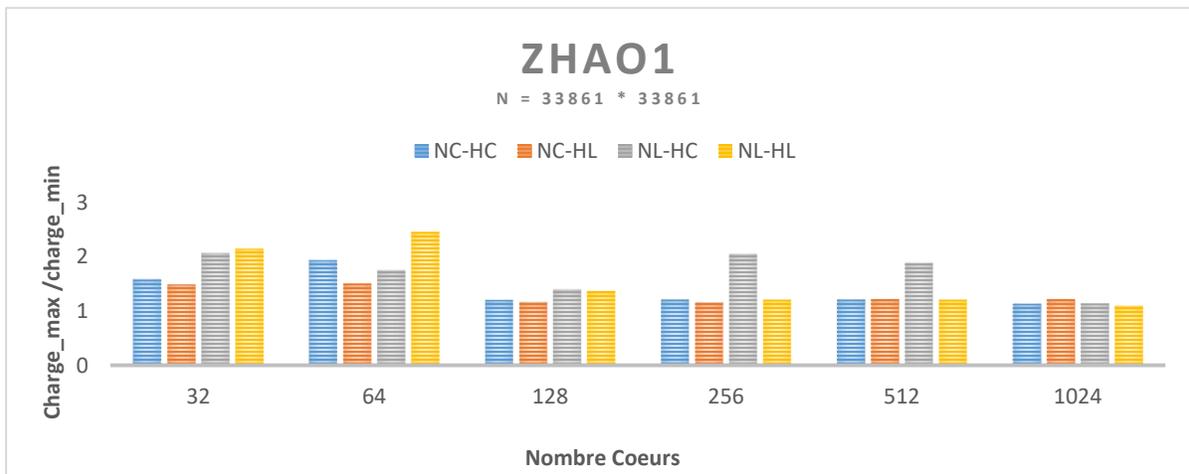

Figure 00.15 : Equilibrage des charges au niveau des cœurs de la matrice «ZHAO1»

**Communication des données (Scatter) :**

L'augmentation du nombre de nœuds induit une augmentation du nombre des fragments traités, induit une diminution de la taille de ces derniers et une augmentation au niveau de nombre de messages communiqués. Ceci influe sur la durée totale de communication des données (envoi des blocs Ak et sous-vecteurs Xk du nœud master aux nœuds slave). Nous remarquons aussi que pour toutes les valeurs de f, c'est toujours la combinaison $NEZGT_L$-$HYPER_L$ assure toujours le temps de communication minimal (70% des cas). Ceci est dû à l'utilisation de la méthode hypergraphe qui optimise les communications. La combinaison $NEZGT_L$-$HYPER_C$ arrive en deuxième place, ensuite le $NEZGT_c$-$HYPER_L$ puis $NEZGT_C$-$HYPER_C$.

Nous concluons alors, que l'utilisation de la méthode $NEZGT_L$ pour la distribution inter-nœuds permet de minimiser le temps nécessaire pour la communication des données par rapport l'usage du $NEZGT_C$.

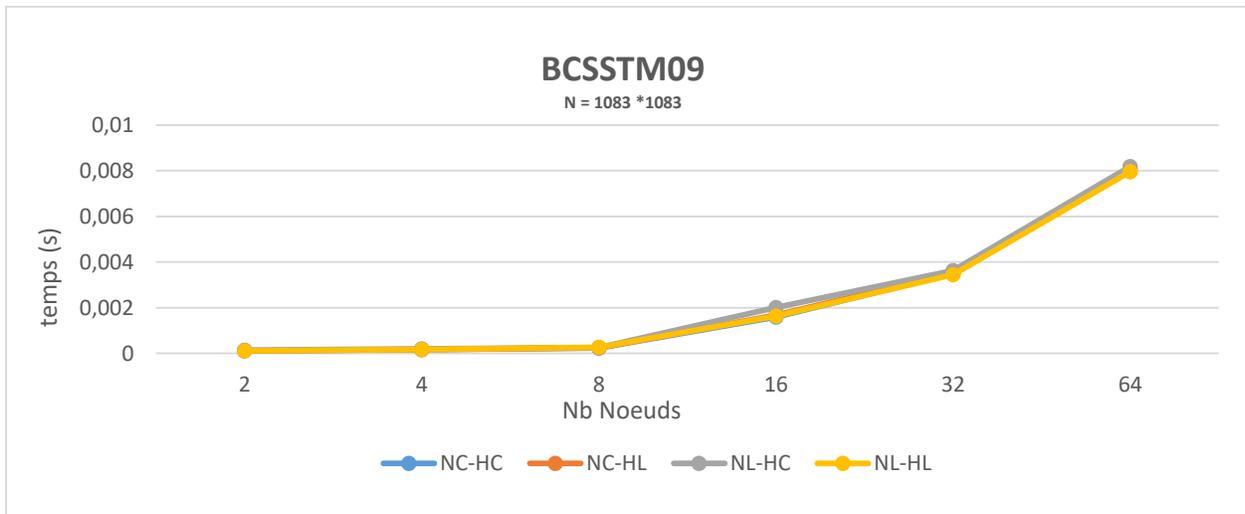

**Figure 00.16 : Effets augmentation Nb_Noeuds sur le temps des communications pour la matrice «BCSSTM09 »**

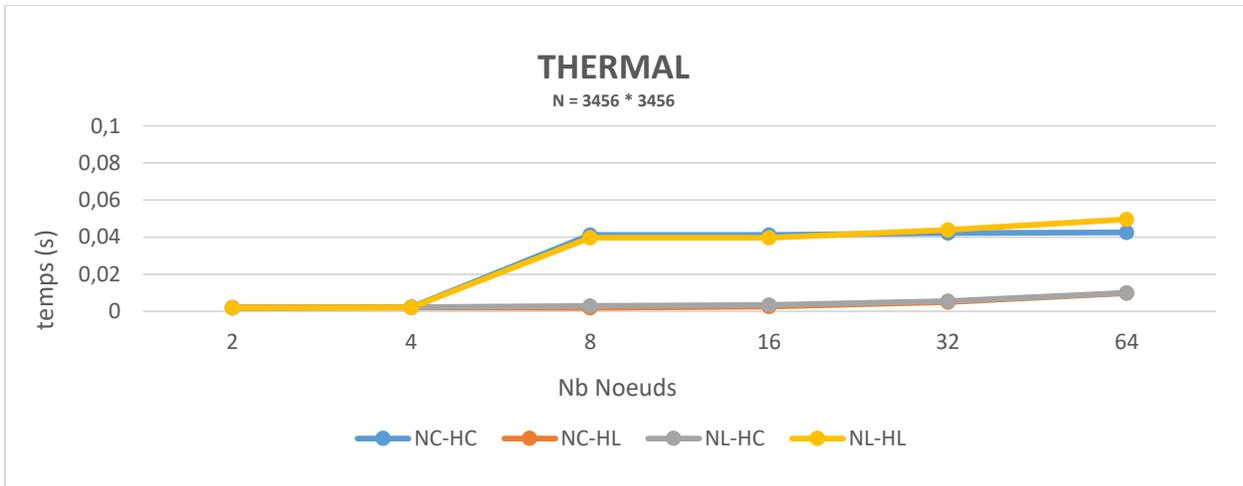

**Figure 0.17 : Effets augmentation Nb_Noeuds sur le temps des communications pour la matrice «THERMAL »**

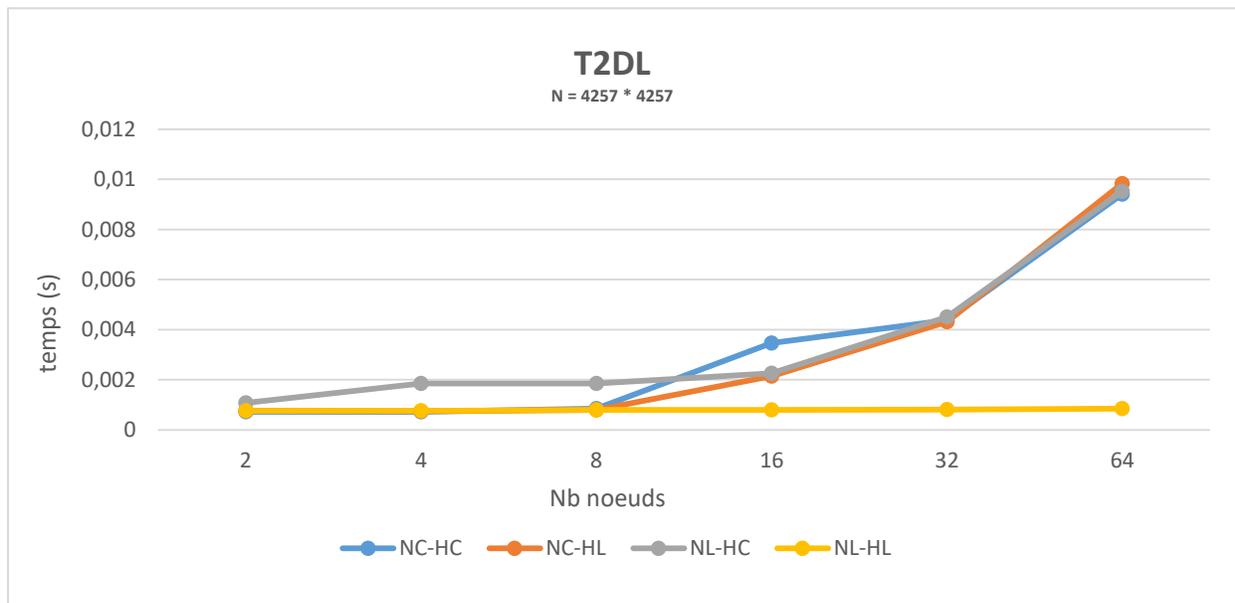

**Figure 0.18 : Effets augmentation Nb_Noeuds sur le temps des communications pour la matrice «T2DL »**

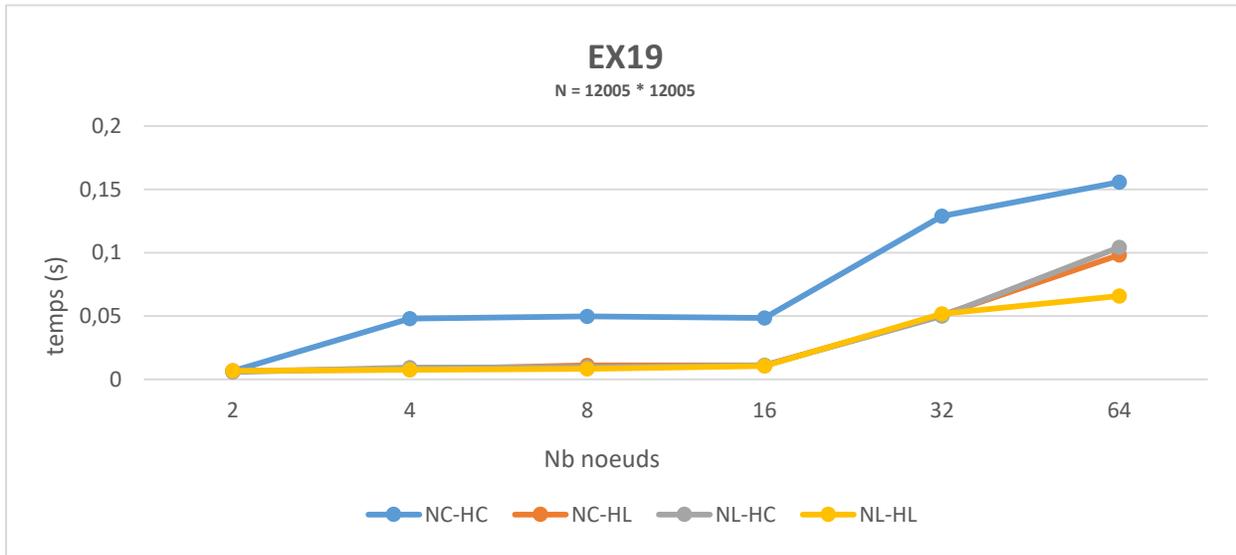

**Figure 00.19 : Effets augmentation Nb_Noeuds sur le temps des communications pour la matrice «EX19»**

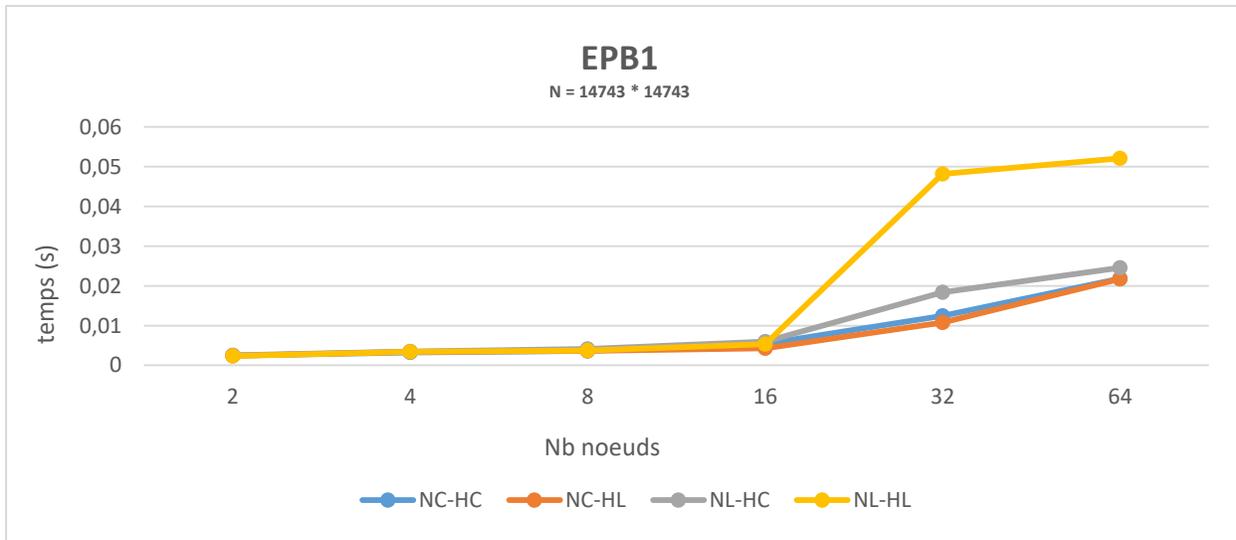

**Figure 0.20 : Effets augmentation Nb_Noeuds sur le temps des communications pour la matrice «EPB1»**

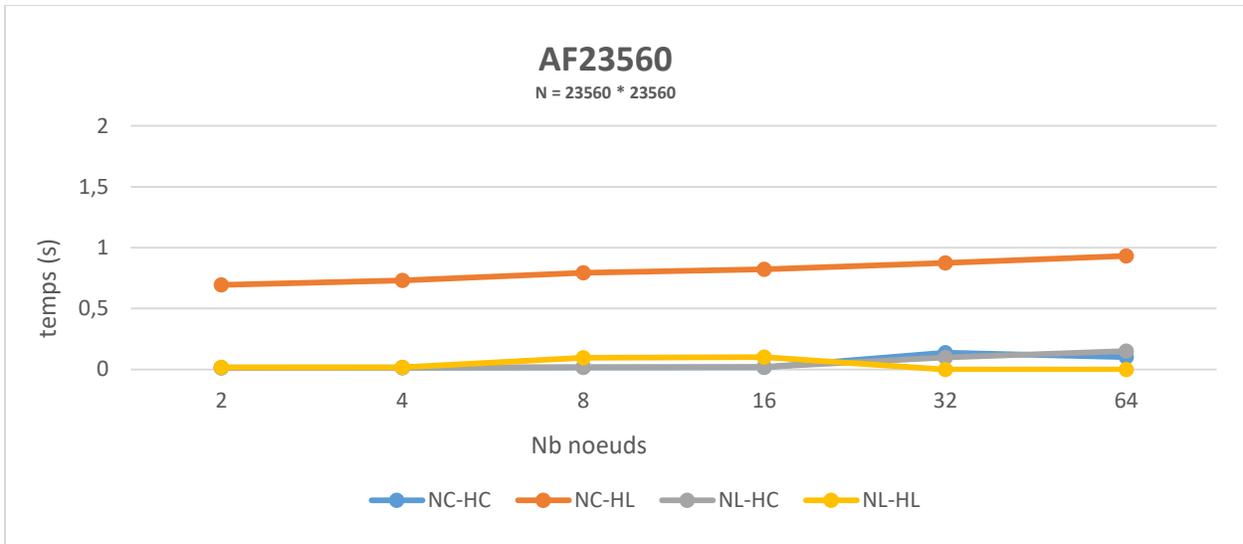

**Figure 00.21 : Effets augmentation Nb_Noeuds sur le temps des communications pour la matrice «AF23560»**

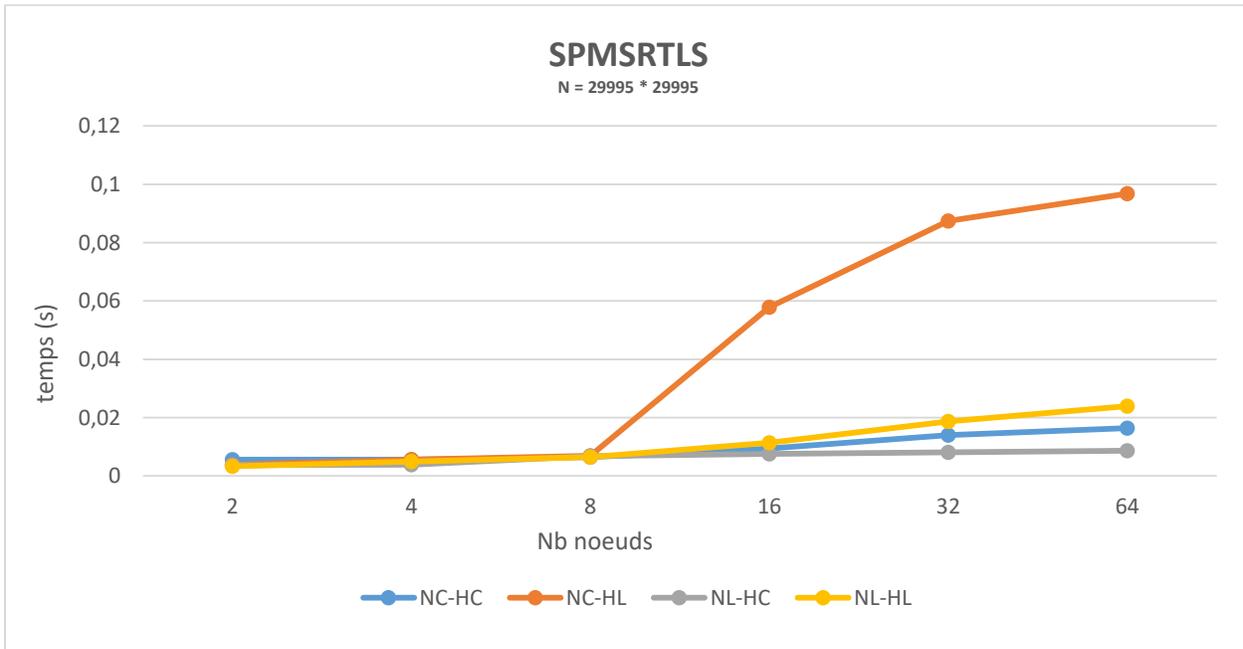

**Figure 00.22 : Effets augmentation Nb_Noeuds sur le temps des communications pour la matrice «SPMSRTLS»**

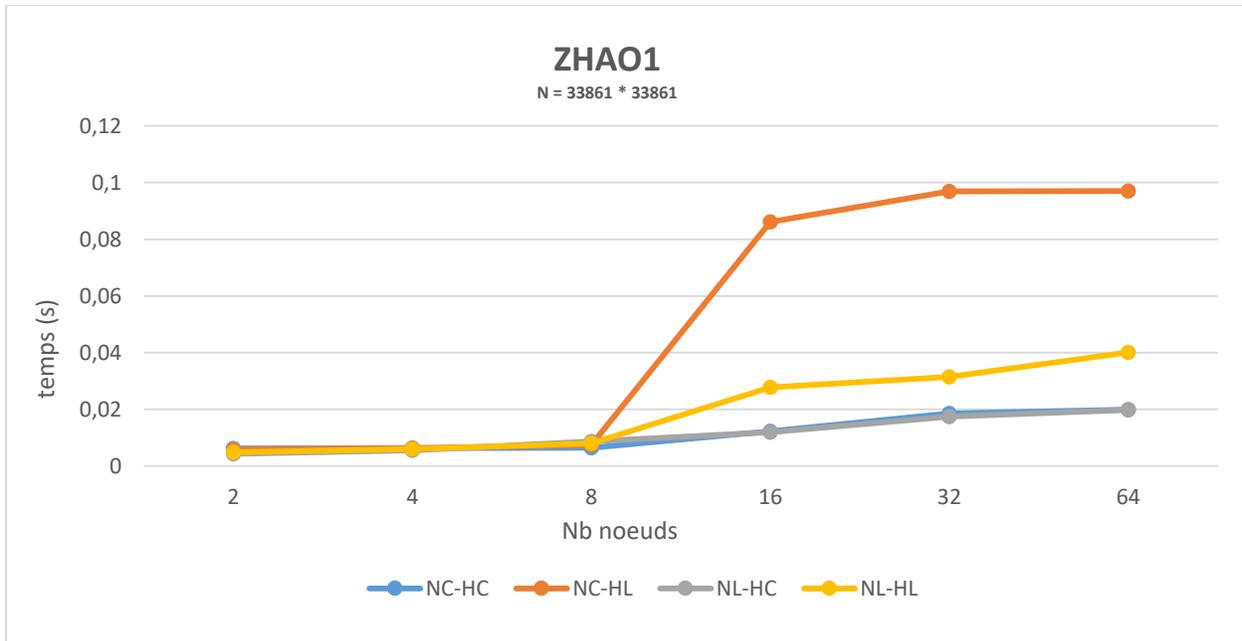

**Figure 00.23 : Effets augmentation Nb_Noeuds sur le temps des communications pour la matrice «ZHAO1»**

## **Temp Calcul Y :**

Dans les figures (**4.24** jusqu'au **4.31**), nous remarquons que lorsque nous augmentons le nombre des nœuds, le temps de calcul de Y pour chaque combinaison (date de fin d'exécution du dernier cœur moins date de début d'exécution du premier cœur) diminue.

Ce résultat peut être justifié par le fait que lorsque le nombre des nœuds croît (par suite, celui des cœurs), le nombre d'éléments non nuls dans chaque fragments diminue, donc la granularité des tâches diminue.

En conséquence, la quantité de données traitée par les calculs décroit, d'où la durée du Makespan diminue aussi.

Nous remarquons aussi, qu'en augmentant le nombre des nœuds, le temps de calcul de Y des quatre combinaisons sont de plus en plus proches, mais dans 62.5% des cas, c'est la méthode $NEZ_L$-$HYP_L$ qui assure le meilleur Makespan puis $NEZ_L$-$HYP_C$ avec 20% puis $NEZ_C$-$NEZ_C$ avec 17.5%.

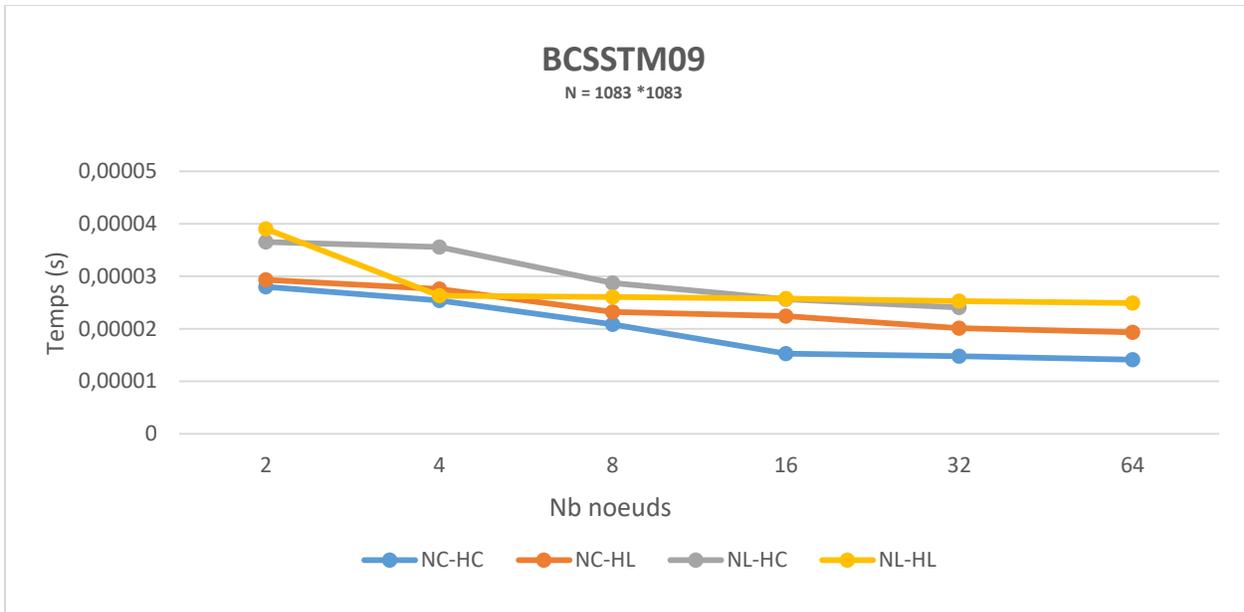

**Figure 00.24 : Effets augmentation Nb_Noeuds sur le Temps de Calcul pour la matrice «BCSSTM09»**

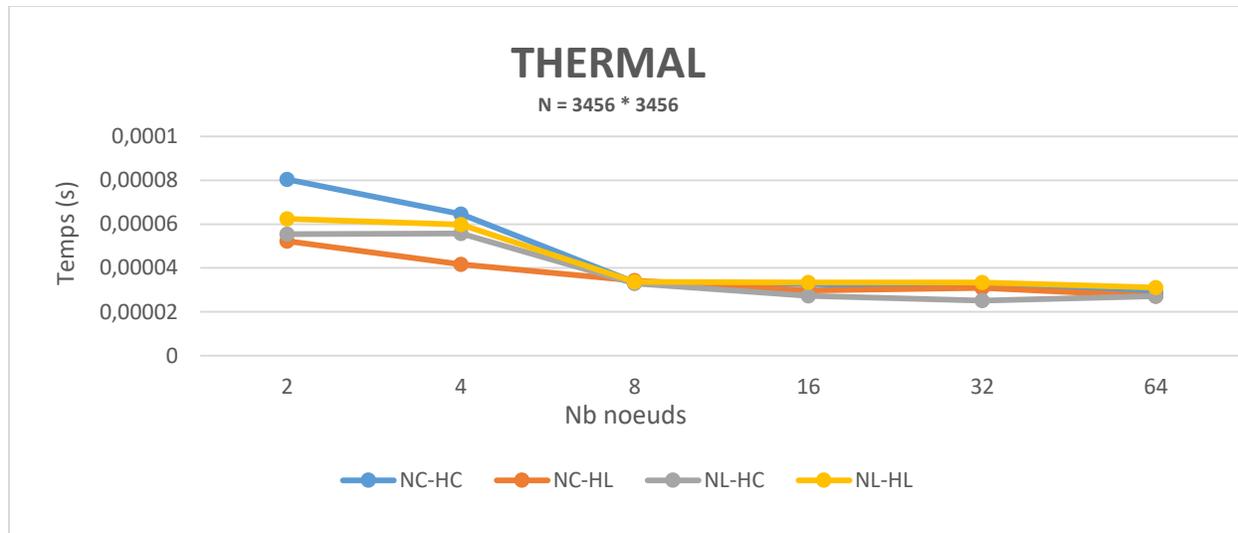

**Figure 0.25 : Effets augmentation Nb_Noeuds sur le Temps de Calcul pour la matrice «THERMAL»**

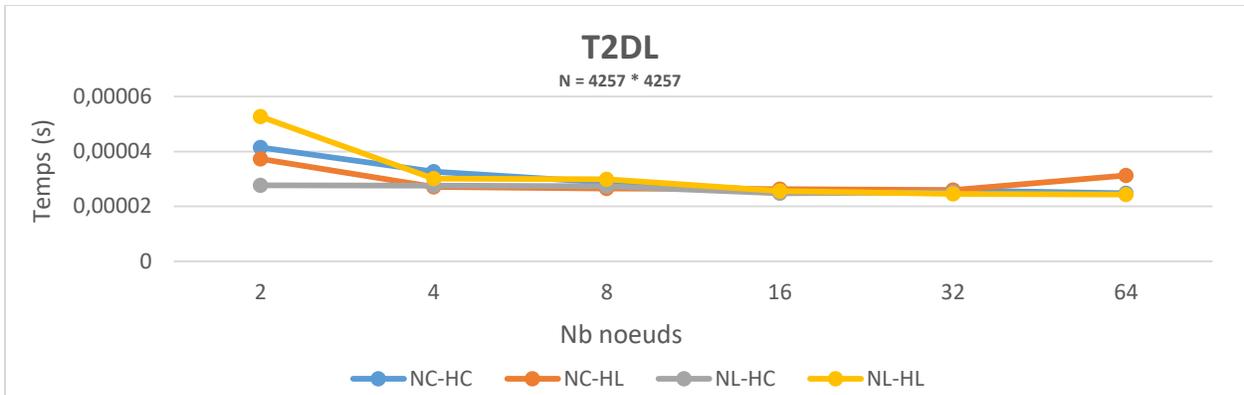

**Figure 00.26 :** Effets augmentation Nb_Noeuds sur le Temps de Calcul pour la matrice « T2DL »

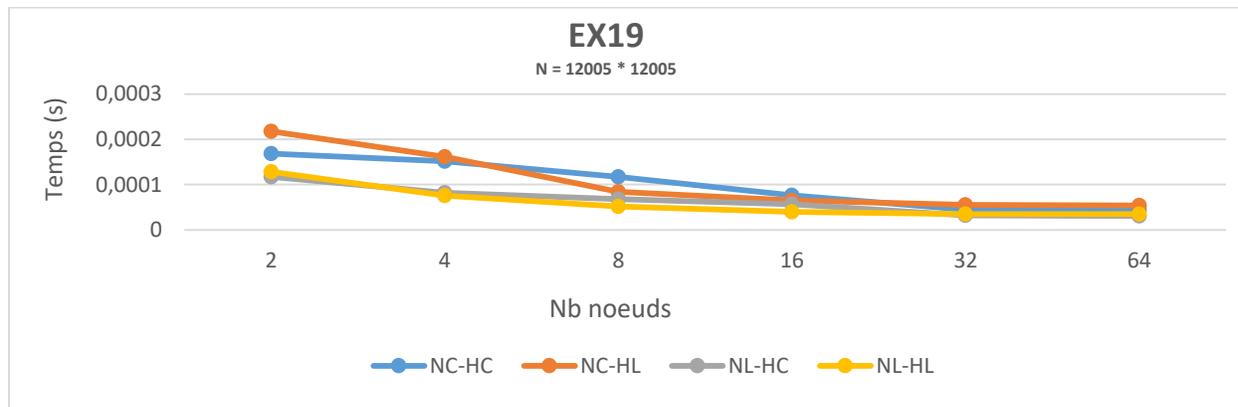

**Figure 00.27 :** Effets augmentation Nb_Noeuds sur le Temps de Calcul pour la matrice « EX19 »

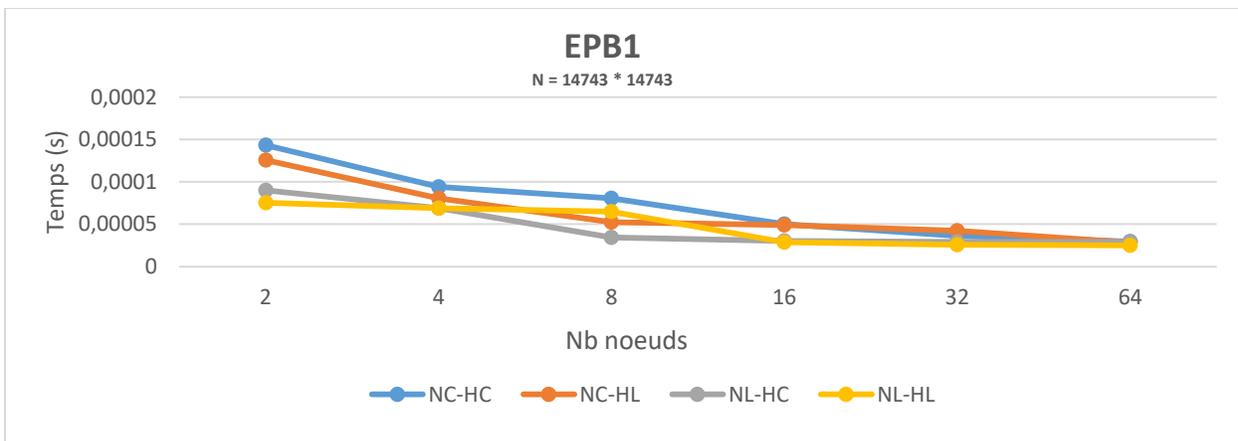

**Figure 0.28 :** Effets augmentation Nb_Noeuds sur le Temps de Calcul pour la matrice « EPB1 »

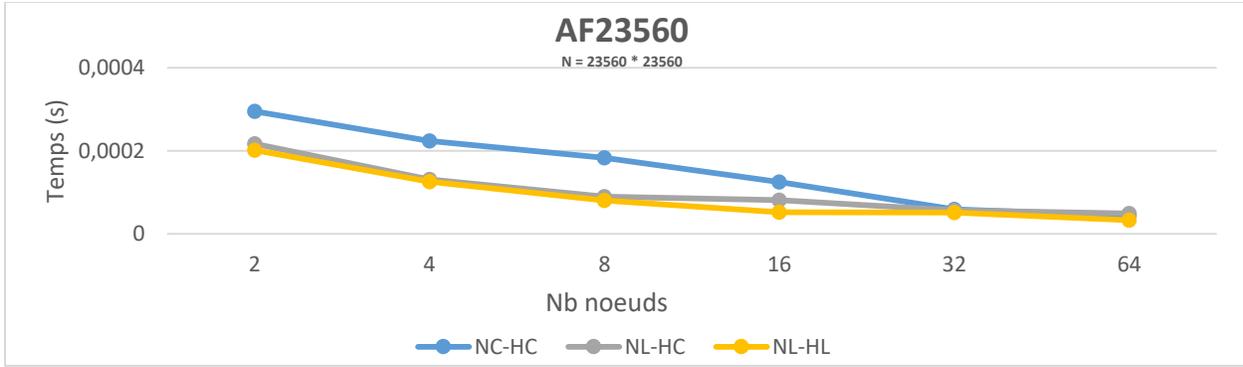

**Figure 00.29 : Effets augmentation Nb_Noeuds sur le Temps de Calcul pour la matrice «AF23560»**

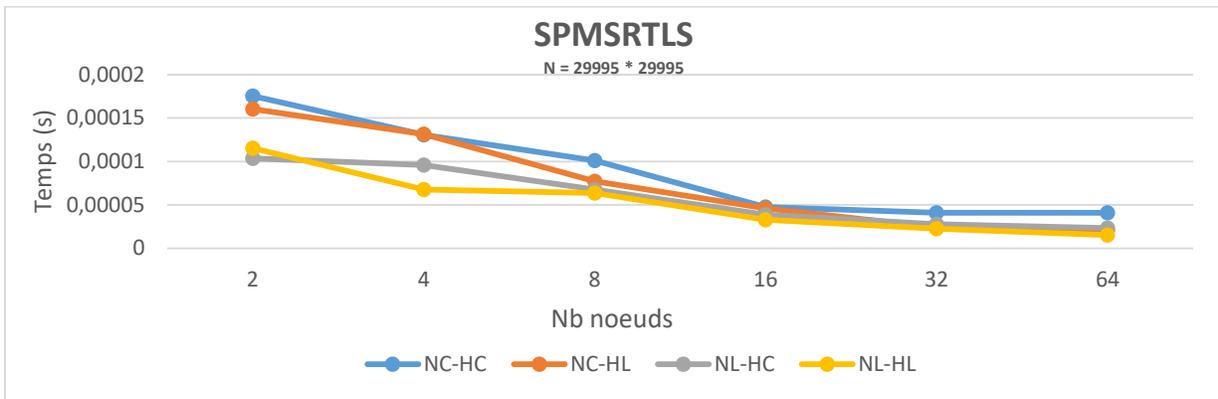

**Figure 0.30 : Effets augmentation Nb_Noeuds sur le Temps de Calcul pour la matrice «SPMSRTLS»**

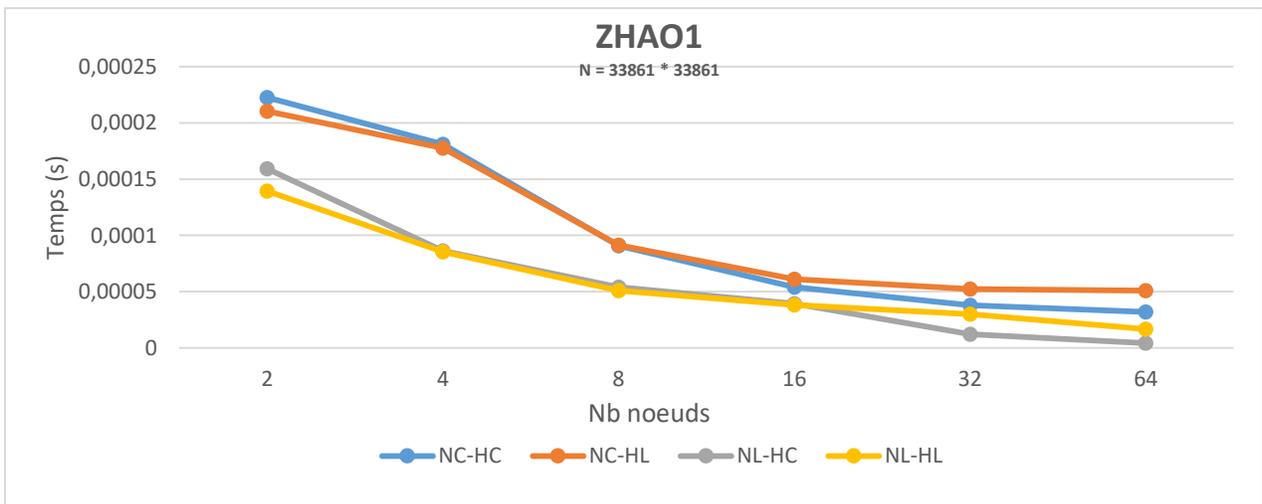

**Figure 0.31 : Effets augmentation Nb_Noeuds sur le Temps de Calcul pour la matrice « ZHAO1 »**

**Temps nécessaire pour la construction du vecteur Y sur le nœud Local :**

Nous remarquons que le temps nécessaire pour la réalisation de cette étape est dû essentiellement à celui nécessaire pour le calcul des Y locaux (Figures **4.32** jusqu'au **4.39**) pour les quatre combinaisons, mais le $NEZ_L$-$HYP_L$ est le plus favoris vue que lors de la construction des vecteurs de petit taille du nœud assure un temps minimale que celle lors de la sommation de plusieurs vecteurs de taille N pour le cas des autres combinaisons.

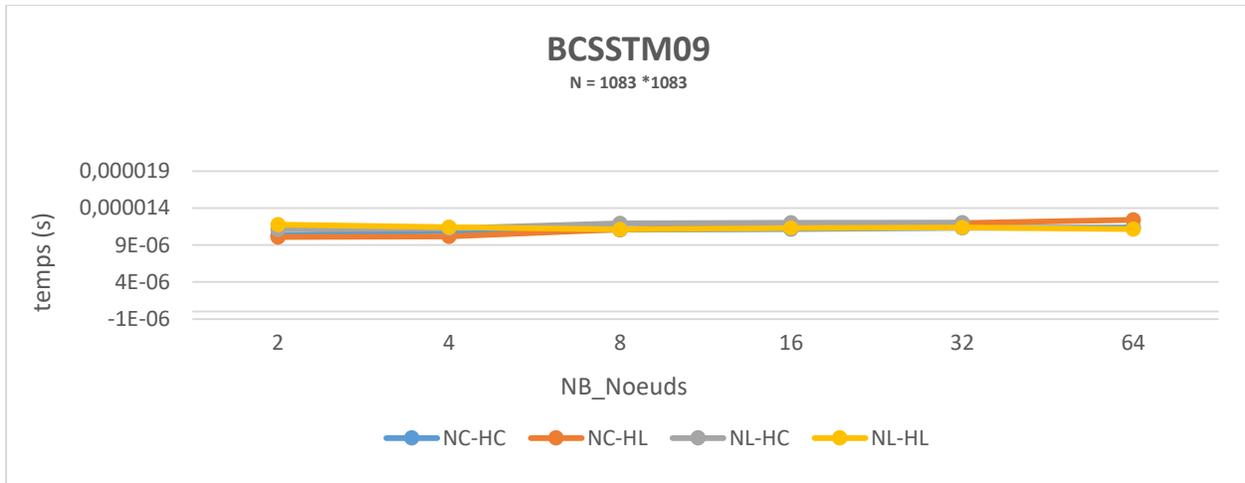

**Figure 0.32 : Temps nécessaire pour la construction du vecteur Y pour la matrice «BCSSTM09»**

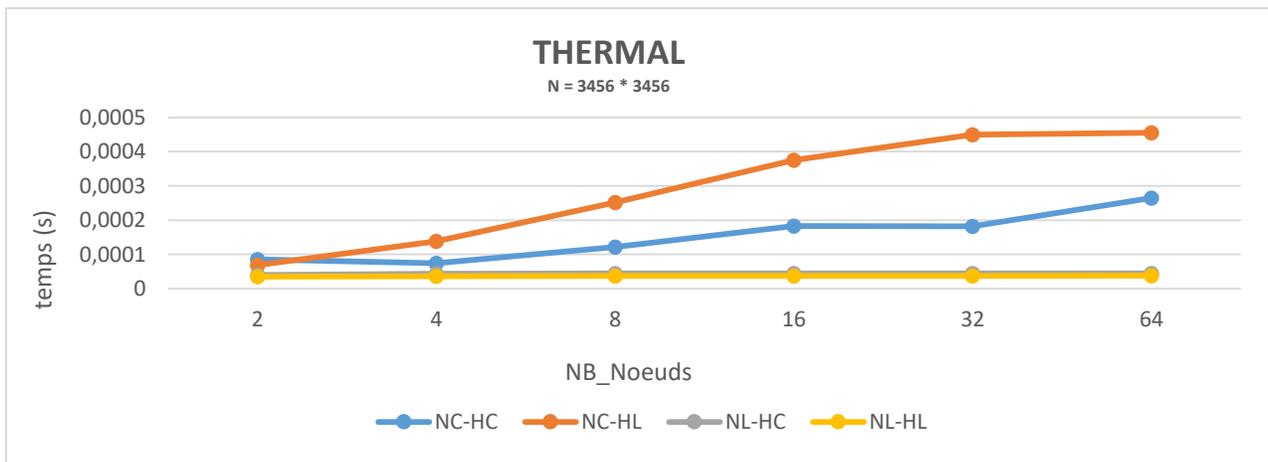

**Figure 0.33 : Temps nécessaire pour la construction du vecteur Y pour la matrice «THERMAL»**

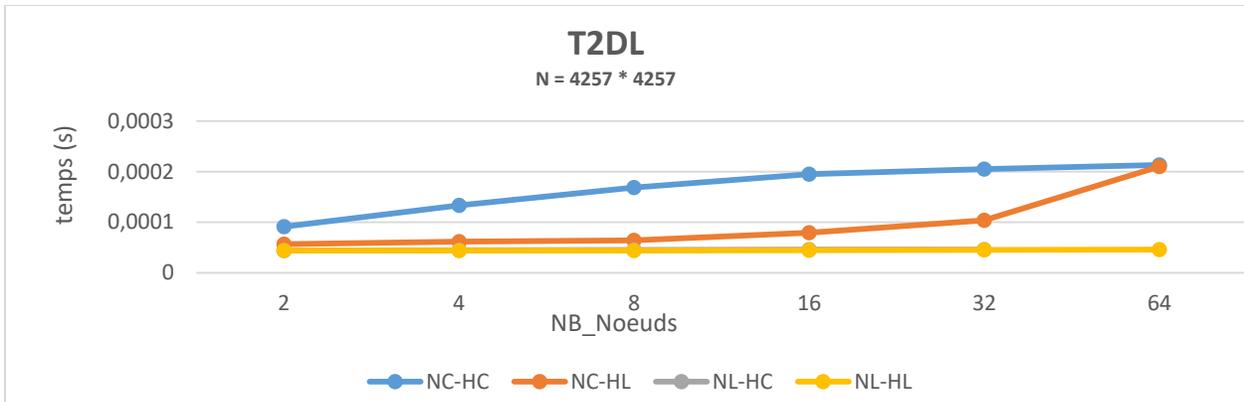

Figure 0.34 : Temps nécessaire pour la construction du vecteur Y pour la matrice «T2DL »

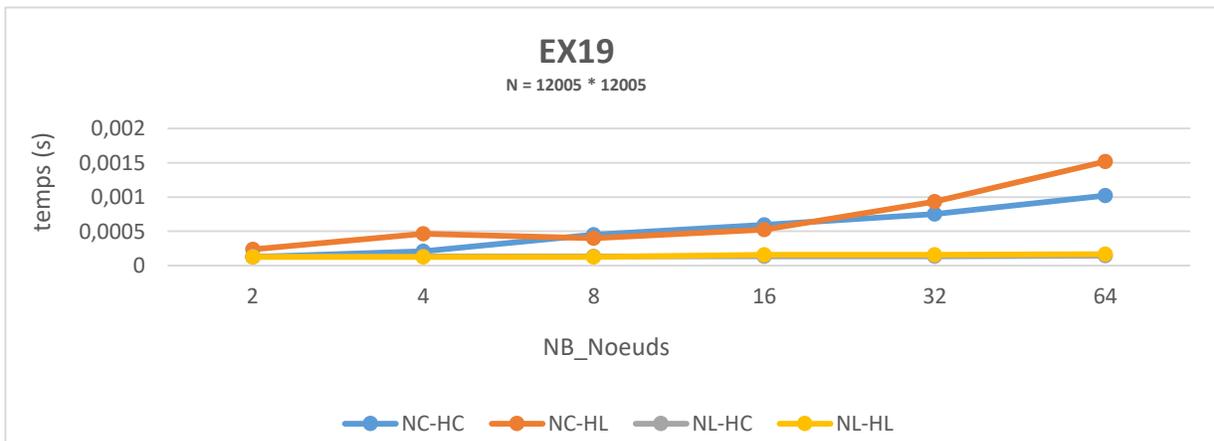

Figure 0.35 : Temps nécessaire pour la construction du vecteur Y pour la matrice «EX19 »

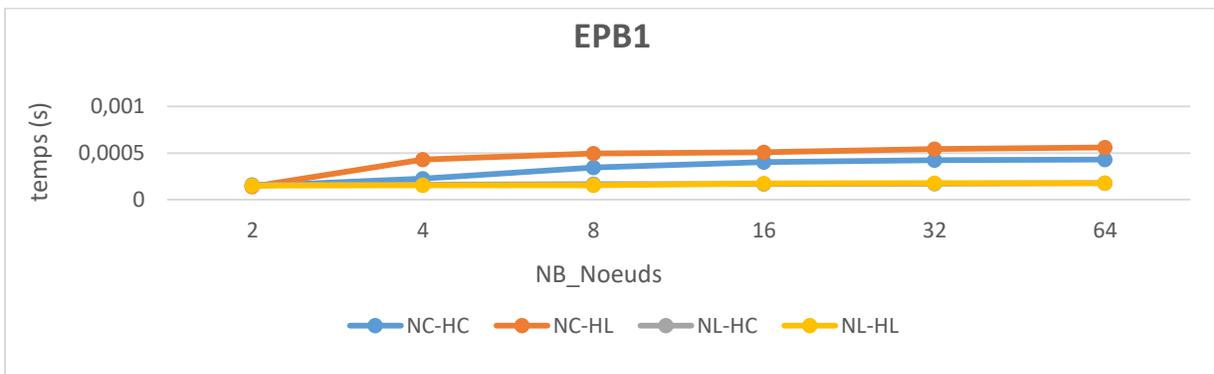

Figure 0.36 : Temps nécessaire pour la construction du vecteur Y pour la matrice «EPB1 »

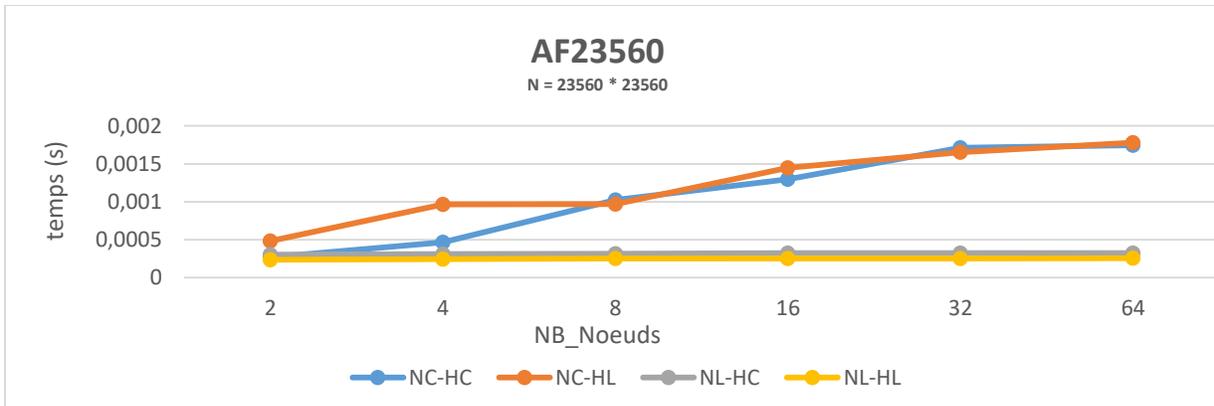

**Figure 0.37 : Temps nécessaire pour la construction du vecteur Y pour la matrice «AF23560»**

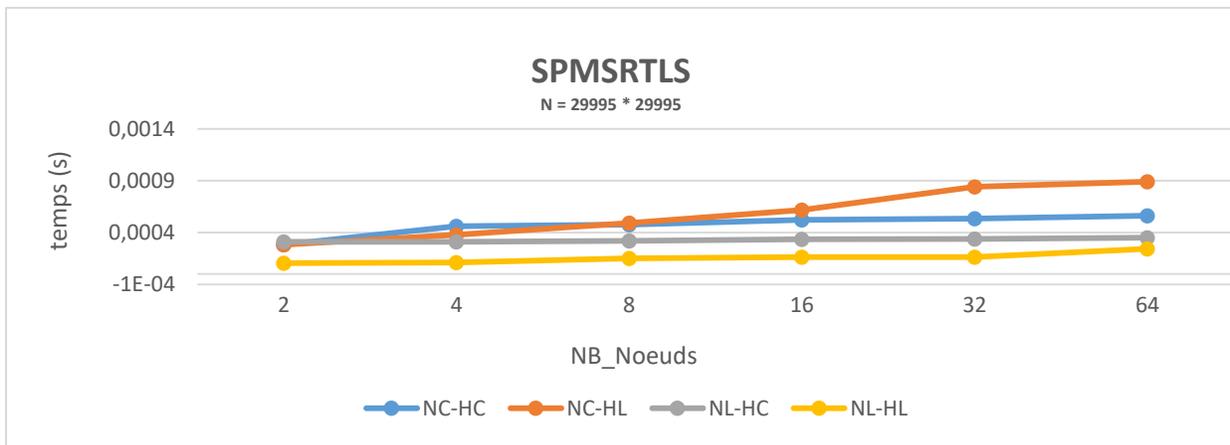

**Figure 00.38 : Temps nécessaire pour la construction du vecteur Y pour la matrice «SPMRLTS»**

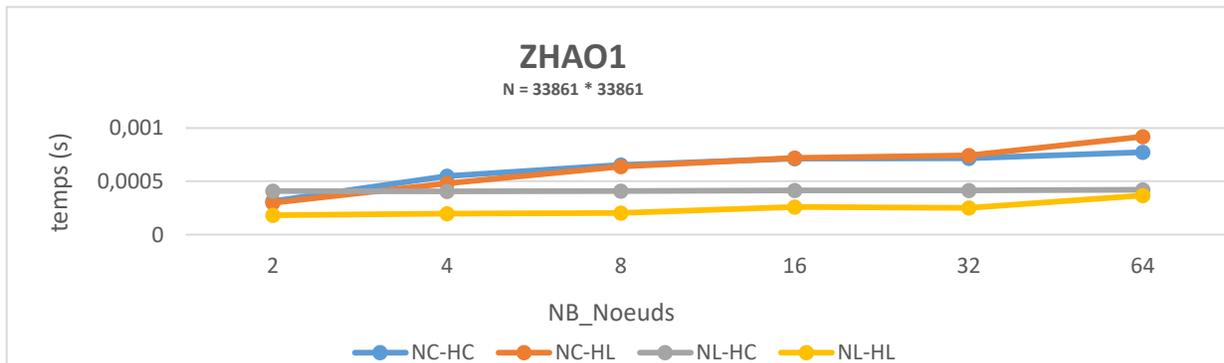

**Figure 0.39 : Temps nécessaire pour la construction du vecteur Y pour la matrice «ZHAO1»**

## Collecte des résultats partiels et construction de Y :

La dernière étape dans le calcul distribué du PMVC est la collecte des résultats partiels à partir des différents nœuds de la grappe pour la construction ensuite du résultat final Y. Nous remarquons que dans la plupart des cas, la combinaison $NEZ_L$-$HYP_L$ fournit le meilleur temps pour la réalisation de cette étape (voir Figure 4.41 jusqu'au 4.47) dans 88% puis avec $NEZ_C$-$HYP_C$ avec 12% ceci est dû au fait que lors du Gather on fait la collection des petit vecteurs de Y alors que pour les autres on fait la collection des vecteurs de taille N chacun d'eux.

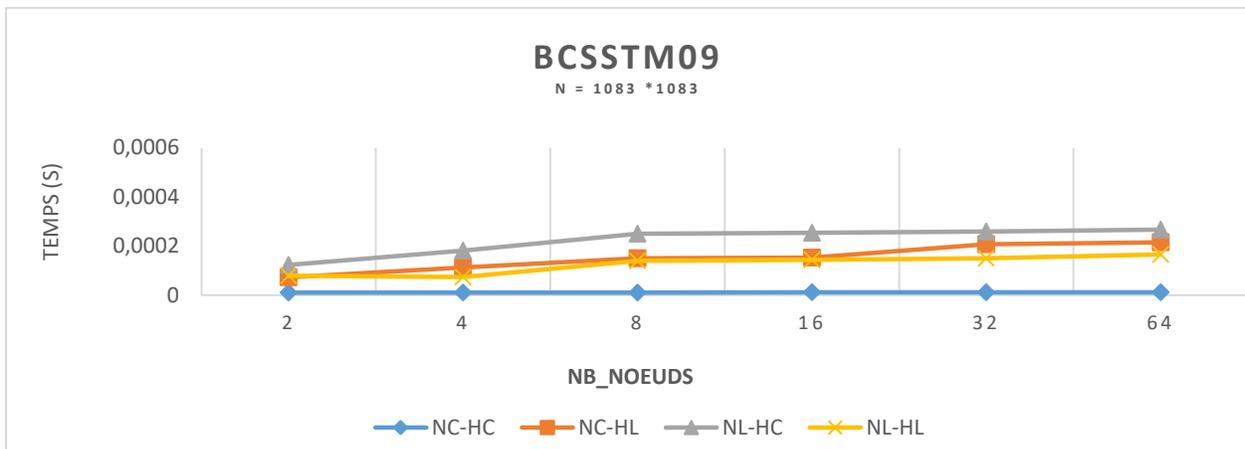

Figure 0.40 : Collecte des résultats et construction de Y pour la matrice « BCSSTM09 »

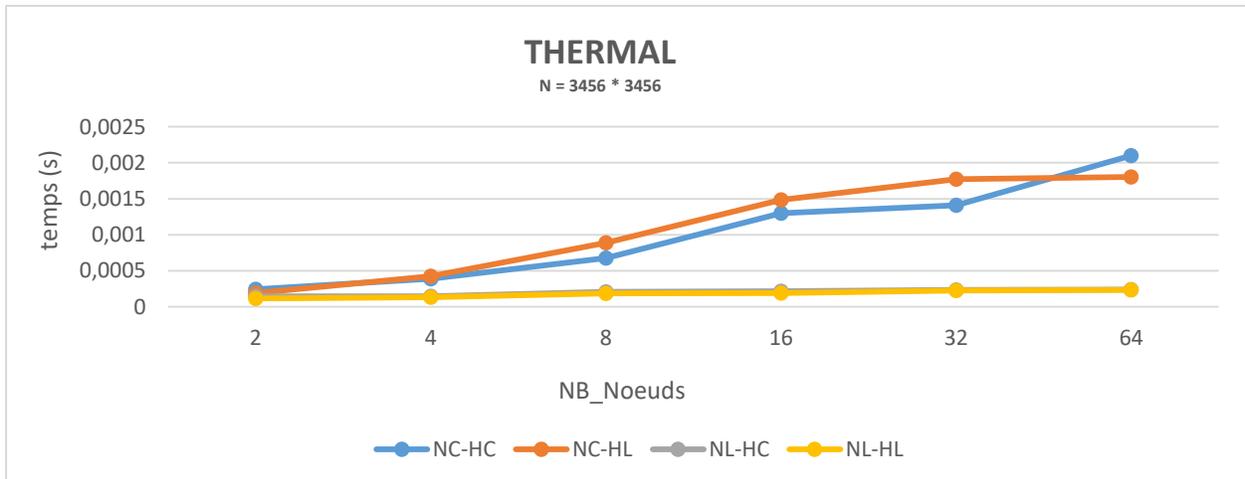

Figure 0.41 : Collecte des résultats et construction de Y pour la matrice « THERMAL»

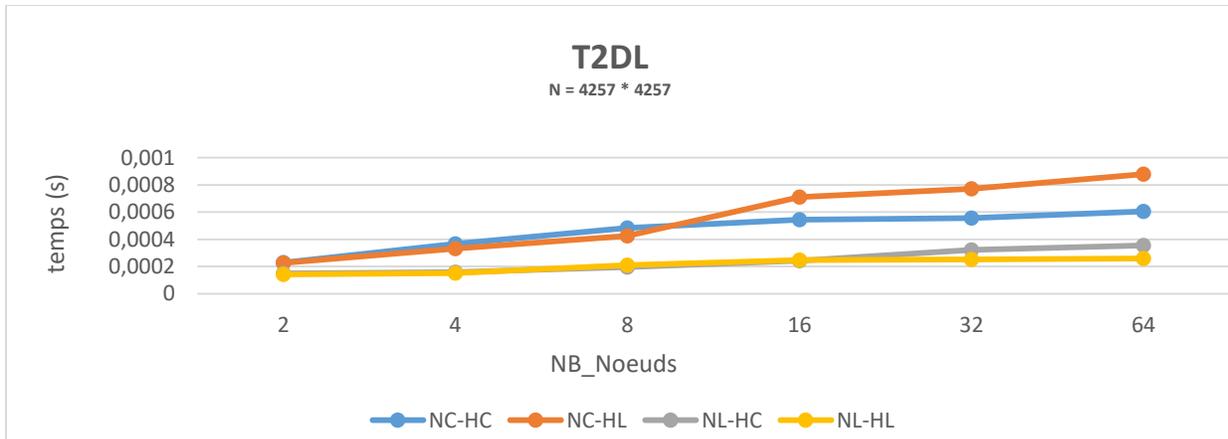

**Figure 0.42 : Collecte des résultats et construction de Y pour la matrice « T2DL»**

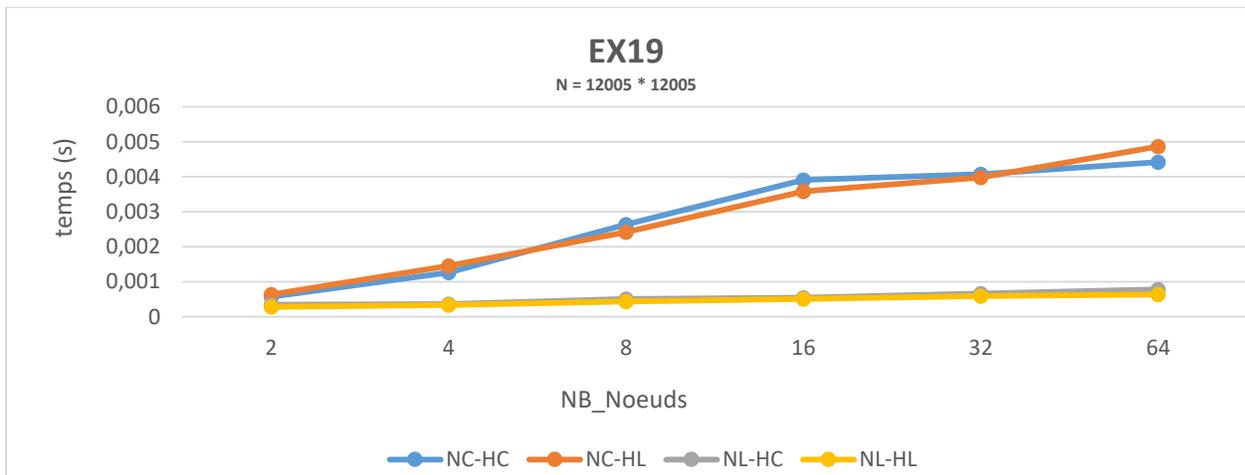

**Figure 0.43 : Collecte des résultats et construction de Y pour la matrice «EX19»**

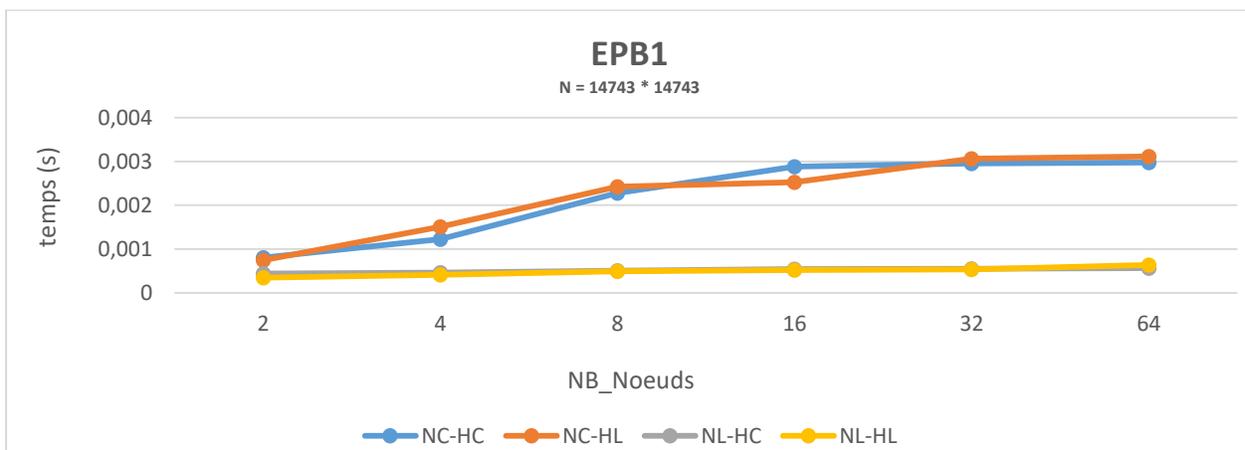

**Figure 0.44 : Collecte des résultats et construction de Y pour la matrice «EPB1»**

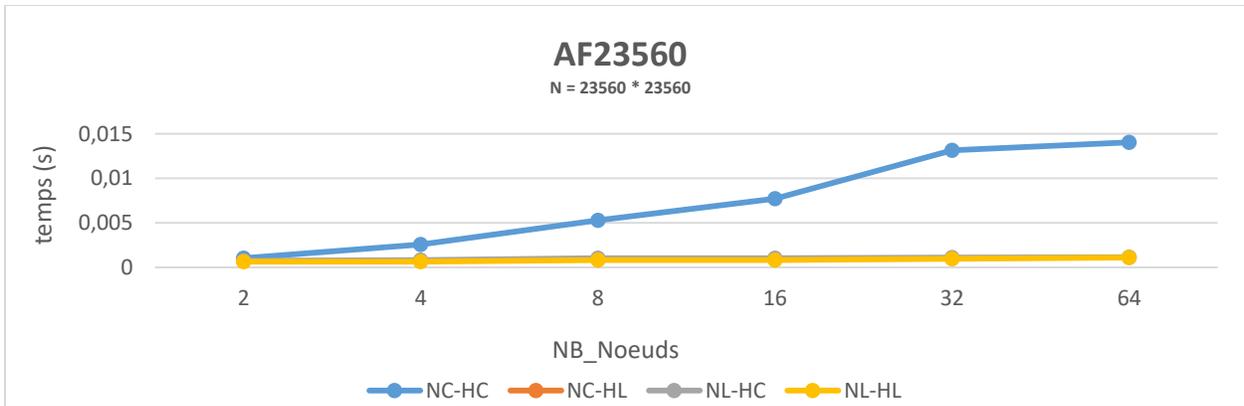

**Figure 0.45 :** Collecte des résultats et construction de Y pour la matrice «AF23560»

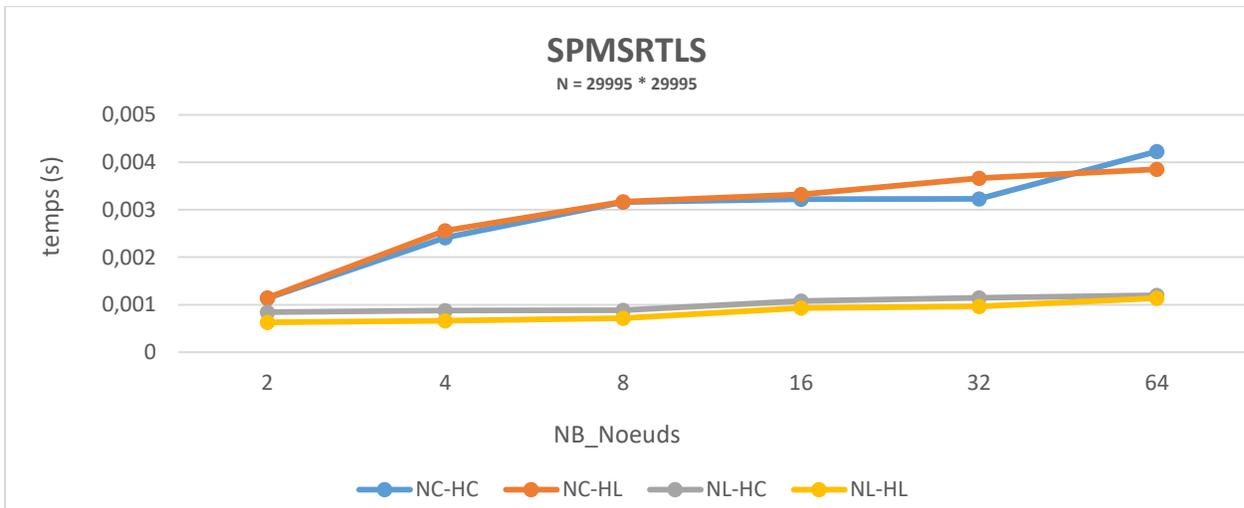

**Figure 0.46 :** Collecte des résultats et construction de Y pour la matrice «SPMRLTS»

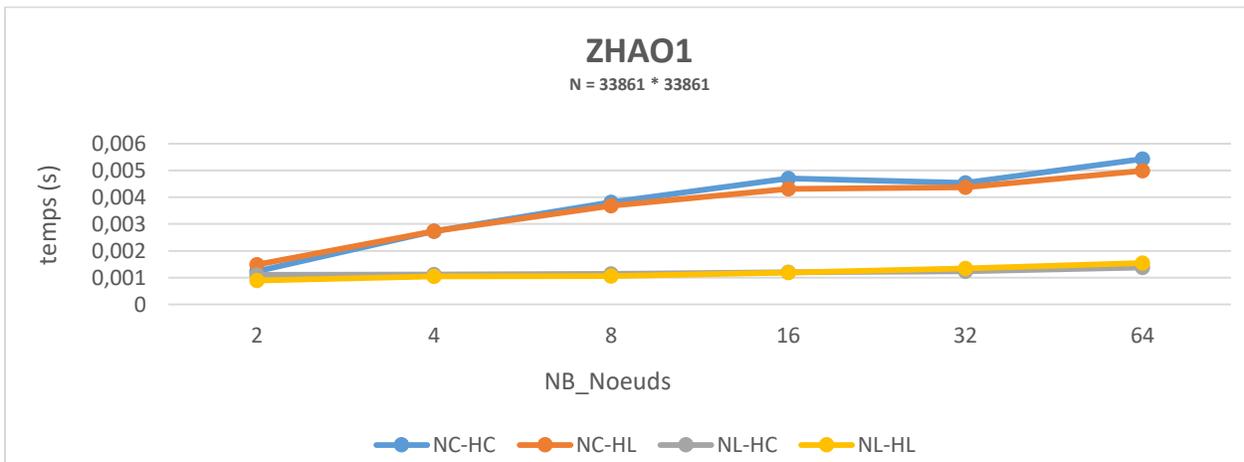

**Figure 0.47 :** Collecte des résultats et construction de Y pour la matrice «ZHAO1»

**Temps total de traitement :**

Pour les quatre combinaisons, nous comparons le temps total, c.-à-d., le temps nécessaire pour le calcul des PFVC auquel on ajoute le temps de construction du résultat finale pour les quatre combinaisons (voir Figure 4.50 jusqu'au 4.57). Dans 62% des cas la combinaison $NEZ_L$-$HYP_L$ puis $NEZ_L$-$HYP_C$ avec 19% et $NEZ_C$-$HYP_C$ avec 19% donnent le meilleur temps, ceci est dû principalement au temps minimal qu'elles assurent pour la collecte des données (les Gather) et la construction du résultat final.

Le Tableau 4.5 présente une synthèse des différents résultats obtenus lors de l'étude de la distribution du PMVC, sur une grappe multicoeurs, en utilisant les différentes combinaisons des méthodes NEZGT et hypergraphe pour la distribution inter et intra-nœud (les matrices étudiées sont celles décrites dans le tableau 4.2).

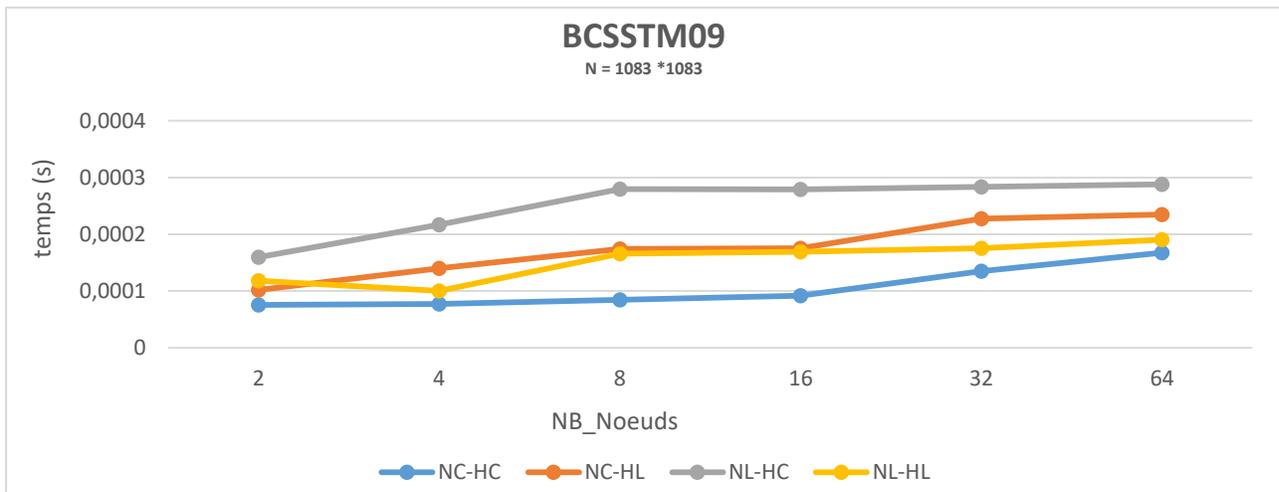

**Figure 0.48 : Temps nécessaire pour le calcul des PFVC et la construction du résultat final de la matrice «BCSSTM09»**

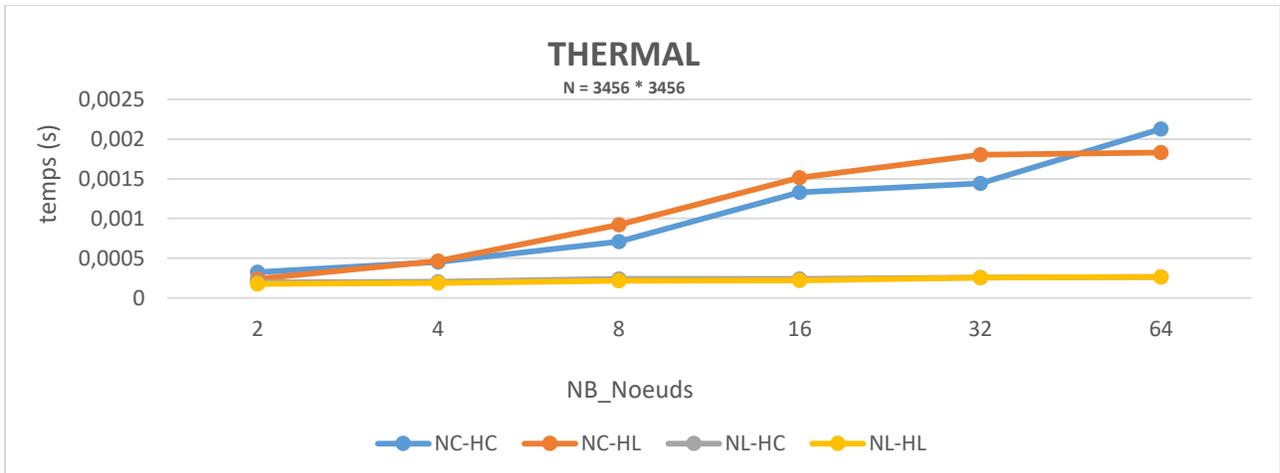

**Figure 0.49 : Temps nécessaire pour le calcul des PFVC et la construction du résultat final de la matrice «THERMAL»**

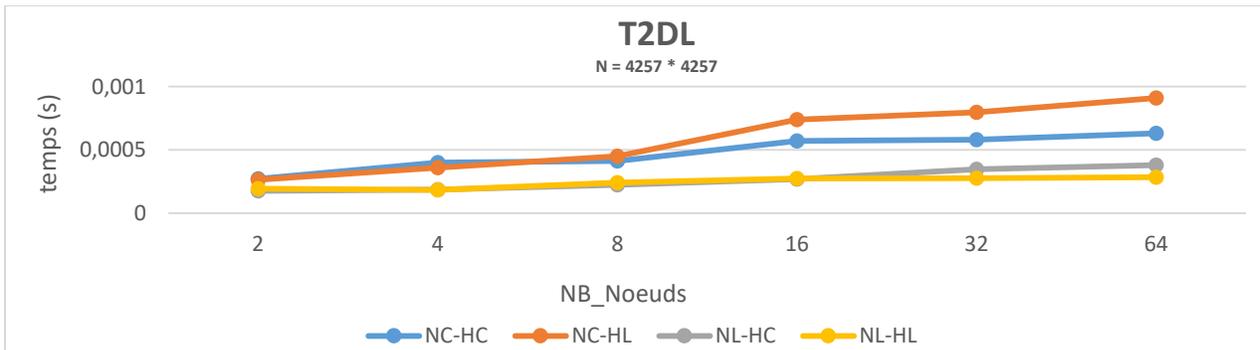

**Figure 00.50 : Temps nécessaire pour le calcul des PFVC et la construction du résultat final de la matrice «T2L»**

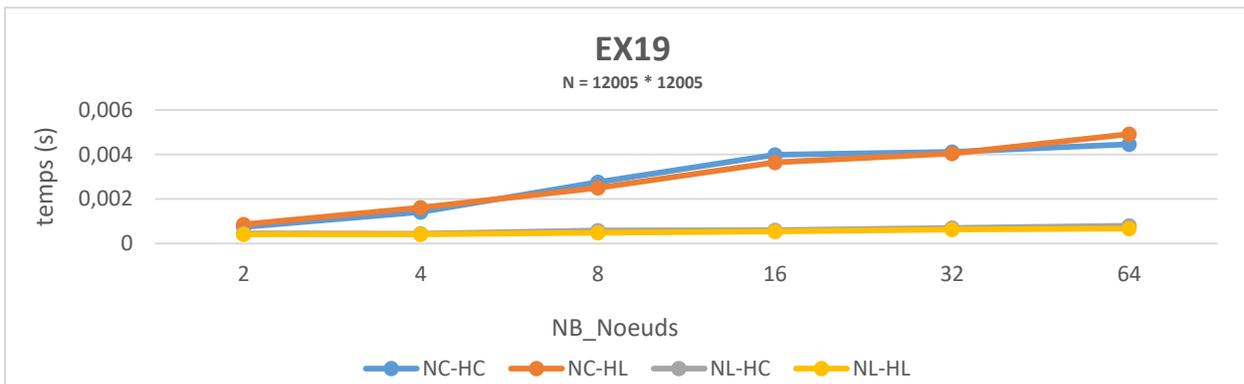

**Figure 00.51 : Temps nécessaire pour le calcul des PFVC et la construction du résultat final de la matrice «EX19»**

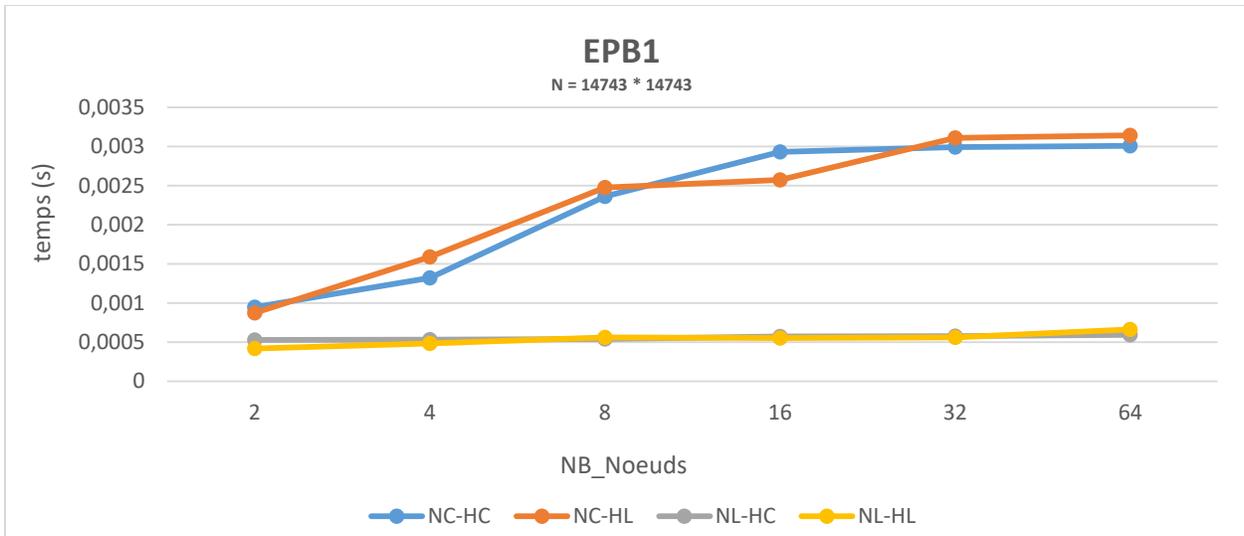

**Figure 00.52 : Temps nécessaire pour le calcul des PFVC et la construction du résultat final de la matrice «EPB1»**

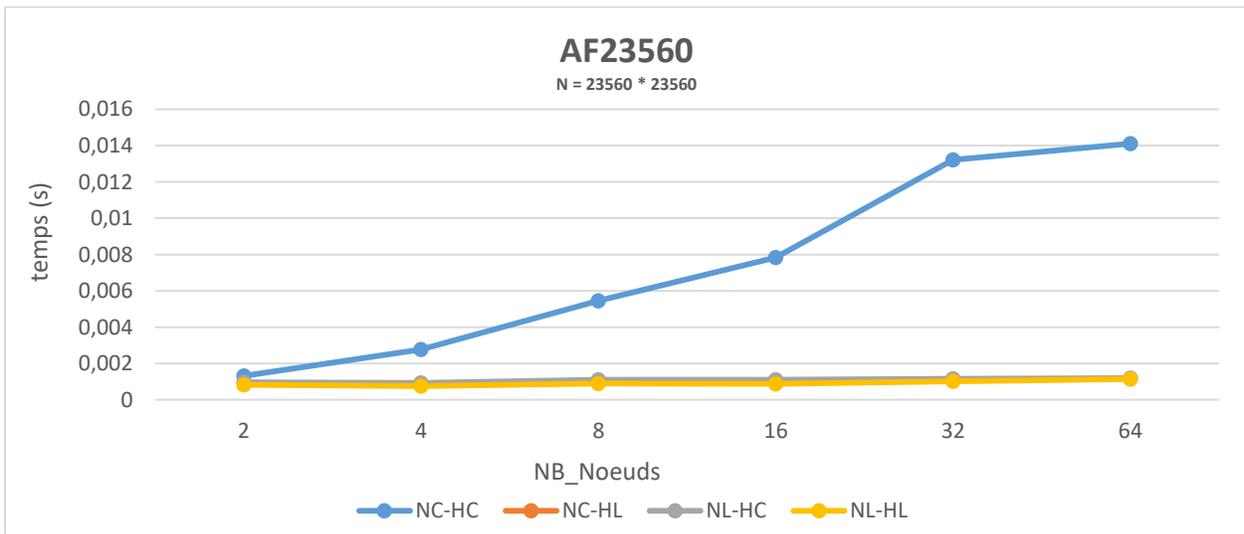

**Figure 00.53 : Temps nécessaire pour le calcul des PFVC et la construction du résultat final de la matrice «AF23560»**

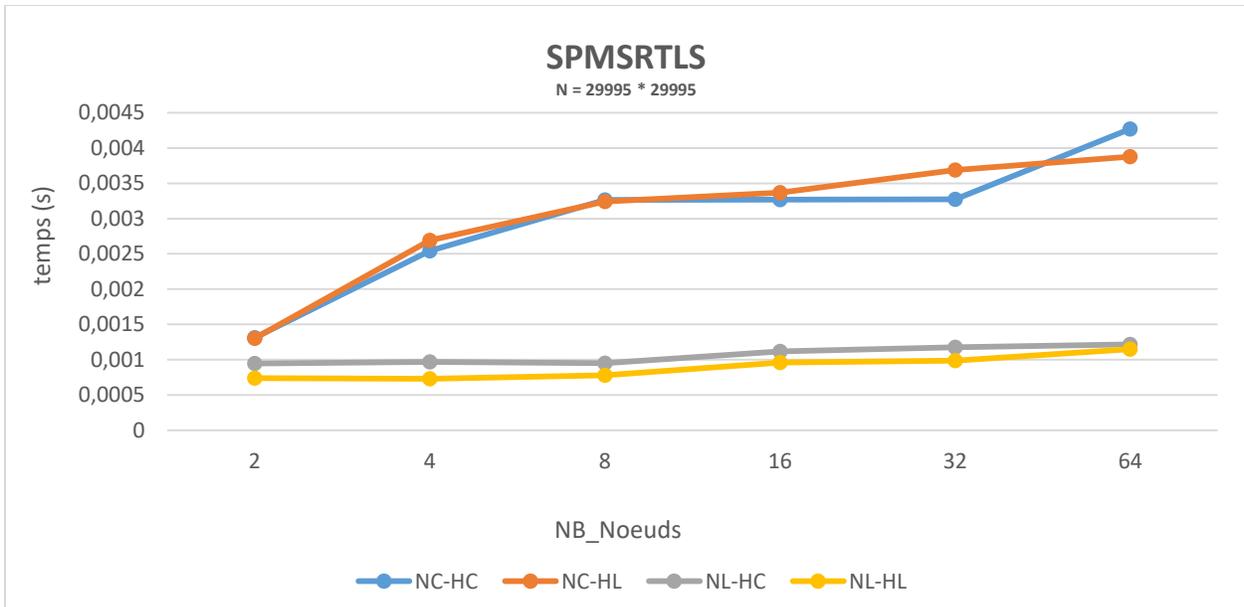

**Figure 00.54 : Temps nécessaire pour le calcul des PFVC et la construction du résultat final de la matrice «SPMSRTLS»**

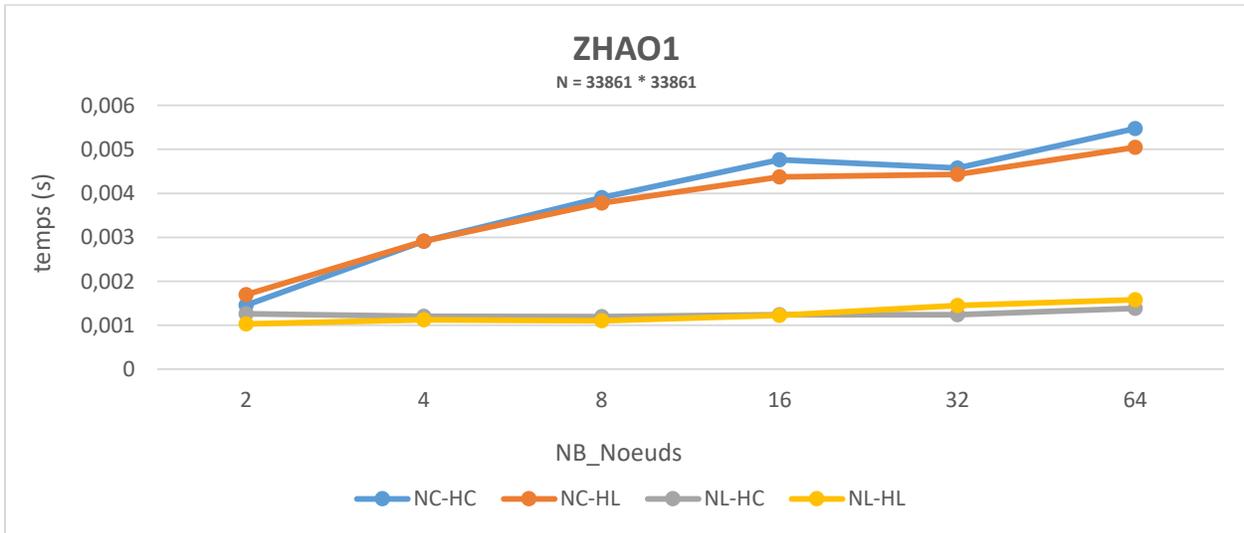

**Figure 00.55 : Temps nécessaire pour le calcul des PFVC et la construction du résultat final de la matrice «ZHAO1»**

# 5. Conclusion

Dans ce chapitre nous avons présenté une évaluation expérimentale de notre approche basée sur les algorithmes NEZGT$_{LIGNE}$, NEZGT$_{COLONNE}$ et hypergraphe pour la distribution du PMVC.

Notre plate-forme cible étant un cluster homogène de la grille de calcul GRID5000. Nous avons testé quatre combinaisons différentes pour la décomposition de la matrice creuse inter et intra-noeud. Nous avons constaté qu'adopter une décomposition NEZGT$_{LIGNE}$ inter-noeuds et HYPER$_{LIGNE}$ permet d'optimiser les communications nécessaires pour la distribution des données. Aussi elle ssure le meilleur équilibrage des charges et fournit un meilleur Makespan dans 62% des cas testés .

Le tableau suivant présente une synthèse comparative des meilleurs résultats obtenus lors de l'étude de la distribution du PMVC, sur une grappe multicoeurs, en utilisant les quatre combinaisons.

|  | NC-HC | NC-HL | NL-HC | NL-HL |
|---|---|---|---|---|
| *Scatter* | - | 13% | 38% | 49% |
| *Temp calcul de Y* | 18% | - | 20% | 62% |
| *Temp Construction de Y* | - | - | - | 100% |
| *Temp de Gather + Tmp Construction du Y* | 12% | - | - | 88% |
| *Temp Totale Traitement* | 19% | - | 19% | 62% |

**Tableau 0.70: Récapitulation des résultats obtenus**

# Conclusion Générale

De nombreuses applications en calcul scientifique, telles que la simulation de la mécanique des structures, l'étude de la dynamique des fluides, le traitement d'image/signal, se ramènent à la résolution de problèmes matriciels creux d'algèbre linaire. Il faut préciser que les deux problèmes fondamentaux en algèbre linaire sont la résolution de systèmes linéaires et le calcul d'éléments propres. Contrairement au second problème pour lequel on ne peut utiliser que les méthodes itératives, les méthodes de RSL peuvent être directes ou itératives. Sur des problèmes de très grande taille, les méthodes directes peuvent devenir prohibitives car elles sont trop gourmandes en termes de mémoire et de calcul.

Sur cette gamme de problème, les méthodes itératives constituent ainsi une alternative intéressante. Le traitement de matrices creuses de grandes tailles par des MI nécessite l'utilisation de superordinateurs de haute performance qui, souvent, peuvent ne pas être disponibles. Dans le cas de certaines applications scientifiques, les systèmes distribués, telle que les grappes de calcul, peuvent constituer une bonne alternative. De nos jours, avec l'avènement des architectures multi-cœurs, la performance d'un processeur devient liée principalement au nombre de cœurs qu'il renferme. D'où, au niveau des grappes de calcul, la tendance n'est plus uniquement d'augmenter le nombre de nœud, mais aussi d'accroitre le nombre de cœurs par processeur à l'intérieur de chaque nœud d'où on a choisi la Grid'5000 comme notre environnement d'expérimentations.

Notons que les méthodes itératives de RSL et de CEP reposent essentiellement sur le calcul du Produit Matrice-Vecteur Creux « PMVC ». Pour la distribution de ce noyau sur une architecture parallèle, on est confronté à deux problèmes principaux : l'équilibrage des charges entre les unités de calcul et l'optimisation des communications. Une corrélation entre ces deux critères fait que généralement l'optimisation de l'un peut dégrader l'autre. Ainsi, pour la distribution du PMVC sur une grappe de nœuds « paravance » du Grid'5000 multi-cœurs, nous avons proposé une solution basée sur la combinaison de deux approches différents pour la décomposition de la matrice creuse sur deux niveaux on a utilisé le NEZGT en version ligne et en version colonne sur la décomposition inter-nœud et l'Hypergraph avec la version 1D ligne et 1D colonne et on a proposé 4 combinaisons et les évaluer et nous avons conclu que la combinaisons $NEZ_L$-$HYP_L$ est la meilleur entre eux.

Notre perspective sera d'essayer de comparer ces combinaisons avec d'autres bibliothèques existantes afin de savoir leurs performances par rapport à ces bibliothèques.

# Bibliographie

# Netographie

**[CSN]** Comparaison de solveurs numériques

https://www.i2m.univ-amu.fr/~herard/Public/crasDFH.pdf , Dernière visite le 04/03/15

**[ALR]** Algèbre linéaire, matrice bandes

http://gilles.dubois10.free.fr/algebre_lineaire/bandes.html , Dernière visite le 07/02/15

**[CHE]** Architectures Parallèles , H.Cherroun

http://perso.lagh-univ.dz/~hcherroun/cours_html/architecture_parallele/Architecture_parallele_02_taxonomie.pdf

Dernière visite le 04/03/15

**[DRI]** Domino Research, processeurs multicoeurs
http://domino.research.ibm.com/comm/research_projects.nsf/pages/multicore.index.html ,
Dernière visite le 05/03/15

**[DTS]** DataSwift, La technologie myrinet

http://www.dataswift.fr/reseaux/myrinet/ , Dernière visite le 05/03/15.

**[ETC]** Exact corporation, AMD Opteron 6276 Box

http://exxactcorp.com/index.php/product/prod_detail/184 ,Dernière visite le 05/03/15

**[ITE]** Itespresso, tilera, précurseur des processeurs à forte densité de cœurs, séduit cisco et samsung

http://www.itespresso.fr/tilera-precurseur-des-processeurs-a-forte-densite-decoeurs-seduit-cisco-et-samsung-40545.html , Dernière visite le 05/03/15

**[LMI]** Le Monde Informatique, AMD opteron

# Annexe

# Annexe Calcul PMVC

# Combinaison NC-HC

La première phase consiste à faire une décomposition de la matrice sur 2 niveaux inter nœuds sur 2 fragments f0 et f1 avec l'heuristique $NEZGT_{colonne}$ et intra nœud sur les cœurs en 4 fragments avec la méthode $Hypergraph_{colonne}$.

|    | 0 | 1  | 2  | 3  | 4  | 5  | 6  | 7  | 8  | 9  | 10 | 11 | 12 | 13 | 14  |
|----|---|----|----|----|----|----|----|----|----|----|----|----|----|----|-----|
| 0  | 1 | .  | .  | 27 | .  | .  | .  | .  | .  | .  | .  | .  | .  | .  | .   |
| 1  | . | 10 | .  | .  | .  | .  | .  | .  | .  | .  | .  | .  | .  | .  | .   |
| 2  | 2 | .  | 18 | .  | 33 | .  | 49 | .  | .  | .  | .  | .  | .  | .  | .   |
| 3  | . | 11 | 19 | 28 | 34 | .  | 50 | 55 | .  | 64 | .  | 78 | 85 | .  | 97  |
| 4  | . | .  | 20 | 29 | .  | .  | .  | .  | .  | .  | 72 | .  | .  | .  | .   |
| 5  | . | .  | .  | .  | 35 | 42 | .  | .  | .  | .  | .  | 79 | .  | 93 | .   |
| 6  | 3 | 12 | 21 | .  | 36 | 43 | 51 | .  | .  | 65 | .  | .  | 86 | .  | .   |
| 7  | 4 | 13 | 22 | 30 | 37 | 44 | 52 | 56 | 59 | 66 | 73 | 80 | 87 | 94 | 98  |
| 8  | 5 | 14 | .  | .  | 38 | .  | 53 | .  | 60 | 67 | 74 | 81 | 88 | .  | 99  |
| 9  | 6 | 15 | 23 | .  | 39 | 45 | .  | 57 | 61 | 68 | 75 | 82 | 89 | .  | 100 |
| 10 | 7 | .  | 24 | .  | 40 | .  | .  | .  | .  | .  | 76 | .  | .  | 95 | 101 |
| 11 | . | 16 | .  | 31 | .  | 46 | .  | 58 | .  | 69 | .  | 83 | .  | .  | 102 |
| 12 | 8 | 17 | 25 | 32 | 41 | 47 | 54 | .  | 62 | 70 | .  | .  | 90 | 96 | 103 |
| 13 | . | .  | .  | .  | .  | .  | .  | .  | .  | .  | .  | .  | 91 | .  | .   |
| 14 | 9 | .  | 26 | .  | .  | 48 | .  | .  | 63 | 71 | 77 | 84 | 92 | .  | 104 |

**Matrice 15*15 & NNZ = 104**

La fragmentation dans la phase suivante sera une fragmentation en intra nœud il s'agit d'une décomposition en blocs de colonnes.

|    | 0 | 4  | 6  | 7  | 8  | 11 | 12 | 13 |
|----|---|----|----|----|----|----|----|----|
| 0  | 1 | .  | .  | .  | .  | .  | .  | .  |
| 1  | . | .  | .  | .  | .  | .  | .  | .  |
| 2  | 2 | 33 | 49 | .  | .  | .  | .  | .  |
| 3  | . | 34 | 50 | 55 | .  | 78 | 85 | .  |
| 4  | . | .  | .  | .  | .  | .  | .  | .  |
| 5  | . | 35 | .  | .  | .  | 79 | .  | 93 |
| 6  | 3 | 36 | 51 | .  | .  | .  | 86 | .  |
| 7  | 4 | 37 | 52 | 56 | 59 | 80 | 87 | 94 |
| 8  | 5 | 38 | 53 | .  | 60 | 81 | 88 | .  |
| 9  | 6 | 39 | .  | 57 | 61 | 82 | 89 | .  |
| 10 | 7 | 40 | .  | .  | .  | .  | .  | 95 |
| 11 | . | .  | .  | 58 | .  | 83 | .  | .  |
| 12 | 8 | 41 | 54 | .  | 62 | .  | 90 | 96 |
| 13 | . | .  | .  | .  | .  | .  | 91 | .  |
| 14 | 9 | .  | .  | .  | 63 | 84 | 92 | .  |

**Fragment 0**

Après cette décomposition en blocs de colonnes pour la matrice la phase suivante consiste à effectuer une décomposition en intra nœuds sur 4 cœurs en blocs de colonnes avec l'Hypergraph$_{colonne}$.

Fragment 0.0:

|    | 6  | 12 |
|----|----|----|
| 0  | .  | .  |
| 1  | .  | .  |
| 2  | 49 | .  |
| 3  | 50 | 85 |
| 4  | .  | .  |
| 5  | .  | .  |
| 6  | 51 | 86 |
| 7  | 52 | 87 |
| 8  | 53 | 88 |
| 9  | .  | 89 |
| 10 | .  | .  |
| 11 | .  | .  |
| 12 | 54 | 90 |
| 13 | .  | 91 |
| 14 | .  | 92 |

Fragment 0.1:

|    | 7  | 11 |
|----|----|----|
| 0  | .  | .  |
| 1  | .  | .  |
| 2  | .  | .  |
| 3  | 55 | 78 |
| 4  | .  | .  |
| 5  | .  | 79 |
| 6  | .  | .  |
| 7  | 56 | 80 |
| 8  | .  | 81 |
| 9  | 57 | 82 |
| 10 | .  | .  |
| 11 | 58 | 83 |
| 12 | .  | .  |
| 13 | .  | .  |
| 14 | .  | 84 |

Fragment 0.2:

|    | 0 | 8  |
|----|---|----|
| 0  | 1 | .  |
| 1  | . | .  |
| 2  | 2 | .  |
| 3  | . | .  |
| 4  | . | .  |
| 5  | . | .  |
| 6  | 3 | .  |
| 7  | 4 | 59 |
| 8  | 5 | 60 |
| 9  | 6 | 61 |
| 10 | 7 | .  |
| 11 | . | .  |
| 12 | 8 | 62 |
| 13 | . | .  |
| 14 | 9 | 63 |

Fragment 0.3:

|    | 4  | 13 |
|----|----|----|
| 0  | .  | .  |
| 1  | .  | .  |
| 2  | 33 | .  |
| 3  | 34 | .  |
| 4  | .  | .  |
| 5  | 35 | 93 |
| 6  | 36 | .  |
| 7  | 37 | 94 |
| 8  | 38 | .  |
| 9  | 39 | .  |
| 10 | 40 | 95 |
| 11 | .  | .  |
| 12 | 41 | 96 |
| 13 | .  | .  |
| 14 | .  | .  |

**Fragment 0.0**   **Fragment 0.1**   **Fragment 0.2**   **Fragment 0.3**

Après effectuer une décomposition sur les 2 niveaux inter et intra nœud, la phase suivante consiste à faire le PMVC pour chaque fragment 0.0 jusqu'au fragment 0.3.

**PMVC partiel Fragment 0.0**

$$\begin{bmatrix} & 6 & 12 \\ 2 & 49 & . \\ 3 & 50 & 85 \\ 6 & 51 & 86 \\ 7 & 52 & 87 \\ 8 & 53 & 88 \\ 9 & . & 89 \\ 12 & 54 & 90 \\ 13 & . & 91 \\ 14 & . & 92 \end{bmatrix} \times \begin{bmatrix} 1 \\ 1 \end{bmatrix} = \begin{bmatrix} Y\_Local\_p0 \\ 49 & 2 \\ 135 & 3 \\ 137 & 6 \\ 139 & 7 \\ 141 & 8 \\ 89 & 9 \\ 144 & 12 \\ 91 & 13 \\ 92 & 14 \end{bmatrix}$$

**PMVC partiel Fragment 0.1**

$$\begin{bmatrix} & 7 & 11 \\ 3 & 55 & 78 \\ 5 & . & 79 \\ 7 & 56 & 80 \\ 8 & . & 81 \\ 9 & 57 & 82 \\ 11 & 58 & 83 \\ 14 & . & 84 \end{bmatrix} \times \begin{bmatrix} 1 \\ 1 \end{bmatrix} = \begin{bmatrix} Y\_Local\_p1 \\ 133 & 3 \\ 79 & 5 \\ 136 & 7 \\ 81 & 8 \\ 139 & 9 \\ 141 & 11 \\ 84 & 14 \end{bmatrix}$$

**PMVC partiel Fragment 0.2**

$$\begin{bmatrix} & 0 & 8 \\ 0 & 1 & . \\ 2 & 2 & . \\ 6 & 3 & . \\ 7 & 4 & 59 \\ 8 & 5 & 60 \\ 9 & 6 & 61 \\ 10 & 7 & . \\ 12 & 8 & 62 \\ 14 & 9 & 63 \end{bmatrix} \times \begin{bmatrix} 1 \\ 1 \end{bmatrix} = \begin{bmatrix} Y\_Local\_p2 \\ 1 & 0 \\ 2 & 2 \\ 3 & 6 \\ 63 & 7 \\ 65 & 8 \\ 67 & 9 \\ 7 & 10 \\ 70 & 12 \\ 72 & 14 \end{bmatrix}$$

**PMVC partiel Fragment 0.3**

$$\begin{bmatrix} & 4 & 13 \\ 2 & 33 & . \\ 3 & 34 & . \\ 5 & 35 & 93 \\ 6 & 36 & . \\ 7 & 37 & 94 \\ 8 & 38 & . \\ 9 & 39 & . \\ 10 & 40 & 95 \\ 12 & 41 & 96 \end{bmatrix} \times \begin{bmatrix} 1 \\ 1 \end{bmatrix} = \begin{bmatrix} Y\_Local\_p3 \\ 33 & 2 \\ 34 & 3 \\ 128 & 5 \\ 36 & 6 \\ 131 & 7 \\ 38 & 8 \\ 39 & 9 \\ 135 & 10 \\ 137 & 12 \end{bmatrix}$$

Cette phase est une phase de collections des résultats en faisant la sommation des vecteurs pour avoir le vecteur issue du premier fragment f0.

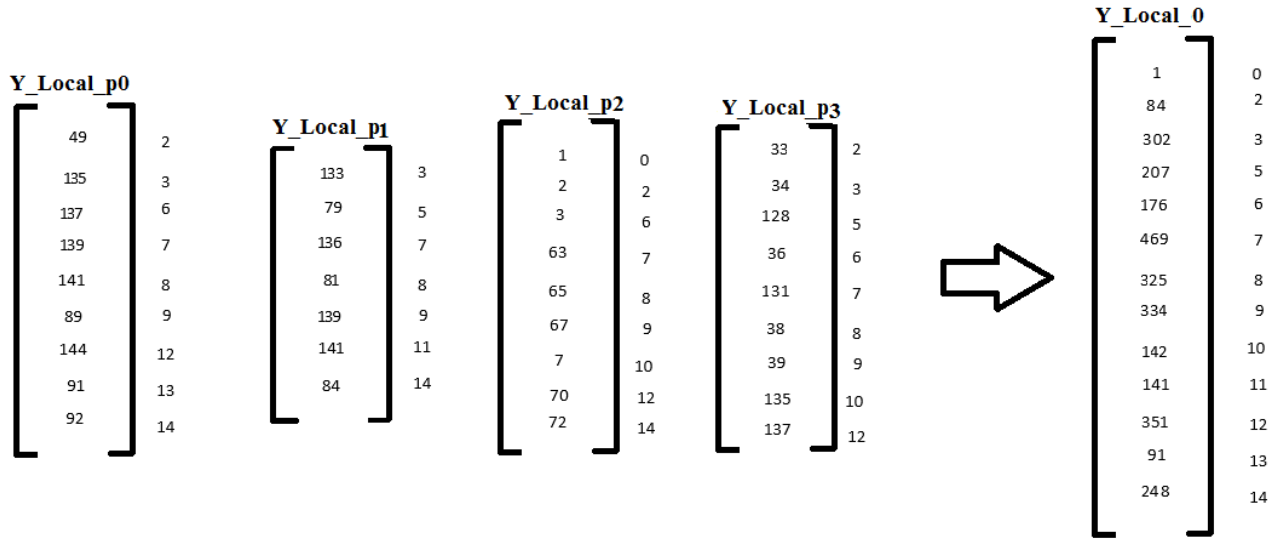

**Collection du résultat du fragment f0**

**Fragment 1**

$$\text{Fragment 1.0} \quad \begin{bmatrix} & 3 & 5 \\ 0 & 27 & . \\ 1 & . & . \\ 2 & . & . \\ 3 & 28 & . \\ 4 & 29 & . \\ 5 & . & 42 \\ 6 & . & 43 \\ 7 & 30 & 44 \\ 8 & . & . \\ 9 & . & 45 \\ 10 & . & . \\ 11 & 31 & 46 \\ 12 & 32 & 47 \\ 13 & . & . \\ 14 & . & 48 \end{bmatrix} \quad \text{Fragment 1.1} \quad \begin{bmatrix} & 10 & 14 \\ 0 & . & . \\ 1 & . & . \\ 2 & . & . \\ 3 & . & 97 \\ 4 & 72 & . \\ 5 & . & . \\ 6 & . & . \\ 7 & 73 & 98 \\ 8 & 74 & 99 \\ 9 & 75 & 100 \\ 10 & 76 & 101 \\ 11 & . & 102 \\ 12 & . & 103 \\ 13 & . & . \\ 14 & 77 & 104 \end{bmatrix} \quad \text{Fragment 1.2} \quad \begin{bmatrix} & 1 & 9 \\ 0 & . & . \\ 1 & 10 & . \\ 2 & . & . \\ 3 & 11 & 64 \\ 4 & . & . \\ 5 & . & . \\ 6 & 12 & 65 \\ 7 & 13 & 66 \\ 8 & 14 & 67 \\ 9 & 15 & 68 \\ 10 & . & . \\ 11 & 16 & 69 \\ 12 & 17 & 70 \\ 13 & . & . \\ 14 & . & 71 \end{bmatrix} \quad \text{Fragment 1.3} \quad \begin{bmatrix} & 2 \\ 0 & . \\ 1 & . \\ 2 & 18 \\ 3 & 19 \\ 4 & 20 \\ 5 & . \\ 6 & 21 \\ 7 & 22 \\ 8 & . \\ 9 & 23 \\ 10 & 24 \\ 11 & . \\ 12 & 25 \\ 13 & . \\ 14 & 26 \end{bmatrix}$$

**Fragment 1.0**  **Fragment 1.1**  **Fragment 1.2**  **Fragment 1.3**

PMVC pour chaque fragment 1.0 jusqu'au fragment 1.3

$$\begin{bmatrix} & 3 & 5 \\ 0 & 27 & . \\ 3 & 28 & . \\ 4 & 29 & . \\ 5 & . & 42 \\ 6 & . & 43 \\ 7 & 30 & 44 \\ 9 & . & 45 \\ 11 & 31 & 46 \\ 12 & 32 & 47 \\ 14 & . & 48 \end{bmatrix} \times \begin{bmatrix} 1 \\ 1 \end{bmatrix} = \begin{bmatrix} & Y\_Local\_p0 \\ 27 & 0 \\ 28 & 3 \\ 29 & 4 \\ 42 & 5 \\ 43 & 6 \\ 74 & 7 \\ 45 & 9 \\ 77 & 11 \\ 79 & 12 \\ 48 & 14 \end{bmatrix} \qquad \begin{bmatrix} & 10 & 14 \\ 3 & . & 97 \\ 4 & 72 & . \\ 7 & 73 & 98 \\ 8 & 74 & 99 \\ 9 & 75 & 100 \\ 10 & 76 & 101 \\ 11 & . & 102 \\ 12 & . & 103 \\ 14 & 77 & 104 \end{bmatrix} \times \begin{bmatrix} 1 \\ 1 \end{bmatrix} = \begin{bmatrix} & Y\_Local\_p1 \\ 97 & 3 \\ 72 & 4 \\ 171 & 7 \\ 173 & 8 \\ 175 & 9 \\ 177 & 10 \\ 102 & 11 \\ 103 & 12 \\ 181 & 14 \end{bmatrix}$$

**PMVC partiel Fragment 1.0**                                **PMVC partiel Fragment 1.1**

$$\begin{bmatrix} 1 & & 9 \\ 10 & & \cdot \\ 11 & & 64 \\ 12 & & 65 \\ 13 & & 66 \\ 14 & & 67 \\ 15 & & 68 \\ 16 & & 69 \\ 17 & & 70 \\ \cdot & & 71 \end{bmatrix} \begin{matrix} 1 \\ 3 \\ 6 \\ 7 \\ 8 \\ 9 \\ 11 \\ 12 \\ 14 \end{matrix} \times \begin{bmatrix} 1 \\ 1 \end{bmatrix} = \begin{bmatrix} 10 \\ 75 \\ 77 \\ 79 \\ 81 \\ 83 \\ 85 \\ 87 \\ 71 \end{bmatrix} \begin{matrix} Y\_Local\_p2 \\ 1 \\ 3 \\ 6 \\ 7 \\ 8 \\ 9 \\ 11 \\ 12 \\ 14 \end{matrix}$$

$$\begin{bmatrix} 2 \\ 18 \\ 19 \\ 20 \\ 21 \\ 22 \\ 23 \\ 24 \\ 25 \\ 26 \end{bmatrix} \begin{matrix} 2 \\ 3 \\ 4 \\ 6 \\ 7 \\ 9 \\ 10 \\ 12 \\ 14 \end{matrix} \times \begin{bmatrix} 1 \end{bmatrix} = \begin{bmatrix} 18 \\ 19 \\ 20 \\ 21 \\ 22 \\ 23 \\ 24 \\ 25 \\ 26 \end{bmatrix} \begin{matrix} Y\_Local\_p3 \\ 2 \\ 3 \\ 4 \\ 6 \\ 7 \\ 9 \\ 10 \\ 12 \\ 14 \end{matrix}$$

**PMVC partiel Fragment 1.2**  **PMVC partiel Fragment 1.3**

Collections des résultats en faisant la sommation des vecteurs pour avoir le vecteur issue du premier fragment f1.

Y_Local_p0: 27(0), 28(3), 29(4), 42(5), 43(6), 74(7), 45(9), 77(11), 79(12), 48(14)

Y_Local_p1: 97(3), 72(4), 171(7), 173(8), 175(9), 177(10), 102(11), 103(12), 181(14)

Y_Local_p2: 10(1), 75(3), 77(6), 79(7), 81(8), 83(9), 85(11), 87(12), 71(14)

Y_Local_p3: 18(2), 19(3), 20(4), 21(6), 22(7), 23(9), 24(10), 25(12), 26(14)

Y_Local_1: 27(0), 10(1), 18(2), 219(3), 121(4), 42(5), 141(6), 346(7), 254(8), 326(9), 201(10), 264(11), 114(12), 326(14)

**Collection des résultats du fragment f1**

Résultat Final :

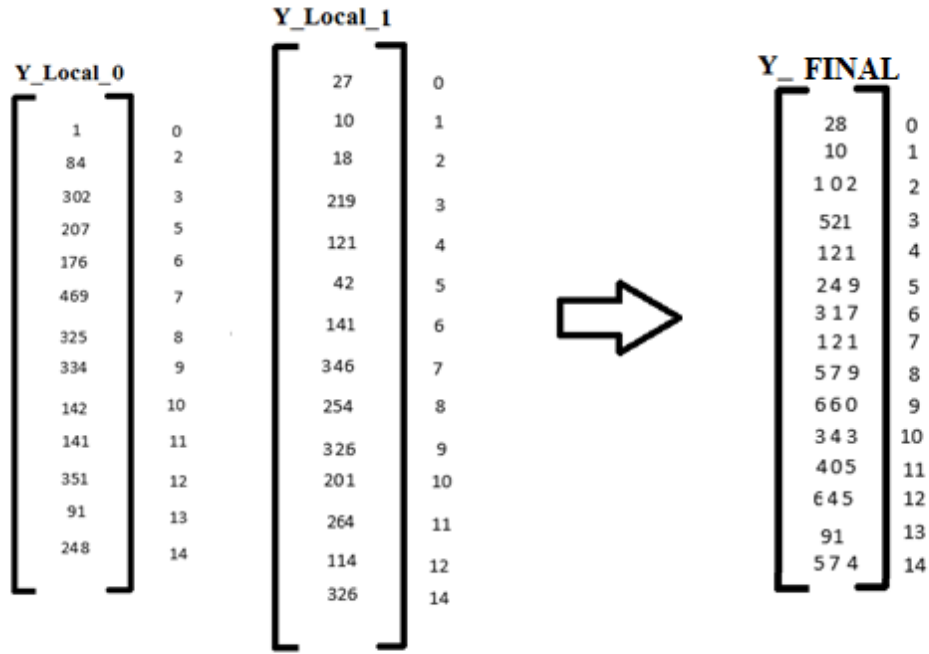

# **Combinaison NC-HL**

Cette phase consiste à décomposer la matrice sur 2 niveaux inter nœuds sur 2 fragments f0 et f1 avec l'heuristique NEZGT$_{colonne}$ et avec la même matrice et la même décomposition effectué dans la combinaison NC-HC en inter nœud et intra nœud sur les cœurs en 4 fragments en blocs de lignes avec la méthode Hypergraph$_{ligne}$

Les mêmes fragments f0 et f1 obtenu avec la combinaison NC-HC seront utilisés.

|    | 0 | 4  | 6  | 7  | 8  | 11 | 12 | 13 |
|----|---|----|----|----|----|----|----|----|
| 0  | 1 | .  | .  | .  | .  | .  | .  | .  |
| 9  | 6 | 39 | .  | 57 | 61 | 82 | 89 | .  |
| 11 | . | .  | .  | 58 | .  | 83 | .  | .  |
| 14 | 9 | .  | .  | .  | 63 | 84 | 92 | .  |

|   | 0 | 4  | 6  | 7  | 8  | 11 | 12 | 13 |
|---|---|----|----|----|----|----|----|----|
| 3 | . | 34 | 50 | 55 | .  | 78 | 85 | .  |
| 7 | 4 | 37 | 52 | 56 | 59 | 80 | 87 | 94 |

**Fragment 0.0**          **Fragment 0.1**

$$\begin{array}{c} & \begin{array}{ccccccc} 0 & 4 & 6 & 7 & 8 & 11 & 12 & 13 \end{array} \\ \begin{array}{c} 8 \\ 12 \\ 13 \end{array} & \left[ \begin{array}{cccccccc} 5 & 38 & 53 & . & 60 & 81 & 88 & . \\ 8 & 41 & 54 & . & 62 & . & 90 & 96 \\ . & . & . & . & . & . & 91 & . \end{array} \right] \end{array}$$

**Fragment 0.2**

$$\begin{array}{c} & \begin{array}{ccccccc} 0 & 4 & 6 & 7 & 8 & 11 & 12 & 13 \end{array} \\ \begin{array}{c} 2 \\ 5 \\ 6 \\ 10 \end{array} & \left[ \begin{array}{cccccccc} 2 & 33 & 49 & . & . & . & . & . \\ . & 35 & . & . & . & 79 & . & 93 \\ 3 & 36 & 51 & . & . & . & 86 & . \\ 7 & 40 & . & . & . & . & . & 95 \end{array} \right] \end{array}$$

**Fragment 0.3**

PMVC pour chaque fragment 0.0 jusqu'au fragment 0.3.

$$\begin{array}{c} & \begin{array}{ccccccc} 0 & 4 & 6 & 7 & 8 & 11 & 12 & 13 \end{array} \\ \begin{array}{c} 0 \\ 9 \\ 11 \\ 14 \end{array} & \left[ \begin{array}{cccccccc} 1 & . & . & . & . & . & . & . \\ 6 & 39 & . & 57 & 61 & 82 & 89 & . \\ . & . & . & 58 & . & 83 & . & . \\ 9 & . & . & . & 63 & 84 & 92 & . \end{array} \right] \end{array} \mathbf{X} \begin{bmatrix} 1 \\ 1 \\ 1 \\ 1 \\ 1 \\ 1 \\ 1 \\ 1 \\ 1 \\ 1 \\ 1 \\ 1 \\ 1 \\ 1 \\ 1 \end{bmatrix} = \begin{array}{c} Y\_Local\_p0 \\ \begin{bmatrix} 1 \\ 334 \\ 141 \\ 248 \end{bmatrix} \begin{array}{c} 0 \\ 9 \\ 11 \\ 14 \end{array} \end{array}$$

$$\begin{array}{c} & \begin{array}{ccccccc} 0 & 4 & 6 & 7 & 8 & 11 & 12 & 13 \end{array} \\ \begin{array}{c} 3 \\ 7 \end{array} & \left[ \begin{array}{cccccccc} . & 34 & 50 & 55 & . & 78 & 85 & . \\ 4 & 37 & 52 & 56 & 59 & 80 & 87 & 94 \end{array} \right] \end{array} \mathbf{X} \begin{bmatrix} 1 \\ 1 \\ 1 \\ 1 \\ 1 \\ 1 \\ 1 \\ 1 \\ 1 \\ 1 \\ 1 \\ 1 \\ 1 \\ 1 \\ 1 \end{bmatrix} = \begin{array}{c} Y\_Local\_p1 \\ \begin{bmatrix} 302 \\ 469 \end{bmatrix} \begin{array}{c} 3 \\ 7 \end{array} \end{array}$$

**PMVC partiel Fragment 0.0**         **PMVC partiel Fragment 0.1**

$$\begin{bmatrix} & 0 & 4 & 6 & 7 & 8 & 11 & 12 & 13 \\ 8 & 5 & 38 & 53 & . & 60 & 81 & 88 & . \\ 12 & 8 & 41 & 54 & . & 62 & . & 90 & 96 \\ 13 & . & . & . & . & . & . & 91 & . \end{bmatrix} \times \begin{bmatrix} 1\\1\\1\\1\\1\\1\\1\\1\\1\\1\\1\\1\\1\\1\\1 \end{bmatrix} = \begin{bmatrix} Y\_Local\_p2 \\ 325 \\ 351 \\ 91 \end{bmatrix} \begin{matrix} 8 \\ 12 \\ 13 \end{matrix}$$

$$\begin{bmatrix} & 0 & 4 & 6 & 7 & 8 & 11 & 12 & 13 \\ 2 & 2 & 33 & 49 & . & . & . & . & . \\ 5 & . & 35 & . & . & . & 79 & . & 93 \\ 6 & 3 & 36 & 51 & . & . & . & 86 & . \\ 10 & 7 & 40 & . & . & . & . & . & 95 \end{bmatrix} \times \begin{bmatrix} 1\\1\\1\\1\\1\\1\\1\\1\\1\\1\\1\\1\\1\\1\\1 \end{bmatrix} = \begin{bmatrix} Y\_Local\_p3 \\ 84 \\ 207 \\ 176 \\ 142 \end{bmatrix} \begin{matrix} 2 \\ 5 \\ 6 \\ 10 \end{matrix}$$

**PMVC partiel Fragment 0.2**   **PMVC partiel Fragment 0.3**

Collection des résultats en faisant la sommation des vecteurs pour avoir le vecteur issue du premier fragment f0.

$$Y\_Local\_p0 \begin{bmatrix} 1 \\ 334 \\ 141 \\ 248 \end{bmatrix} \begin{matrix} 0\\9\\11\\14 \end{matrix} \quad Y\_Local\_p1 \begin{bmatrix} 302 \\ 469 \end{bmatrix} \begin{matrix} 3\\7 \end{matrix} \quad Y\_Local\_p2 \begin{bmatrix} 325 \\ 351 \\ 91 \end{bmatrix} \begin{matrix} 8\\12\\13 \end{matrix} \quad Y\_Local\_p3 \begin{bmatrix} 84 \\ 207 \\ 176 \\ 142 \end{bmatrix} \begin{matrix} 2\\5\\6\\10 \end{matrix} \Rightarrow Y\_Local\_0 \begin{bmatrix} 1 \\ 84 \\ 302 \\ 207 \\ 176 \\ 469 \\ 325 \\ 334 \\ 142 \\ 141 \\ 351 \\ 91 \\ 248 \end{bmatrix} \begin{matrix} 0\\2\\3\\5\\6\\7\\8\\9\\10\\11\\12\\13\\14 \end{matrix}$$

**Collection des résultats du fragment f0**

PMVC pour chaque fragment 1.0 jusqu'au fragment 1.3.

$$\begin{bmatrix} & 1 & 2 & 3 & 5 & 9 & 10 & 14 \\ 0 & . & . & 27 & . & . & . & . \\ 3 & 11 & 19 & 28 & . & 64 & . & 97 \\ 12 & 17 & 25 & 32 & 47 & 70 & . & 103 \end{bmatrix} \qquad \begin{bmatrix} & 1 & 2 & 3 & 5 & 9 & 10 & 14 \\ 4 & . & 20 & 29 & . & . & 72 & . \\ 7 & 13 & 22 & 30 & 44 & 66 & 73 & 98 \\ 11 & 16 & . & 31 & 46 & 69 & . & 102 \end{bmatrix}$$

**Fragment 1.0**                                              **Fragment 1.1**

$$\text{Fragment 1.2} = \begin{bmatrix} & 1 & 2 & 3 & 5 & 9 & 10 & 14 \\ 1 & 10 & . & . & . & . & . & . \\ 2 & . & 18 & . & . & . & . & . \\ 5 & . & . & . & 42 & . & . & . \\ 6 & 12 & 21 & . & 43 & 65 & . & . \\ 14 & . & 26 & . & 48 & 71 & 77 & 104 \end{bmatrix}$$

**Fragment 1.2**

$$\text{Fragment 1.3} = \begin{bmatrix} & 1 & 2 & 3 & 5 & 9 & 10 & 14 \\ 8 & 14 & . & . & . & 67 & 74 & 99 \\ 9 & 15 & 23 & . & 45 & 68 & 75 & 100 \\ 10 & . & 24 & . & . & . & 76 & 101 \end{bmatrix}$$

**Fragment 1.3**

PMVC pour chaque fragment 1.0 jusqu'au fragment 1.3.

$$\begin{bmatrix} & 1 & 2 & 3 & 5 & 9 & 10 & 14 \\ 0 & . & . & 27 & . & . & . & . \\ 3 & 11 & 19 & 28 & . & 64 & . & 97 \\ 12 & 17 & 25 & 32 & 47 & 70 & . & 103 \end{bmatrix} \times \begin{bmatrix} 1 \\ 1 \\ 1 \\ 1 \\ 1 \\ 1 \\ 1 \\ 1 \\ 1 \\ 1 \\ 1 \\ 1 \\ 1 \\ 1 \end{bmatrix} = \begin{matrix} Y\_Local\_p0 \\ \begin{bmatrix} 27 \\ 219 \\ 264 \end{bmatrix} \begin{matrix} 0 \\ 3 \\ 12 \end{matrix} \end{matrix}$$

**PMVC partiel Fragment 1.0**

$$\begin{bmatrix} & 1 & 2 & 3 & 5 & 9 & 10 & 14 \\ 4 & . & 20 & 29 & . & . & 72 & . \\ 7 & 13 & 22 & 30 & 44 & 66 & 73 & 98 \\ 11 & 16 & . & 31 & 46 & 69 & . & 102 \end{bmatrix} \times \begin{bmatrix} 1 \\ 1 \\ 1 \\ 1 \\ 1 \\ 1 \\ 1 \\ 1 \\ 1 \\ 1 \\ 1 \\ 1 \\ 1 \\ 1 \end{bmatrix} = \begin{matrix} Y\_Local\_p1 \\ \begin{bmatrix} 121 \\ 346 \\ 264 \end{bmatrix} \begin{matrix} 4 \\ 7 \\ 11 \end{matrix} \end{matrix}$$

**PMVC partiel Fragment 1.1**

$$\begin{bmatrix} & 1 & 2 & 3 & 5 & 9 & 10 & 14 \\ 1 & 10 & . & . & . & . & . & . \\ 2 & . & 18 & . & . & . & . & . \\ 5 & . & . & . & 42 & . & . & . \\ 6 & 12 & 21 & . & 43 & 65 & . & . \\ 14 & . & 26 & . & 48 & 71 & 77 & 104 \end{bmatrix} \times \begin{bmatrix} 1 \\ 1 \\ 1 \\ 1 \\ 1 \\ 1 \\ 1 \\ 1 \\ 1 \\ 1 \\ 1 \\ 1 \\ 1 \\ 1 \end{bmatrix} = \begin{matrix} Y\_Local\_p2 \\ \begin{bmatrix} 10 \\ 18 \\ 42 \\ 141 \\ 326 \end{bmatrix} \begin{matrix} 1 \\ 2 \\ 5 \\ 6 \\ 14 \end{matrix} \end{matrix}$$

**PMVC partiel Fragment 1.2**

$$\begin{bmatrix} & 1 & 2 & 3 & 5 & 9 & 10 & 14 \\ 8 & 14 & . & . & . & 67 & 74 & 99 \\ 9 & 15 & 23 & . & 45 & 68 & 75 & 100 \\ 10 & . & 24 & . & . & . & 76 & 101 \end{bmatrix} \times \begin{bmatrix} 1 \\ 1 \\ 1 \\ 1 \\ 1 \\ 1 \\ 1 \\ 1 \\ 1 \\ 1 \\ 1 \\ 1 \\ 1 \\ 1 \end{bmatrix} = \begin{matrix} Y\_Local\_p3 \\ \begin{bmatrix} 254 \\ 326 \\ 201 \end{bmatrix} \begin{matrix} 8 \\ 9 \\ 10 \end{matrix} \end{matrix}$$

**PMVC partiel Fragment 1.3**

Collection des résultats en faisant la sommation des vecteurs pour avoir le vecteur issue du premier fragment f1.

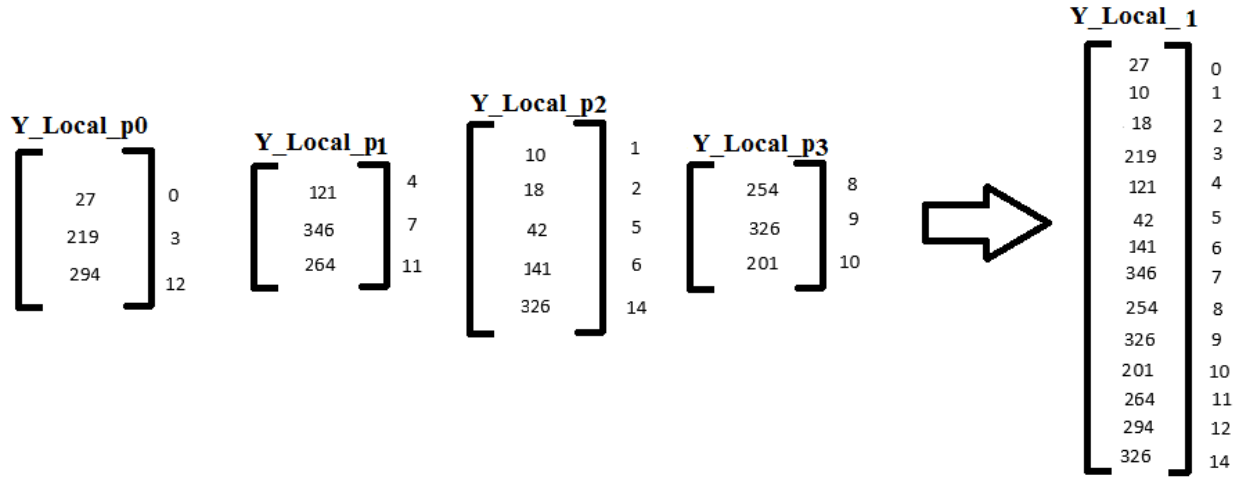

**Collection des résultats du fragment f1**

# Combinaison NL-HL

Cette phase consiste à décomposer la matrice sur 2 niveaux inter nœuds sur 2 fragments f0 et f1 avec l'heuristique NEZGT$_{ligne}$ et intra nœud sur les cœurs en 4 fragments avec la méthode Hypergraph$_{ligne}$.

|   | 0 | 1 | 2 | 3 | 4 | 5 | 6 | 7 | 8 | 9 | 10 | 11 | 12 | 13 | 14 |
|---|---|---|---|---|---|---|---|---|---|---|----|----|----|----|----|
| 0 | 1 | . | . | 27 | . | . | . | . | . | . | . | . | . | . | . |
| 2 | 2 | . | 18 | . | 33 | . | 49 | . | . | . | . | . | . | . | . |
| 3 | . | 11 | 19 | 28 | 34 | . | 50 | 55 | . | 64 | . | 78 | 85 | . | 97 |
| 5 | . | . | . | . | 35 | 42 | . | . | . | . | . | 79 | . | 93 | . |
| 6 | 3 | 12 | 21 | . | 36 | 43 | 51 | . | . | 65 | . | . | 86 | . | . |
| 7 | 4 | 13 | 22 | 30 | 37 | 44 | 52 | 56 | 59 | 66 | 73 | 80 | 87 | 94 | 98 |
| 14 | 9 | . | 26 | . | . | 48 | . | . | 63 | 71 | 77 | 84 | 92 | . | 104 |

**Fragment f0**

$$\begin{array}{c} \phantom{0}\phantom{0}\phantom{0}0\phantom{00}1\phantom{00}2\phantom{00}3\phantom{00}4\phantom{00}5\phantom{00}6\phantom{00}7\phantom{00}8\phantom{00}9\phantom{0}10\phantom{0}11\phantom{0}12\phantom{0}13\phantom{0}14 \\ \begin{array}{c} 0 \\ 3 \end{array}\left[\begin{array}{ccccccccccccccc} 1 & . & . & 27 & . & . & . & . & . & . & . & . & . & . & . \\ . & 11 & 19 & 28 & 34 & . & 50 & 55 & . & 64 & . & 78 & 85 & . & 97 \end{array}\right] \end{array}$$

**Fragment 0.0**

$$\begin{array}{c} \phantom{0}\phantom{0}\phantom{0}0\phantom{00}1\phantom{00}2\phantom{00}3\phantom{00}4\phantom{00}5\phantom{00}6\phantom{00}7\phantom{00}8\phantom{00}9\phantom{0}10\phantom{0}11\phantom{0}12\phantom{0}13\phantom{0}14 \\ 7\left[\begin{array}{ccccccccccccccc} 4 & 13 & 22 & 30 & 37 & 44 & 52 & 56 & 59 & 66 & 73 & 80 & 87 & 94 & 98 \end{array}\right] \end{array}$$

**Fragment 0.1**

$$\begin{array}{c} \phantom{0}\phantom{0}\phantom{0}0\phantom{00}1\phantom{00}2\phantom{00}3\phantom{00}4\phantom{00}5\phantom{00}6\phantom{00}7\phantom{00}8\phantom{00}9\phantom{0}10\phantom{0}11\phantom{0}12\phantom{0}13\phantom{0}14 \\ \begin{array}{c} 2 \\ 6 \end{array}\left[\begin{array}{ccccccccccccccc} 2 & . & 18 & . & 33 & . & 49 & . & . & . & . & . & . & . & . \\ 3 & 12 & 21 & . & 36 & 43 & 51 & . & . & 65 & . & . & 86 & . & . \end{array}\right] \end{array}$$

**Fragment 0.2**

$$\begin{array}{c} \phantom{0}\phantom{0}\phantom{0}0\phantom{00}1\phantom{00}2\phantom{00}3\phantom{00}4\phantom{00}5\phantom{00}6\phantom{00}7\phantom{00}8\phantom{00}9\phantom{0}10\phantom{0}11\phantom{0}12\phantom{0}13\phantom{0}14 \\ \begin{array}{c} 5 \\ 14 \end{array}\left[\begin{array}{ccccccccccccccc} . & . & . & . & 35 & 42 & . & . & . & . & . & 79 & . & 93 & . \\ 9 & . & 26 & . & . & 48 & . & . & 63 & 71 & 77 & 84 & 92 & . & 104 \end{array}\right] \end{array}$$

**Fragment 0.3**

$$\begin{array}{c} 0 \\ 3 \end{array}\left[\begin{array}{ccccccccccccccc} 1 & . & . & 27 & . & . & . & . & . & . & . & . & . & . & . \\ . & 11 & 19 & 28 & 34 & . & 50 & 55 & . & 64 & . & 78 & 85 & . & 97 \end{array}\right] \mathbf{X} \begin{bmatrix} 1 \\ 1 \\ 1 \\ 1 \\ 1 \\ 1 \\ 1 \\ 1 \\ 1 \\ 1 \\ 1 \\ 1 \\ 1 \\ 1 \\ 1 \end{bmatrix} = \begin{bmatrix} 28 \\ 521 \end{bmatrix} \begin{array}{c} 0 \\ 3 \end{array}$$

Y_Local_p0

**PMVC partiel 0.0**

$$7 \begin{bmatrix} 0 & 1 & 2 & 3 & 4 & 5 & 6 & 7 & 8 & 9 & 10 & 11 & 12 & 13 & 14 \\ 4 & 13 & 22 & 30 & 37 & 44 & 52 & 56 & 59 & 66 & 73 & 80 & 87 & 94 & 98 \end{bmatrix} \times \begin{bmatrix} 1 \\ 1 \\ 1 \\ 1 \\ 1 \\ 1 \\ 1 \\ 1 \\ 1 \\ 1 \\ 1 \\ 1 \\ 1 \\ 1 \\ 1 \\ 1 \\ 1 \\ 1 \end{bmatrix} = \begin{bmatrix} \text{Y\_Local\_p1} \\ 1\,2\,1 \end{bmatrix} 7$$

**PMVC partiel 0.1**

$$\begin{matrix} 2 \\ 6 \end{matrix} \begin{bmatrix} 0 & 1 & 2 & 3 & 4 & 5 & 6 & 7 & 8 & 9 & 10 & 11 & 12 & 13 & 14 \\ 2 & . & 18 & . & 33 & . & 49 & . & . & . & . & . & . & . & . \\ 3 & 12 & 21 & . & 36 & 43 & 51 & . & . & 65 & . & . & 86 & . & . \end{bmatrix} \times \begin{bmatrix} 1 \\ 1 \\ 1 \\ 1 \\ 1 \\ 1 \\ 1 \\ 1 \\ 1 \\ 1 \\ 1 \\ 1 \\ 1 \\ 1 \\ 1 \\ 1 \\ 1 \\ 1 \end{bmatrix} = \begin{bmatrix} \text{Y\_Local\_p2} \\ 1\,0\,2 \\ 3\,1\,7 \end{bmatrix} \begin{matrix} 2 \\ 6 \end{matrix}$$

**PMVC partiel 0.2**

$$\begin{matrix} 5 \\ 14 \end{matrix} \begin{bmatrix} 0 & 1 & 2 & 3 & 4 & 5 & 6 & 7 & 8 & 9 & 10 & 11 & 12 & 13 & 14 \\ . & . & . & . & 35 & 42 & . & . & . & . & . & 79 & . & 93 & . \\ 9 & . & 26 & . & . & 48 & . & 63 & 71 & 77 & 84 & 92 & . & 104 \end{bmatrix} \times \begin{bmatrix} 1 \\ 1 \\ 1 \\ 1 \\ 1 \\ 1 \\ 1 \\ 1 \\ 1 \\ 1 \\ 1 \\ 1 \\ 1 \\ 1 \\ 1 \\ 1 \\ 1 \\ 1 \end{bmatrix} = \begin{bmatrix} \text{Y\_Local\_p3} \\ 2\,4\,9 \\ 5\,7\,4 \end{bmatrix} \begin{matrix} 5 \\ 14 \end{matrix}$$

**PMVC partiel 0.3**

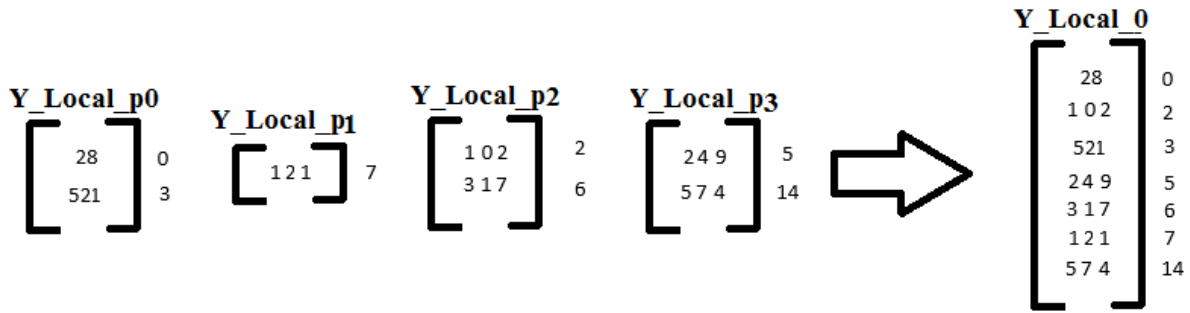

Collection des résultats du fragment f0

|    | 0 | 1  | 2  | 3  | 4  | 5  | 6  | 7  | 8  | 9  | 10 | 11 | 12 | 13 | 14  |
|----|---|----|----|----|----|----|----|----|----|----|----|----|----|----|-----|
| 1  | . | 10 | .  | .  | .  | .  | .  | .  | .  | .  | .  | .  | .  | .  | .   |
| 4  | . | .  | 20 | 29 | .  | .  | .  | .  | .  | .  | 72 | .  | .  | .  | .   |
| 8  | 5 | 14 | .  | .  | 38 | .  | 53 | .  | 60 | 67 | 74 | 81 | 88 | .  | 99  |
| 9  | 6 | 15 | 23 | .  | 39 | 45 | .  | 57 | 61 | 68 | 75 | 82 | 89 | .  | 100 |
| 10 | 7 | .  | 24 | .  | 40 | .  | .  | .  | .  | .  | 76 | .  | .  | 95 | 101 |
| 11 | . | 16 | .  | 31 | .  | 46 | .  | 58 | .  | 69 | .  | 83 | .  | .  | 102 |
| 12 | 8 | 17 | 25 | 32 | 41 | 47 | 54 | .  | 62 | 70 | .  | .  | 90 | 96 | 103 |
| 13 | . | .  | .  | .  | .  | .  | .  | .  | .  | .  | .  | .  | 91 | .  | .   |

Fragment 1

|    | 0 | 1  | 2  | 3 | 4  | 5  | 6 | 7  | 8  | 9  | 10 | 11 | 12 | 13 | 14  |
|----|---|----|----|---|----|----|---|----|----|----|----|----|----|----|-----|
| 9  | 6 | 15 | 23 | . | 39 | 45 | . | 57 | 61 | 68 | 75 | 82 | 89 | .  | 100 |
| 13 | . | .  | .  | . | .  | .  | . | .  | .  | .  | .  | .  | 91 | .  | .   |

Fragment 1.0

|    | 0 | 1  | 2  | 3  | 4 | 5  | 6 | 7  | 8 | 9  | 10 | 11 | 12 | 13 | 14  |
|----|---|----|----|----|---|----|---|----|---|----|----|----|----|----|-----|
| 1  | . | 10 | .  | .  | . | .  | . | .  | . | .  | .  | .  | .  | .  | .   |
| 4  | . | .  | 20 | 29 | . | .  | . | .  | . | .  | 72 | .  | .  | .  | .   |
| 11 | . | 16 | .  | 31 | . | 46 | . | 58 | . | 69 | .  | 83 | .  | .  | 102 |
| 13 | . | .  | .  | .  | . | .  | . | .  | . | .  | .  | .  | 91 | .  | .   |

Fragment 1.1

$$12 \begin{bmatrix} 0 & 1 & 2 & 3 & 4 & 5 & 6 & 7 & 8 & 9 & 10 & 11 & 12 & 13 & 14 \\ 8 & 17 & 25 & 32 & 41 & 47 & 54 & . & 62 & 70 & . & . & 90 & 96 & 103 \end{bmatrix}$$

**Fragment 1.2**

$$\begin{matrix} & 0 & 1 & 2 & 3 & 4 & 5 & 6 & 7 & 8 & 9 & 10 & 11 & 12 & 13 & 14 \\ 8 & 5 & 14 & . & . & 38 & . & 53 & . & 60 & 67 & 74 & 81 & 88 & . & 99 \\ 10 & 7 & . & 24 & . & 40 & . & . & . & . & . & 76 & . & . & 95 & 101 \end{matrix}$$

**Fragment 1.3**

$$9 \begin{bmatrix} 0 & 1 & 2 & 3 & 4 & 5 & 6 & 7 & 8 & 9 & 10 & 11 & 12 & 13 & 14 \\ 6 & 15 & 23 & . & 39 & 45 & . & 57 & 61 & 68 & 75 & 82 & 89 & . & 100 \end{bmatrix} \times \begin{bmatrix} 1 \\ 1 \\ 1 \\ 1 \\ 1 \\ 1 \\ 1 \\ 1 \\ 1 \\ 1 \\ 1 \\ 1 \\ 1 \\ 1 \\ 1 \end{bmatrix} = \begin{bmatrix} Y\_Local\_p0 \\ 660 \end{bmatrix}$$

$$\begin{matrix} & 0 & 1 & 2 & 3 & 4 & 5 & 6 & 7 & 8 & 9 & 10 & 11 & 12 & 13 & 14 \\ 1 & . & 10 & . & . & . & . & . & . & . & . & . & . & . & . & . \\ 4 & . & . & 20 & 29 & . & . & . & . & . & . & 72 & . & . & . & . \\ 11 & . & 16 & . & 31 & . & 46 & . & 58 & . & 69 & . & 83 & . & . & 102 \\ 13 & . & . & . & . & . & . & . & . & . & . & . & 91 & . & . & . \end{matrix} \times \begin{bmatrix} 1 \\ \vdots \\ 1 \end{bmatrix} = \begin{bmatrix} Y\_Local\_p1 \\ 10 \\ 121 \\ 405 \\ 91 \end{bmatrix} \begin{matrix} 1 \\ 4 \\ 11 \\ 13 \end{matrix}$$

**PMVC partiel Fragment 1.0**         **PMVC partiel Fragment 1.1**

$$12 \begin{bmatrix} 0 & 1 & 2 & 3 & 4 & 5 & 6 & 7 & 8 & 9 & 10 & 11 & 12 & 13 & 14 \\ 8 & 17 & 25 & 32 & 41 & 47 & 54 & . & 62 & 70 & . & . & 90 & 96 & 103 \end{bmatrix} \times \begin{bmatrix} 1 \\ \vdots \\ 1 \end{bmatrix} = \begin{bmatrix} Y\_Local\_p2 \\ 645 \end{bmatrix}$$

$$\begin{matrix} & 0 & 1 & 2 & 3 & 4 & 5 & 6 & 7 & 8 & 9 & 10 & 11 & 12 & 13 & 14 \\ 8 & 5 & 14 & . & . & 38 & . & 53 & . & 60 & 67 & 74 & 81 & 88 & . & 99 \\ 10 & 7 & . & 24 & . & 40 & . & . & . & . & . & 76 & . & . & 95 & 101 \end{matrix} \times \begin{bmatrix} 1 \\ \vdots \\ 1 \end{bmatrix} = \begin{bmatrix} Y\_Local\_p3 \\ 579 \\ 343 \end{bmatrix} \begin{matrix} 8 \\ 10 \end{matrix}$$

**PMVC partiel Fragment 1.2**         **PMVC partiel Fragment 1.3**

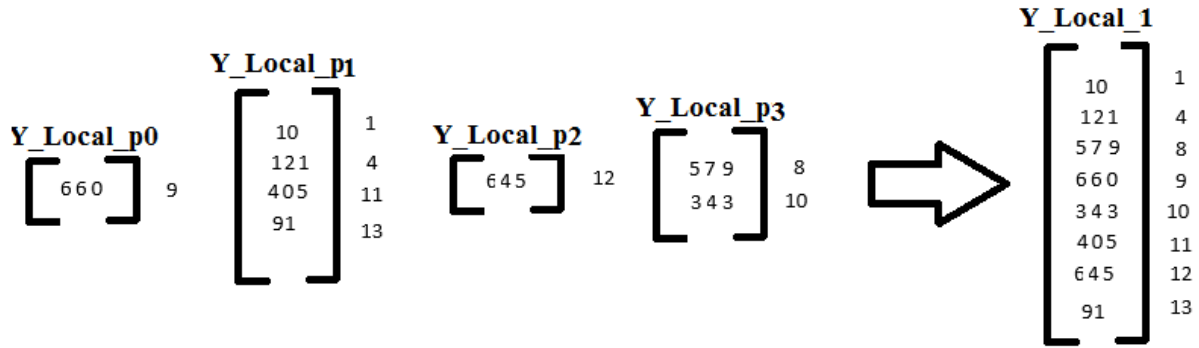

**Collection des résultats du fragment f1**

# Combinaison NL-HC

Cette phase consiste à décomposer la matrice sur 2 niveaux inter nœuds sur 2 fragments f0 et f1 avec l'heuristique $NEZGT_{ligne}$ avec la même matrice et la même décomposition effectué dans la combinaison NL-HL en inter nœud et en intra nœud sur les cœurs en 4 fragments en blocs de colonnes avec la méthode $Hypergraph_{colonne}$

Les mêmes fragments f0 et f1 obtenu avec la combinaison NL-HL seront utilisés dans cette combinaison.

|    | 0 | 1  | 2  | 3  | 4  | 5  | 6  | 7  | 8  | 9  | 10 | 11 | 12 | 13 | 14 |
|----|---|----|----|----|----|----|----|----|----|----|----|----|----|----|-----|
| 0  | 1 | .  | .  | 27 | .  | .  | .  | .  | .  | .  | .  | .  | .  | .  | .   |
| 2  | 2 | .  | 18 | .  | 33 | .  | 49 | .  | .  | .  | .  | .  | .  | .  | .   |
| 3  | . | 11 | 19 | 28 | 34 | .  | 50 | 55 | .  | 64 | .  | 78 | 85 | .  | 97  |
| 5  | . | .  | .  | .  | 35 | 42 | .  | .  | .  | .  | .  | 79 | .  | 93 | .   |
| 6  | 3 | 12 | 21 | .  | 36 | 43 | 51 | .  | .  | 65 | .  | .  | 86 | .  | .   |
| 7  | 4 | 13 | 22 | 30 | 37 | 44 | 52 | 56 | 59 | 66 | 73 | 80 | 87 | 94 | 98  |
| 14 | 9 | .  | 26 | .  | .  | 48 | .  | .  | 63 | 71 | 77 | 84 | 92 | .  | 104 |

**Fragment 0**

$$\text{Fragment 0.0} \quad \begin{bmatrix} & 0 & 3 & 4 \\ 0 & 1 & 27 & . \\ 2 & 2 & . & 33 \\ 3 & . & 28 & 34 \\ 5 & . & . & 35 \\ 6 & 3 & . & 36 \\ 7 & 4 & 30 & 37 \\ 14 & 9 & . & . \end{bmatrix}$$

$$\text{Fragment 0.1} \quad \begin{bmatrix} & 1 & 5 & 11 & 13 \\ 3 & 11 & . & 78 & . \\ 5 & . & 42 & 79 & 93 \\ 6 & 12 & 43 & . & . \\ 7 & 13 & 44 & 80 & 94 \\ 14 & . & 48 & 84 & . \end{bmatrix}$$

$$\text{Fragment 0.2} \quad \begin{bmatrix} & 2 & 6 & 7 & 8 \\ 2 & 18 & 49 & . & . \\ 3 & 19 & 50 & 55 & . \\ 6 & 21 & 51 & . & . \\ 7 & 22 & 52 & 56 & 59 \\ 14 & 26 & . & . & 63 \end{bmatrix}$$

$$\text{Fragment 0.3} \quad \begin{bmatrix} & 9 & 10 & 12 & 14 \\ 3 & 64 & . & 85 & 97 \\ 6 & 65 & . & 86 & . \\ 7 & 66 & 73 & 87 & 98 \\ 14 & 71 & 77 & 92 & 104 \end{bmatrix}$$

**PMVC partiel 0.0**

$$\begin{bmatrix} 1 & 27 & . \\ 2 & . & 33 \\ . & 28 & 34 \\ . & . & 35 \\ 3 & . & 36 \\ 4 & 30 & 37 \\ 9 & . & . \end{bmatrix} \times \begin{bmatrix} 1 \\ 1 \\ 1 \end{bmatrix} = \begin{bmatrix} 28 \\ 35 \\ 62 \\ 35 \\ 39 \\ 71 \\ 9 \end{bmatrix} \begin{matrix} 0 \\ 2 \\ 3 \\ 5 \\ 6 \\ 7 \\ 14 \end{matrix} \; Y\_Local\_p0$$

**PMVC partiel 0.1**

$$\begin{bmatrix} 11 & . & 78 & . \\ . & 42 & 79 & 93 \\ 12 & 43 & . & . \\ 13 & 44 & 80 & 94 \\ . & 48 & 84 & . \end{bmatrix} \times \begin{bmatrix} 1 \\ 1 \\ 1 \\ 1 \end{bmatrix} = \begin{bmatrix} 89 \\ 214 \\ 55 \\ 231 \\ 132 \end{bmatrix} \begin{matrix} 3 \\ 5 \\ 6 \\ 7 \\ 14 \end{matrix} \; Y\_Local\_p1$$

**PMVC partiel 0.2**

$$\begin{bmatrix} 18 & 49 & . & . \\ 19 & 50 & 55 & . \\ 21 & 51 & . & . \\ 22 & 52 & 56 & 59 \\ 26 & . & . & 63 \end{bmatrix} \times \begin{bmatrix} 1 \\ 1 \\ 1 \\ 1 \end{bmatrix} = \begin{bmatrix} 67 \\ 124 \\ 72 \\ 189 \\ 89 \end{bmatrix} \begin{matrix} 2 \\ 3 \\ 6 \\ 7 \\ 14 \end{matrix} \; Y\_Local\_p2$$

**PMVC partiel 0.3**

$$\begin{bmatrix} 64 & . & 85 & 97 \\ 65 & . & 86 & . \\ 66 & 73 & 87 & 98 \\ 71 & 77 & 92 & 104 \end{bmatrix} \times \begin{bmatrix} 1 \\ 1 \\ 1 \\ 1 \end{bmatrix} = \begin{bmatrix} 246 \\ 151 \\ 324 \\ 344 \end{bmatrix} \begin{matrix} 3 \\ 6 \\ 7 \\ 14 \end{matrix} \; Y\_Local\_p3$$

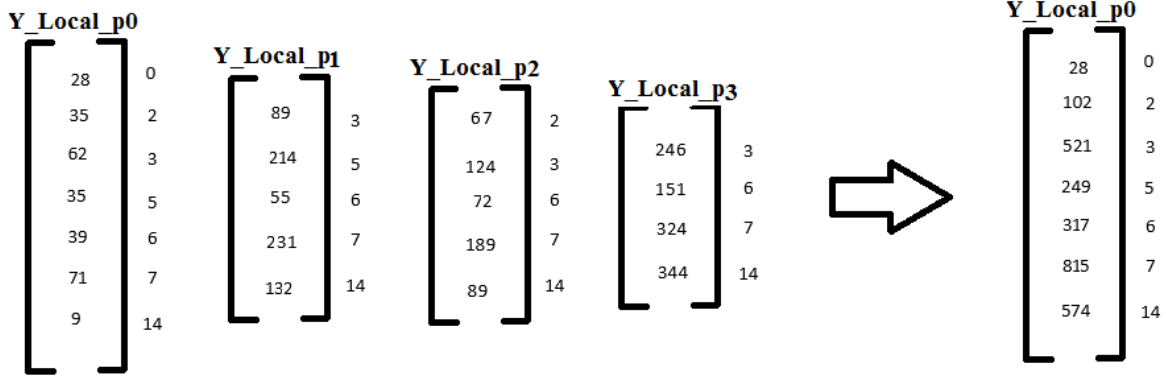

Collection des résultats du fragment f0

|  | 0 | 1 | 2 | 3 | 4 | 5 | 6 | 7 | 8 | 9 | 10 | 11 | 12 | 13 | 14 |
|---|---|---|---|---|---|---|---|---|---|---|---|---|---|---|---|
| 1 | . | 10 | . | . | . | . | . | . | . | . | . | . | . | . | . |
| 4 | . | . | 20 | 29 | . | . | . | . | . | . | 72 | . | . | . | . |
| 8 | 5 | 14 | . | . | 38 | . | 53 | . | 60 | 67 | 74 | 81 | 88 | . | 99 |
| 9 | 6 | 15 | 23 | . | 39 | 45 | . | 57 | 61 | 68 | 75 | 82 | 89 | . | 100 |
| 10 | 7 | . | 24 | . | 40 | . | . | . | . | . | 76 | . | . | 95 | 101 |
| 11 | . | 16 | . | 31 | . | 46 | . | 58 | . | 69 | . | 83 | . | . | 102 |
| 12 | 8 | 17 | 25 | 32 | 41 | 47 | 54 | . | 62 | 70 | . | . | 90 | 96 | 103 |
| 13 | . | . | . | . | . | . | . | . | . | . | . | . | 91 | . | . |

Fragment 1

|  | 1 | 2 | 3 | 5 | 9 | 10 | 14 |
|---|---|---|---|---|---|---|---|
| 0 | . | . | 27 | . | . | . | . |
| 3 | 11 | 19 | 28 | . | 64 | . | 97 |
| 12 | 17 | 25 | 32 | 47 | 70 | . | 103 |

|  | 1 | 2 | 3 | 5 | 9 | 10 | 14 |
|---|---|---|---|---|---|---|---|
| 4 | . | 20 | 29 | . | . | 72 | . |
| 7 | 13 | 22 | 30 | 44 | 66 | 73 | 98 |
| 11 | 16 | . | 31 | 46 | 69 | . | 102 |

Fragment 1.0                                                                                       Fragment 1.1

$$\text{Fragment 1.2} \quad \begin{bmatrix} & 1 & 2 & 3 & 5 & 9 & 10 & 14 \\ 1 & 10 & . & . & . & . & . & . \\ 2 & . & 18 & . & . & . & . & . \\ 5 & . & . & . & 42 & . & . & . \\ 6 & 12 & 21 & . & 43 & 65 & . & . \\ 14 & . & 26 & . & 48 & 71 & 77 & 104 \end{bmatrix}$$

**Fragment 1.2**

$$\text{Fragment 1.3} \quad \begin{bmatrix} & 1 & 2 & 3 & 5 & 9 & 10 & 14 \\ 8 & 14 & . & . & . & 67 & 74 & 99 \\ 9 & 15 & 23 & . & 45 & 68 & 75 & 100 \\ 10 & . & 24 & . & . & . & 76 & 101 \end{bmatrix}$$

**Fragment 1.3**

**PMVC partiel fragment 1.0**

**PMVC partiel fragment 1.1**

**PMVC partiel fragment 1.2**

**PMVC partiel Fragment 1.3**

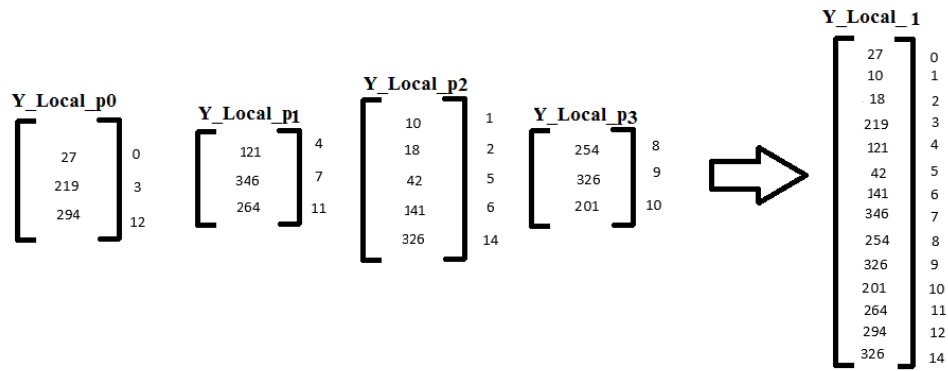

**Collection des résultats du fragment f1**

# Exploitation du Grid5000

Comment connecter et exploiter le cluster du test ?

On connecte à la Grid5000 à partir du terminal en utilisant la syntaxe suivant :

ssh [login@access.site.grid5000.fr](login@access.site.grid5000.fr)

Maintenant on se connecte à un site particulier en utilisant cette commande :

ssh nom_site

Pour utiliser la Grid on peut soit faire une réservation d'avance en tapant :

**oarsub** -r '*date heure*' –l nodes=*nb-noeuds*, walltime=*durée* -p 'cluster="*nom_cluster*"' -t deploy

Un environnement dans Kadeploy 3 est un ensemble de fichier décrivant un système d'exploitation entièrement fonctionnel.

Donc pour déployer cet environnement on utilise la commande suivante :

**kadeploy3** -e *nom_env* -f $OAR_FILE_NODES –k

Remarque : $OAR_FILE_NODES contient la liste des nœuds réservés.

Afin de lancer un script on utilise la commande :

sh script0.sh

**Génération Clé ssh**

SSH (Secure Shell) est un protocole de réseau et des applications généralement utilisé pour accéder à un compte shell sur une machine distante. Il est le principal outil utilisé pour accéder au banc d'essai Grille 5000.

Lors de la connexion à un ordinateur distant, le moyen standard pour authentifier est via le login et mot de passe du compte distant. Toutefois, ce schéma d'authentification est désactivé sur Grille 5000 pour des raisons de sécurité (comme il est souvent soumis à des attaques bruteforcer). Au lieu de cela, les utilisateurs doivent utiliser Grid'5000 authentification par clé publique, comme suit:

Vous (l'utilisateur) de générer une paire de clés en utilisant ssh-keygen. La clé publique peut alors être utilisée pour chiffrer les messages que seule la clé privée pourra déchiffrer.

ssh-keygen -t dsa

*Création d'un nouvel environnement*

*Installation des bibliothèques NUMA*

root@paravance-5:~# sudo apt-get install numactl

root@paravance-5:~# sudo apt-get install libnuma-dev

## *Exemple réservation*

user@frennes:~$ oarsub -r '2015-11-20 22:00:00' –l nodes=64, walltime=8:00:00 -p
'cluster="paravance"' -t deploy

⇨ Réserver 64 nœuds le 20 Novembre 2015 du site paravance, pour une période 8 heures à partir de 22h

user@frennes:~$ kadeploy3 -e wheezy-x64-NFS -f $OAR_FILE_NODES –k

⇨ Déploiement de l'environnement crée sur les nœuds réservés

user@frennes-3:~$ cd Work/sfe/